\documentclass[11pt]{article}

\usepackage{amsfonts}
\usepackage{amssymb}
\usepackage{amsmath}
\usepackage[margin=1in]{geometry}
\usepackage{fancyhdr}
\usepackage{graphicx}
\usepackage{epsf}
\usepackage{epsfig}
\usepackage[bottom]{footmisc}
\usepackage{natbib}
\usepackage{setspace}
\usepackage{rotating}
\usepackage{threeparttable}
\usepackage{booktabs}
\usepackage{caption}
\usepackage{array}
\usepackage{float}
\usepackage{forest}
\usepackage{tikz}
\usepackage{pgfkeys}
\usepackage{comment}
\usepackage{xcolor}
\usepackage{multirow}
\usepackage{mathtools}
\usepackage{pdflscape}
\usepackage{subcaption}

\usepackage[colorlinks, urlcolor=blue, linkcolor=blue,citecolor=blue,backref=page]{hyperref}

\usepackage{esint}
\usepackage{amstext}
\usepackage{graphics,xspace,endnotes}
\usepackage{lineno}
\usepackage{amsthm}
\usepackage{physics}
\usepackage{verbatim}
\usepackage{algorithm}
\usepackage{algorithmic}
\usepackage[utf8]{inputenc} 
\usepackage[T1]{fontenc} 
\usepackage{pgfplots}
\pgfplotsset{compat=1.18}

\newcommand{\YY}{\mathcal{Y}}
\newcommand{\Wone}{W_1}

\theoremstyle{plain}
\newtheorem{prop}{Proposition}
\numberwithin{equation}{section}
\newtheorem{assumption}{Assumption}
\newtheorem{theorem}{Theorem}
\newtheorem{lemma}{Lemma}

\newtheorem{corollary}{Corollary}
\theoremstyle{definition} 
\newtheorem{definition}{Definition}

\newcommand{\assinterp}[1]{\par\smallskip{\normalfont #1}\par\smallskip}

\DeclareMathOperator{\diam}{diam}

\usetikzlibrary{patterns}

\setcitestyle{sort&compress}

\title{Bounds on inequality with incomplete data}

\author{James Banks\thanks{Department of Economics, University of Manchester.\ james.banks@manchester.ac.uk.}  \and
Thomas Glinnan\thanks{Department of Economics, LSE.  t.m.glinnan@lse.ac.uk.}
\and 
Tatiana Komarova\thanks{Corresponding author. Faculty of Economics, University of Cambridge.\ tk670@cam.ac.uk.}}

\date{\today \\ \vspace{.5cm}}
\begin{document}
\maketitle

\singlespacing

\begin{abstract}
We study inequality measures when outcomes are observed only in intervals, as in historical tabulations, privacy-protected grouped data, and modern surveys. We develop a nonparametric framework for sharp identification and inference with grouped and interval-valued data, covering brackets and overlapping intervals. For a class of inequality indices, sharp bounds are attained by discrete distributions with finite support, reducing the problem to optimization; linear-fractional indices, including the Gini and quantile ratios, yield linear or quadratic programs. Plug-in bound endpoints have a $\sqrt{n}$ asymptotic distribution, using an $m$-out-of-$n$ bootstrap. Applications to wealth and historical income data compare identified sets with imputation-based estimates.

\vspace{0.05in}

\begin{description}
\item[Keywords:] Inequality indices, Incomplete data, Grouped data, Interval-valued data, Partial identification, Linear-fractional programming, Wealth inequality, Historic income inequality.

\end{description}
\end{abstract}

\newpage 

\onehalfspacing

\section{Introduction}

Measuring economic inequality is central to empirical work, yet in many applications the underlying outcomes are only partially observed. Information on historical income distributions survives only as published frequency tables. Statistical surveys and administrative datasets protect respondent privacy by releasing grouped records. Household surveys reduce item nonresponse and respondent burden by using bracketed response options or unfolding brackets, which place each value in a respondent-specific interval rather than recording it exactly. In all these settings standard inequality indices are partially identified, which means that the available data are consistent with a set of values for the index, not a single point.

The conventional response in empirical practice is to impute. Researchers assign midpoints, bracket means, hot-deck or parametric draws to interval observations and proceed as if the resulting dataset were complete. Imputation is convenient, but any imputation rule necessarily takes a stand on where, within each observed interval, the unobserved outcome lies. This is a stand that the interval data alone cannot verify. The imputation approach conflates two distinct sources of uncertainty. Sampling uncertainty, which shrinks with the square root of sample size, is conceptually separate from identification uncertainty, which captures the irreducible ambiguity about within-interval location and does not diminish as more observations are collected. By collapsing both into a single point estimate and its standard error, imputation-based methods produce a point estimate whose apparent precision reflects the imputation rule chosen, not the data alone. The resulting confidence intervals reflect only sampling variation around the chosen imputation rule and say nothing about how much the answer would change if a different rule, equally consistent with the data, had been used instead.

This paper develops a nonparametric framework that delivers upper and lower bounds for a broad class of inequality indices when outcomes are observed  through interval restrictions, possibly combined with auxiliary restrictions such as subgroup means, income shares, or points on the Lorenz curve. These bounds are sharp as every value in the reported set is generated by some distribution consistent with the observed restrictions, and no value outside the set is consistent with those restrictions. Sharpness means the set cannot be shrunk without imposing assumptions beyond the observed restrictions and it does not mean that every value inside the set is equally plausible, or that the endpoints are more likely locations for the true value than the interior. We do not argue or claim that any particular imputation rule is wrong. Rather, we develop bounds that can measure how much additional identifying information any point-identifying rule (midpoint, hot-deck, or a fitted parametric family) is adding beyond what the data alone deliver. A narrow gap between an imputation-based estimate and the sharp bounds indicates that results are not particularly sensitive to the choice of imputation rule; a wide gap indicates that the rule's apparent precision rests heavily on an assumption the data cannot test. 

To structure our analytical framework we divide the main settings encountered in practice into two observational scenarios. Scenario~1, the grouped-data case, has common non-overlapping brackets, as in historical tabulations, data reported in brackets, or privacy-protected grouped records. Scenario~2, the overlapping-interval case, has observation-specific overlapping intervals. Such intervals arise in modern surveys that use unfolding brackets, as in our application, and also when the inequality variable is a sum or other function of components that are each observed as grouped data. 

These two cases are not variants of the same problem. They differ fundamentally in the geometry of the feasible set of unobserved outcome distributions consistent with the data, in the form of the distributions that attain the lower and upper bounds, and in the algorithms required to compute those bounds. In the grouped-data case, unobserved outcomes can be completely ordered ex ante before any analysis is conducted, which simplifies the finite-dimensional characterization. The overlapping-interval case, however, only guarantees a partial ordering, so the Scenario~1 ordering and linear-programming reductions no longer apply. Scenario~2 therefore requires a separate finite-support characterization based on the overlap structure of the intervals. 

Our analysis provides three substantive contributions:

First, for the broad class of inequality measures satisfying the Pigou-Dalton transfer principle (mathematically expressed as Schur-convexity), we show that sharp bounds are attained by discrete distributions with finite support, reducing infinite-dimensional optimization to finite-dimensional programs. We then use the fact that a large subclass of indices can be written, after sorting, as ratios of two linear functions of the unobserved outcome vector. This class of linear-fractional indices includes the Gini coefficient, quantile ratios, top shares, the Palma ratio, and the Bonferroni index. For the non-overlapping grouped data setting, the bounds for these indices reduce to linear programs. For the overlapping-interval setting, we focus our computational development on the Gini index, which reduces to a quadratic-fractional program (similar developments can be pursued for other indices). The finite-support characterizations are the broadest part of the analysis; the computational and inferential results impose the additional objective-specific conditions stated below.

In Scenario~1, the grouped-data case with non-overlapping brackets, existing work has studied several fixed information structures.  \cite{gastwirth1972} bounds the Gini under known subgroup means and obtains non-sharp bounds, \cite{mehran1975} attains sharp bounds using known Lorenz-curve points, \cite{murray1978} gives a computational approach under a known mean, and \cite{Stoye2010} provides sharp theoretical identified sets for the Gini and several spread parameters without auxiliary information. Our Scenario~1 results below give sharp bounds under auxiliary linear equality and inequality restrictions, including subgroup means, income shares, Lorenz-curve points, and combinations of these restrictions, and they pair the computation with the inference theory developed in Section~\ref{sec:asymptotics}. 

Scenario~2, the overlapping-interval case, has a different structure. Overlapping intervals, as generated by unfolding-bracket surveys or by sums of separately bracketed components, are not grouped data with a finer common partition. In Scenario~1, every observation belongs to one of finitely many common, mutually exclusive brackets. This induces a global ordering: all observations in bracket $d$ lie between all observations in brackets $d-1$ and $d+1$. The ordering is what makes the linear-fractional formulation useful. In Scenario~2, there is no common partition, no ex ante ordering, and no group structure over which counts can be aggregated. Two individuals may have intervals [\$5,000, \$20,000] and [\$1,000, \$50,000], respectively. These intervals overlap, and the feasible set of unobserved outcome distributions consistent with them has a different geometry. It is characterized by inequality constraints on the mass allocated to overlapping subintervals rather than by group membership counts. Optimizing an inequality index over this set therefore requires a different characterization from the grouped-data case. We develop the corresponding computation directly from the overlap structure of the intervals. 

The asymptotic theory establishes a \(\sqrt n\) distribution for the estimated bound endpoints and a generally valid bootstrap procedure for inference. The key step is to represent each bound endpoint as the value function of a constrained optimization problem over probability measures and to show that this value function is Hadamard directionally differentiable. A directional delta-method argument, using \citet[Theorem~2.1]{FangSantos2019}, combined with a functional central limit theorem for the estimated parameters indexing the constraint set, delivers the asymptotic distribution. Because the value function is generally directionally, rather than fully, differentiable, the confidence intervals use the Fang--Santos bootstrap for directionally differentiable functions, implemented here with the \(m\)-out-of-\(n\) bootstrap of \citet{Shao1994}; when the value function is fully Hadamard differentiable and the ordinary bootstrap consistently estimates the first-stage distribution, the standard nonparametric bootstrap is valid.

The existing inference literature for inequality bounds is narrow. \cite{MCDONALD1981} and \citet{gastwirth_et_al_estimation} derive asymptotic distributions for specific closed-form Gini bounds under the grouped-data setup of \cite{gastwirth1972}. \cite{DedduwakumaraPrendergast2019} propose a parametric bootstrap when subgroup means are available. Our approach requires neither closed-form bounds nor a specific configuration of auxiliary information. It covers both the grouped-data and overlapping-interval cases and applies to the objective classes verified below, including smooth moment-based inequality indices and the Gini and Hoover indices under the stated local objective conditions. Quantile-ratio bounds are covered computationally throughout; formal inference for quantile ratios requires the additional quantile-stability condition stated in the asymptotic section.

Finally, we present two applications that demonstrate both the practical reach of the framework and the costs of ignoring identification uncertainty. The first uses cross-sectional data from the English Longitudinal Study of Ageing (ELSA) to document inequality in liquid savings wealth among older households. A substantial fraction of respondents report exact savings balances in the survey, but in order to minimise non-response the remainder are routed through unfolding brackets that generate respondent-specific, potentially overlapping intervals. This is a Scenario~2 setting in which the overlap structure, rather than a common group partition, determines the sharp bounds. We compute sharp Gini bounds of [0.714, 0.792] for a narrow single-component savings measure and [0.686, 0.767] for a broader three-component composite. Several standard imputation methods all produce point estimates that fall within the identified set, which is an interval in this application, but they cluster in a much narrower range, roughly one-third as wide as the identified set, with bootstrap confidence intervals that do not come close to spanning the sharp bounds. These results illustrate a general principle that the apparent precision of imputation estimates is a property of the imputation rule, not of the data.

The second application uses published U.S. income distribution tables from 1929 to 1971 to construct time-series bounds for the Gini coefficient and the 90/50 percentile ratio of household income. These data tables, originally produced by Selma Goldsmith at the Department of Commerce and later by the Office of Business Economics, are the primary source for historical U.S. inequality series but have not previously been used to compute formally valid bounds. Early years, when only eight income brackets are available, yield wide bounds that nonetheless confirm a sharp decline in inequality between 1929 and the early 1940s. From 1964 onward, when subgroup means are reported alongside finer bracket counts, the bounds narrow substantially, illustrating how auxiliary linear information functions as a direct complement to interval structure. The lower bound on the Gini for 1929 lies above current point estimates for recent decades, consistent with other evidence on the exceptional concentration of the pre-Depression income distribution.

The paper proceeds as follows. Section~\ref{sec:scenarios} formalizes the two observational cases and the class of inequality indices. Sections~\ref{sec:Scenario1} and~\ref{sec:Sc2} develop computation for Scenarios~1 and~2, respectively. Section~\ref{sec:asymptotics} establishes inference. Section~\ref{sec:applications} presents the applications. Section~\ref{sec:conclusion} concludes.

\vspace{0.2cm}
 
\noindent \paragraph{Related literature.} The closest literature studies inequality measurement under grouped data with common, non-overlapping brackets, corresponding to Scenario~1, the grouped-data case.
Early contributions by \cite{gastwirth1972} and \cite{mehran1975} derive closed-form bounds for the Gini coefficient under specific auxiliary information, while \cite{murray1978}  discusses computation of 
Gini bounds under a known mean. These studies fix particular informational settings, such as known means, subgroup means, or Lorenz-curve points, and do not cover combined auxiliary restrictions or other inequality measures in a single computation and inference framework. \cite{Cowell1991} characterizes solution forms under grouped data with either known means or subgroup means for general Schur-convex inequality measures, but does not address computation or inference. \cite{Stoye2010} provides characterizations of the identified sets for the expectation and several spread parameters, including the Gini coefficient, but does not consider auxiliary information, computational methods, or statistical inference. The Scenario~1 results below accommodate auxiliary linear restrictions, give tractable computation for widely used inequality measures, and support formal inference.

Scenario~2 is a fundamentally different setting because intervals are observation-specific and may overlap arbitrarily, so there is no common partition or ex ante ordering of observations. The Scenario~1 arguments do not apply to this informational environment because the feasible set of unobserved outcome distributions has a different geometric structure, and techniques based on Lorenz-curve interpolation or group-level aggregation are no longer directly applicable. This setting raises distinct conceptual and computational challenges. The treatment below gives sharp identification, computation, and inference for the objective classes considered in this overlapping-interval environment. \cite{ManskiTamer2002} is tangentially related in studying interval data, but focuses on regression models with an interval-valued regressor and on identification of conditional mean functions rather than distributional inequality measures. Sharp bounds for global distributional measures require a different geometric, computational, and inferential treatment.

This paper also connects to the literature on partial identification with incomplete, aggregated, or combined data sources (the auxiliary information in our settings may come from a different data source than the interval data), including work on ecological inference and data combination (e.g., \cite{CrossManski2002}, \cite{Pacini2019}, \cite{DH2025}). While this literature studies identification under limited information, our focus is on inequality measures and on exploiting their structure to obtain sharp, computationally tractable bounds with valid statistical inference.

More broadly, our analysis belongs to the partial-identification tradition associated with weak data and informational restrictions, including \cite{Manski1995,Manski2003}. A general language for many such problems is provided by the random-set approach developed in econometrics by \cite{BeresteanuMolinari2008} and \cite{BeresteanuMolchanovMolinari2012}. Although we do not use the random-set formalism directly, our perspective is closely related.

\section{Data Scenarios and Inequality Indices}
\label{sec:scenarios}

We study settings where a variable of interest $Y$, such as income, wealth, consumption, or wages, is not observed exactly but only through interval restrictions. Formally, for each unit $i=1,\ldots,n$ there is an unobserved outcome $y_i$, and we observe an interval $\mathcal{I}_i=[\underline{a}_i,\overline{a}_i]$ known to contain it, i.e.\ $y_i\in\mathcal{I}_i$, where $\underline{a}_i \leq \overline{a}_i$ are (possibly estimated) finite interval limits. The collection $\mathcal{I}:=\bigl(\mathcal{I}_1,\ldots,\mathcal{I}_n\bigr)$ summarizes the observed information, while $\mathbf{y}=(y_1,\ldots,y_n)\top$ denotes the unobserved outcome vector. We focus on two scenarios that cover many applications.

\begin{definition}[Scenario~1] 
There exists a finite collection of non-overlapping intervals $\mathcal{G}_d = [\underline{a}_d,\overline{a}_d]$, $d=1,\ldots,D$,
with $\underline{a}_d \leq \overline{a}_d$, such that for each unit $i$ there exists $d$ with $\mathcal{I}_i=\mathcal{G}_d$. Thus the econometrician observes which common bracket contains $Y$, rather than observing $Y$ exactly. The boundaries $\underline{a}_d,\overline{a}_d$ may be deterministic (e.g.\ fixed survey brackets) or stochastic (e.g.\ sample quantiles), with stochastic boundaries required to admit an asymptotically linear estimator.\footnote{As the bracket boundaries are often finite-dimensional, this condition is usually quite weak. An example is when the bracket boundaries are given by quantiles of the empirical distribution. These estimate the population quantiles under standard conditions. Whether the bracket boundaries are deterministic or not makes no difference to computation of the bounds -- only to statistical inference}.

Without loss of generality, intervals $\mathcal{G}_d$ are ordered such that
\(\underline{a}_d \le \overline{a}_d \le \underline{a}_{d+1}\) for \(d=1,\ldots,D-1\)\footnote{When boundaries coincide (e.g.\ one bracket is ``$\le 20{,}000$'' and the next is ``$>20{,}000$''), we interpret the brackets as half-open so that membership is unambiguous. This convention is immaterial for our results.}. In addition to interval information, auxiliary information may be imposed as linear restrictions, expressed as $C_n \mathbf{y} = f_n$ or $C_n \mathbf{y} \leq f_n$, where $C_n$ and $f_n$ may be estimated from data.\footnote{Later, in our computational treatment of this scenario, we will express these restrictions in this form for an ordered unobserved outcome vector.} 
\end{definition}

Auxiliary information can be incorporated either by refining the bracket structure or by adding linear restrictions. For instance, if the sample median $Q_{0.5}(\mathbf{y})$ is known and lies in bracket $\mathcal{G}_{d_0}$, one may split $\mathcal{G}_{d_0}$ into $[\underline{a}_{d_0},\,Q_{0.5}(\mathbf{y})]$ and $\big(Q_{0.5}(\mathbf{y}),\,\overline{a}_{d_0}\big]$ and update the bracket counts accordingly. Thus percentile information can be incorporated by refining the intervals $\mathcal{G}_d$ rather than by adding optimization restrictions. 

The additional linear-restriction setup covers many cases of interest, including known subgroup means or income-share restrictions. For example, knowledge of the overall sample mean $\widehat{\mu}$ can be imposed through $\bar{y}=\widehat{\mu}$, while knowledge that the mean in group $d$ equals $\widehat{\mu}_d$ adds the restriction $\bar{y}_d=\widehat{\mu}_d$.  Likewise, if a point $\big(\sum_{d=1}^{h}\widehat{s}_d,\,\ell_h\big)$ on the Lorenz curve is known for some $h$, where $\ell_h := \bar{y}^{-1}\!\sum_{j=1}^{h}\widehat{s}_j \bar{y}_j$, then one can impose the linear restriction $\sum_{j=1}^{h}\widehat{s}_j \bar{y}_j - \ell_h\,\bar{y}=0$. While prior work typically treats such cases in isolation and rarely develops inference, our framework accommodates a broad class of auxiliary information that can be expressed through the linear equality/inequality restrictions studied in this paper. It also allows such information to be combined across multiple data sources within a unified computational and inferential framework.

A fixed number of non-overlapping brackets is common in historical tables and privacy-protected grouped data, such as the historical income distribution tables reproduced in Figures A.1 and A.2 in the online supplement. We study both the simple case with no auxiliary information (Scenario~1A), in which the observed restrictions are the indicators \(\mathbf 1[y_i\in\mathcal G_d]\), \(i=1,\ldots,n\), \(d=1,\ldots,D\), and the richer case (Scenario~1B), where interval indicators are supplemented with linear constraints. We allow \(\underline a_d=\overline a_d\), which yields mixed exact-value and interval-valued data with common brackets. In many applications, however, the data consist only of intervals. 

By contrast, Scenario~2, the overlapping-interval case, allows intervals to differ across observations, as in interval-response surveys that ask sequences of follow-up threshold questions.

\begin{definition}[Scenario~2] 
Each unit has an observation-specific interval $\mathcal{I}_i$, $i=1,\ldots,n$. Intervals may overlap arbitrarily and need not admit a common ordering.
\end{definition}

This design also allows $\underline a_i=\overline a_i$ for some units, producing mixed exact-value and interval-valued observations. Section~\ref{app:wealth} studies this case using data from the English Longitudinal Study of Ageing. Auxiliary linear information can be added by the same logic as in the passage from Scenario~1A to Scenario~1B, but we keep Scenario~2 in its baseline form because this is the case commonly encountered in interval-response survey data.

In both scenarios we characterize the identified set through a finite-dimensional optimization problem, which yields both fast computation and a useful statistical theory. 

A key ingredient is that many common inequality indices can be written, after sorting, as ratios of two linear functions of the outcome vector, which we call \textit{linear-fractional} functions:
\[
G_n(\mathbf{y})=\frac{r_1(n)^\top \mathbf{y}}{r_2(n)^\top \mathbf{y}},
\]
where $r_1(n),r_2(n)\in\mathbb{R}^n$ are known vectors and $\mathbf{y}$ is understood to be sorted so that
\begin{equation}
\label{ordering}
y_1\le \cdots \le y_n.\footnote{All inequality indices considered here are invariant to permutations of the sample and therefore depend only on the ordered vector.}
\end{equation}
 
To ensure these indices are well-defined, we assume throughout that denominators are strictly positive, i.e.\ $r_2(n)^\top \mathbf{y}>0$ for all feasible $\mathbf{y}$. Our two leading examples are the following:

 \textbf{Gini coefficient.} Under \eqref{ordering}, this takes the form of 
 $G_n(\mathbf{y}) = \frac{1}{n^2 \bar{y}} \sum_{i=1}^n (2i - n - 1)y_i$, where $\bar{y} = \frac{1}{n} \sum_{i=1}^n y_i$. This fits the above equation with $r_1(n) = (1-n, 3-n, \ldots, n-1)^\top$ and $r_2(n) = n \iota_n$, where for any $k\in\mathbb{N}$, $\iota_k$ denotes the $k$-vector of ones.

\textbf{Quantile ratio.} The sample quantile ratio $y_{\lceil \tau_2 n \rceil} / y_{\lceil \tau_1 n \rceil}$ for quantile indices $\tau_1, \tau_2$ (e.g., $\tau_1 = 0.5, \tau_2 = 0.9$) fits our setup by taking $r_1(n)=e_{\lceil \tau_2 n \rceil}$ and $r_2(n)=e_{\lceil \tau_1 n \rceil}$, where $e_j$ denotes the $j$th canonical basis vector in $\mathbb{R}^n$. 

Other linear-fractional statistics include weighted or generalized Gini indices, the top-$p$ income share, the Palma ratio, percentile ratios, and the Bonferroni index. While our main computational results apply to the linear-fractional family, the finite-support characterization of the identified set applies more broadly to continuous and Schur-convex inequality functions in the settings considered below. Beyond the linear-fractional examples, this includes the Generalized Entropy family (including the Theil and mean-log-deviation indices), as well as the Atkinson, Elteto--Frigyes, Kolm, and Zenga indices, and the Herfindahl--Hirschman index when applied to income shares. For special cases such as the Hoover index, linear programming ideas still yield fast computation even though the index itself is not linear-fractional; the discussion before Proposition~\ref{prop:Sc1BsolLF} gives the details. Some indices require additional domain or smoothness conditions. Log-based indices such as the Theil index require all \(y_i\) to be bounded away from \(0\). Quantile-based indices such as the 90/50 percentile ratio are covered computationally, while the formal asymptotic treatment uses the finite-branch quantile stability condition stated in Section~\ref{sec:asymptotics}.

\section{Computation in Scenario~1}
\label{sec:Scenario1}

Because the feasible set is compact and connected, and $G_n(\cdot)$ is continuous, the image of the feasible set under $G_n$ is a compact connected subset of $\mathbb R$, hence a (possibly degenerate) closed interval. Therefore the identified set is fully characterized by its upper and lower bounds, obtained by minimizing and maximizing $G_n(\mathbf y)$ over the feasible set.

\subsection{Charnes--Cooper transformation for optimizing linear-fractional inequality indices}

To compute sharp nonparametric bounds for a linear-fractional $G_n(\mathbf{y})$, we solve 
\begin{equation}
 \label{LF1}
 \max_{\mathbf{y}} \; (\text{or } \min_{\mathbf{y}}) \;
 \frac{r_1(n)^\top\mathbf{y}}{r_2(n)^\top\mathbf{y}} ,
\end{equation}
subject to \eqref{ordering}, any additional linear restrictions, written compactly as
\begin{equation}\label{LF4}
C_n \mathbf{y} \le f_n,
\end{equation}
where equality restrictions are included by adding both $c^\top\mathbf y\le f$ and $-c^\top\mathbf y\le -f$, and letting $n_\ell$ denote the number of observations in group $\ell$:
\begin{equation} \label{LF3}
y_i \in [\underline{a}_d, \overline{a}_d], \quad
i=\sum_{\ell=1}^{d-1} {n_{\ell}}+1, \ldots, \sum_{\ell=1}^{d} {n_{\ell}}, \quad d=1, \ldots, D.\end{equation}

Optimization of \eqref{LF1} subject to \eqref{ordering}, \eqref{LF4}, and \eqref{LF3} is a linear-fractional problem and can be converted to a linear program using the Charnes--Cooper transformation \citep{charnescooper}. The matrix form of the constraints is as follows. The ordering \eqref{ordering} 
can be written as $E_n \, \mathbf{y} \leq 0,$
where $E_n$ is the $(n-1)\times n$ matrix of first differences whose elements $(r,r)$, $r=1, \ldots, n-1$, are 1, elements $(r,r+1)$, $r=1, \ldots, n-1$, are -1, and all the other elements are 0. The constraints $y_i \in [\underline{a}_d, \overline{a}_d]$ can be written as 
$I_{n}\mathbf{y} \le b_{U,n}$ and $-I_{n}\mathbf{y} \le -b_{L,n}$, where 
\[
b_{U,n} := (\overline{a}_1 \iota_{n_1}^\top, \ldots, \overline{a}_D \iota_{n_D}^\top)^\top,
\qquad
b_{L,n} := (\underline{a}_1 \iota_{n_1}^\top, \ldots, \underline{a}_D \iota_{n_D}^\top)^\top,
\]
and $I_n$ is the identity matrix of size $n$. Overall, if we denote
\[
H_n := [E_n^\top,\ I_n,\ -I_n,\ C_n^\top]^\top,
\qquad
b_n := (0_{(n-1)\times 1}^\top,\ b_{U,n}^\top,\ -b_{L,n}^\top,\ f_n^\top)^\top,
\]
then the constraint set can be written as $H_n \mathbf{y} \le b_n$.

\paragraph{Charnes--Cooper transformation.} Our linear-fractional problem can be solved as the following linear program: 
\[\max_{z,t} \, r_1(n)^\top \, z \qquad 
\text{subject to } \qquad H_n z -b_n t \leq 0, \quad r_2(n)^\top z=1, \quad 
 t \geq 0.
\]
This reformulation of the linear-fractional program is obtained by the well-known Charnes--Cooper transformation with 
$z=\frac{1}{r_2(n)^\top\mathbf{y}} \cdot \mathbf{y}$ and $t = \frac{1}{r_2(n)^\top\mathbf{y}}$. 
Although the displayed LP uses the weak constraint $t\ge0$, any feasible solution relevant for the original bounded problem has $t>0$: if $t=0$, then $H_nz\le0$ would put $z$ in the recession cone of the bounded feasible set, forcing $z=0$ and contradicting $r_2(n)^\top z=1$. The optimal solution $(z^*,t^*)$ for $(z,t)$ therefore yields the solution of the original problem as $\mathbf{y}^*=\frac{z^*}{t^*}$. This reformulation is attractive because it reduces the problem to linear programming, and also naturally allows one to incorporate a variety of additional information through the constraints $C_n \mathbf{y} \leq f_n$. If $n$ is very large (say, in millions), one can utilize the population-size invariance property of an inequality metric to scale down the computational problem to a more feasible one, at the cost of a small approximation error.

\subsection{Solution form for linear-fractional inequality indices} 

The form of the optimal solution $\mathbf{y}^*$ depends on the additional linear constraints $C_n\mathbf{y} = (\leq) f_n$. Propositions~\ref{prop:base_cornersol} and~\ref{prop:Sc1BsolLF} characterize the solution. Proposition~\ref{prop:base_cornersol} treats Scenario~1A, first for general linear-fractional measures and then for strictly Schur-convex ones.

\begin{prop}
\label{prop:base_cornersol}
Consider the linear-fractional endpoint problem in \eqref{LF1} under the maintained condition that $r_2(n)^\top\mathbf y>0$ for every feasible vector. Suppose the feasible set defined by \eqref{ordering} and \eqref{LF3} is nonempty. Then there exists an optimal solution to each endpoint problem with
$y_i^*\in\{\underline a_{d(i)},\overline a_{d(i)}\}$, where $d(i)$ is the group index of the interval $\mathcal G_{d(i)}=[\underline a_{d(i)},\overline a_{d(i)}]$ containing $y_i$.

If, in addition, $G_n(\mathbf y)=r_1(n)^\top\mathbf y/r_2(n)^\top\mathbf y$ is strictly Schur-convex, the endpoint solutions can be chosen as follows.
\begin{itemize}
  \item[(a)] For the minimizer $\mathbf y^*_{\min}$, there exists $d_0\in\{0,\ldots,D\}$ such that $y^*_{i,\min}=\overline a_{d(i)}$ for all $i$ in groups with $d(i)\le d_0$ and $y^*_{i,\min}=\underline a_{d(i)}$ for all $i$ in groups with $d(i)>d_0$. The cases $d_0=0$ and $d_0=D$ mean, respectively, that all groups are assigned to their lower or upper endpoints.
  \item[(b)] For the maximizer $\mathbf y^*_{\max}$, there exists $d_0\in\{1,\ldots,D\}$ such that $y^*_{i,\max}=\underline a_{d(i)}$ for $d(i)<d_0$, $y^*_{i,\max}\in\{\underline a_{d_0},\overline a_{d_0}\}$ for $d(i)=d_0$, and $y^*_{i,\max}=\overline a_{d(i)}$ for $d(i)>d_0$. Thus at most one group may contain both endpoint values.
\end{itemize}
\end{prop}

Proposition \ref{prop:base_cornersol} reformulates the optimization problem over a $D$-dimensional parameter $\hat{\mathbf{p}} = (\hat{p}_1, \ldots, \hat{p}_D)^\top$, where $\hat{p}_d \in \{0, 1/n_d, \ldots, 1\}$ represents the proportion of units in group $d$ assigned to $\underline{a}_d$, with $1 - \hat{p}_d$ assigned to $\overline{a}_d$. The exact reformulation depends on the vectors $r_1(n)$ and $r_2(n)$ in the linear-fractional definition. For the Gini coefficient, the objective can be written as 
\begin{equation}
  \label{LFP_alter}
  G(\hat{\mathbf{p}}, \hat{\mathbf{s}}) = \frac{0.5 (\hat{\mathbf{p}}^\top, \iota_D^\top - \hat{\mathbf{p}}^\top) A(\hat{\mathbf{s}}) (\hat{\mathbf{p}}^\top, \iota_D^\top - \hat{\mathbf{p}}^\top)^\top}{(\hat{\mathbf{p}}^\top, \iota_D^\top - \hat{\mathbf{p}}^\top) b(\hat{\mathbf{s}})},
\end{equation}
where $\hat{s}_d = n_d / n$, $\hat{\mathbf{s}}=(\hat{s}_1, \ldots, \hat{s}_D)^\top$, $A(\hat{\mathbf{s}})$ is a $2D \times 2D$ symmetric matrix of differences between interval boundaries, and $b(\hat{\mathbf{s}})$ is a $2D \times 1$ vector of weighted boundaries: 
\begin{align*} 
(b)_{\ell}(\hat{\mathbf{s}})  = \hat s_{\ell} \underline{a}_{\ell}, & \;\; (b)_{\ell+D}(\hat{\mathbf{s}})  = \hat s_{\ell} \overline{a}_{\ell}, \;\; (A)_{\ell, \ell}(\hat{\mathbf{s}}) = 0, \;\; (A)_{\ell, \ell+D}(\hat{\mathbf{s}}) = \hat s_{\ell}^2 (\overline{a}_{\ell} -\underline{a}_{\ell}), \;\;  \ell=1, \ldots, D, \\
(A)_{\ell, \ell'}(\hat{\mathbf{s}}) &= \hat s_{\ell} \hat s_{\ell'}(\underline{a}_{\ell'} -\underline{a}_{\ell}), \quad (A)_{\ell+D, \ell'+D}(\hat{\mathbf{s}}) = \hat s_{\ell} \hat s_{\ell'}(\overline{a}_{\ell'} -\overline{a}_{\ell}), \quad \ell <\ell' \leq D,\\
(A)_{\ell, \ell'+D}(\hat{\mathbf{s}}) &= \hat s_{\ell} \hat s_{\ell'} (\overline{a}_{\ell'} -\underline{a}_{\ell}), \quad (A)_{\ell+D, \ell'}(\hat{\mathbf{s}}) = \hat s_{\ell} \hat s_{\ell'} (\underline{a}_{\ell'} -\overline{a}_{\ell}), \quad \ell <\ell' \leq D, 
\end{align*}
The sharp endpoint is obtained by optimizing this function over $\mathcal{P}_n \equiv \times_{d=1}^D \{0,1/n_d,2/n_d,\ldots,1-1/n_d,1\}$.

An analogous finite-dimensional representation in $\hat{\mathbf p}$ applies to any linear-fractional index once the endpoint pattern is fixed. The particular quadratic-fractional formula in \eqref{LFP_alter} is the Gini specialization. Non-Schur indices, such as quantile ratios, are handled separately below. Proposition~\ref{prop:base_cornersol} implies that, for strictly Schur-convex linear-fractional indices, a lower-bound minimizer has a single switch point,
\(\hat{\mathbf{p}}^*_{\min}=(0,\ldots,0,1,\ldots,1)\). The lower endpoint is therefore found by evaluating
\(G(\hat{\mathbf{p}},\hat{\mathbf{s}})\) at the $D+1$ switch vectors whose first $m$ entries are zero and whose remaining $D-m$ entries are one, for $m=0,\ldots,D$, and selecting the minimum. For the upper bound, a maximizer has the form
\((1,\ldots,1,\hat p_{d_0},0,\ldots,0)\), where $d_0\in\{1,\ldots,D\}$ and \(\hat p_{d_0}\in[0,1]\).\footnote{The atomic nature of the argmax and argmin may suggest that 
imposing smoothness on the c.d.f.\ of the unobserved variable $Y$ could sharpen the bounds. However, smoothness alone is unlikely to help, since the step-function c.d.f.s corresponding to the argmax and argmin can be approximated arbitrarily closely by smooth c.d.f.s. Restrictions on the upper or lower bounds of the p.d.f.\ of $Y$, if it exists, can tighten the bounds.}

The observation that our sharp inequality bounds can be taken as dependent only on a finite-dimensional parameter $\hat{\mathbf{p}}$\footnote{This exact $D$-dimensional representation is for fixed $D$. The asymptotic theory later requires bounded effective support, not literally a fixed number of observed intervals. Thus Scenario~2 remains covered when the survey design generates a bounded set of possible interval endpoints, or more generally when Assumption~\ref{ass:finite_complexity} holds.} will be a cornerstone of our statistical theory; analogous finite-dimensional reductions appear in the more general settings we consider.

We now outline a general approach to maximizing or minimizing the quadratic fractional objective in (\ref{LFP_alter}). While minimization is unnecessary in Scenario~1A by the results of Proposition~\ref{prop:base_cornersol}, developing this framework prepares us for Scenario~2. 

Consider the maximization of (\ref{LFP_alter}). Following \citet{dinkelbach67}, we introduce a family of subproblems indexed by $\lambda$: 
\[f_{\max,n}(\lambda)= \max_{\hat{\mathbf{p}} \in \mathcal{P}} \left(\frac{1}{2} (\hat{\mathbf{p}}^\top, \iota_D^\top-\hat{\mathbf{p}}^\top ) \; A(\hat{\mathbf{s}}) \; (\hat{\mathbf{p}}^\top, \iota_D^\top-\hat{\mathbf{p}}^\top )^\top - \lambda (\hat{\mathbf{p}}^\top, \iota_D^\top-\hat{\mathbf{p}}^\top) \; b(\hat{\mathbf{s}})\right)\]
where $\mathcal{P} := [0,1]^D$, and we let $f_{\min,n}$ define the analogous $\min$ problem. By \citet{dinkelbach67}, the function above is continuous, strictly decreasing on any interval where $B(\hat{\mathbf p}) := (\hat{\mathbf{p}}^\top, \iota_D^\top-\hat{\mathbf{p}}^\top) \; b(\hat{\mathbf{s}})>0$, and it is the supremum of affine functions of $\lambda$, hence convex. For the Gini computations below, feasible outcomes are nonnegative and the feasible mean is positive, so $G\in[0,1]$. Under the resulting sign conditions $f_{\max,n}(0)\ge 0$ and $f_{\max,n}(1)\le 0$, there is a unique zero $\lambda^\star=\max_{\hat{\mathbf p}\in\mathcal{P}} G(\hat{\mathbf p},\hat{\mathbf s})$. For other objectives or domains, the bisection interval should be replaced by any known bracket for the relevant endpoint.

We solve for $\lambda^\star$ using bisection on $[0,1]$ in the nonnegative-outcome Gini case: at each step evaluate $f_{\max,n}$ at the midpoint and update the bracket according to the sign; this is given in Algorithm~\ref{alg:baseUP}, which iteratively locates the solution with arbitrary tolerance $\varepsilon$ (e.g., $\varepsilon=10^{-6}$). This differs from the classical \citet{dinkelbach67} procedure (and also procedures in subsequent literature), and achieves geometric convergence since $|\lambda_{i+1}-\lambda_i| = 2^{-(i+1)}$, and $\lambda^*$ is always contained within whichever of the intervals $[\lambda_i, \lambda_i + 2^{-(i+1)}]$ and $[\lambda_i - 2^{-(i+1)}, \lambda_i]$ are selected in the $i$-th iteration of the algorithm. The optimization in $f_{\max,n}$ is over the full set $\mathcal{P}$, not the grid $\mathcal{P}_n$. This reduces computation. Under the regularity conditions used for inference in Section~\ref{sec:asymptotics}, Lemma~\ref{lem:quant} shows that replacing the mass grid by its continuous relaxation creates only an $O(1/n)$ approximation error. Optimizing over $\mathcal{P}$ also permits computation using only the sample shares in each interval, since the sample size is no longer needed to construct $\mathcal{P}_n$.  Computational details are in Algorithm \ref{alg:baseUP}.

\begin{algorithm}
\caption{Upper Bound on $G(\hat{\mathbf{p}}, \hat{\mathbf{s}})$ in \eqref{LFP_alter} \label{alg:baseUP}}
\begin{algorithmic}[1]
\STATE Initialize $\lambda_1 = 0.5$. 
\REPEAT
 \STATE Set $\widetilde{\lambda} = \lambda_i$ and compute $f_{\max,n}(\widetilde{\lambda})$. 
 \IF{$0 \geq f_{\max,n}(\widetilde{\lambda}) > -\varepsilon$}
  \STATE Stop: $\widetilde{\lambda} = \max_{\hat{\mathbf{p}} \in \mathcal{P}} G(\hat{\mathbf{p}}, \hat{\mathbf{s}})$. 
 \ELSIF{$f_{\max,n}(\widetilde{\lambda}) > 0$}
  \STATE Update $\lambda_{i+1} = \lambda_i + 2^{-(i+1)}$. 
 \ELSE
  \STATE Update $\lambda_{i+1} = \lambda_i - 2^{-(i+1)}$. 
 \ENDIF
\UNTIL{convergence}
\end{algorithmic}
\end{algorithm}

The lower bound $\min_{\hat{\mathbf{p}} \in \mathcal{P}} G(\hat{\mathbf{p}}, \hat{\mathbf{s}})$ can be computed by adapting Algorithm \ref{alg:baseUP}: replace $f_{\max,n}(\widetilde{\lambda})$ with $f_{\min,n}(\widetilde{\lambda})$ (substitute maximization with minimization), keep the same sign update rule as in the maximization case (increase $\lambda$ if $f_{\min,n}(\widetilde{\lambda})>0$ and decrease $\lambda$ if $f_{\min,n}(\widetilde{\lambda})<0$), and use the stopping condition  $\varepsilon \geq f_{\min, n}(\widetilde{\lambda}) \geq 0$ to ensure a conservative outer approximation. Since  the Gini numerator is concave, the inner subproblem $f_{\max, n}$ is a convex optimization problem whereas $f_{\min, n}$ is generally a nonconvex global optimization problem. However, in Scenario 1A, this minimization is bypassed entirely by the exact switch-point characterization in Proposition~\ref{prop:base_cornersol}. 
\smallskip 

\noindent \textbf{Quantile ratio.} This inequality index is not strictly Schur-convex. In Scenario~1A, sharp bounds for the sample quantile ratio are immediate: the upper bound is $\overline{a}_{d_2}/\underline{a}_{d_1}$ and the lower bound is $\max(1,\underline{a}_{d_2}/\overline{a}_{d_1})$, where $\sum_{d=1}^{d_j}\hat{s}_d\ge \tau_j$ but $\sum_{d=1}^{d_j-1}\hat{s}_d<\tau_j$ for $j=1,2$.\footnote{We interpret the upper bound as $+\infty$ if $\underline{a}_{d_1}=0$.} Hence, a maximizer (minimizer) $\mathbf{y}^*$ can be taken at the interval boundaries. Since only $y^*_{\lceil\tau_2 n\rceil}$ and $y^*_{\lceil\tau_1 n\rceil}$ matter for the objective, the optimal solution is not unique.

\smallskip 

\noindent \textbf{Hoover index.} Although the Hoover index is not linear-fractional, it can still be bounded efficiently by considering each possible location of the sample mean in the ordered income vector. For each $k=0,\ldots,n$, consider the objective $\frac{1}{2n\bar y}
\left[
\sum_{i=1}^{k}(\bar y-y_i)
+
\sum_{i=k+1}^{n}(y_i-\bar y)
\right]$,
where an empty sum is understood to equal zero. Optimize this objective subject to \eqref{ordering}, \eqref{LF3}, and, in Scenario 1B, \eqref{LF4}, together with the following branch-specific constraints: (i) $y_1\geq \bar y$ when $k=0$; (ii) $y_k\leq \bar y\leq y_{k+1}$ when $1\leq k\leq n-1$; (iii) $y_n\leq \bar y$ when $k=n$. These conditions ensure that the first $k$ ordered observations lie weakly below the mean and the remaining $n-k$ observations lie weakly above it. The  branches $k=0$ and $k=n$ are feasible only when an equal-income vector satisfies the remaining restrictions, in which case the Hoover index is correctly equal to zero. Some branches may be infeasible. The lower and upper bounds are obtained by taking, respectively, the smallest and largest objective values over all feasible branches.

Proposition~\ref{prop:Sc1BsolLF} characterizes the solution in Scenario~1B. With a fixed number of constraints, the optimal solution is supported on finitely many values (possibly sample-dependent) and the proposition bounds how the number of distinct values can grow as constraints are added. This result is used in the asymptotic theory.

\begin{prop}
\label{prop:Sc1BsolLF}
Let $G_n(\mathbf y)=r_1(n)^\top\mathbf y/r_2(n)^\top\mathbf y$ be a linear-fractional measure under \eqref{ordering}, and suppose $r_2(n)^\top\mathbf y>0$ on the feasible set. The additional constraints are $C^{(1)}_n\mathbf y=f^{(1)}_n$ and $C^{(2)}_n\mathbf y\le f^{(2)}_n$, where $C^{(1)}_n$ has $q_1$ rows and $C^{(2)}_n$ has $q_2$ rows. Assume that the feasible set defined by these constraints, \eqref{ordering}, and \eqref{LF3} is nonempty. For each $d$, let $q_1(d)$ and $q_2(d)$ denote the number of rows of $C^{(1)}_n$ and $C^{(2)}_n$ that involve at least one variable from group $d$, and suppose $q_1(d)+q_2(d)\le n_d-2$.

Then there exists an upper endpoint solution $\mathbf y^*_{\max}$ and a lower endpoint solution $\mathbf y^*_{\min}$ to \eqref{LF1}, subject to \eqref{ordering}, \eqref{LF3}, $C^{(1)}_n\mathbf y=f^{(1)}_n$, and $C^{(2)}_n\mathbf y\le f^{(2)}_n$, such that the following hold.
\begin{itemize}
\item[(a)] For each group $d$, the components of $\mathbf y^*_{\max}$ and $\mathbf y^*_{\min}$ in group $d$ take at most $q_1(d)+q_2(d)+2$ distinct values.
\item[(b)] Across all groups, $\mathbf y^*_{\max}$ and $\mathbf y^*_{\min}$ take at most $q_1+q_2+2D$ distinct values.
\end{itemize}
\end{prop}

\subsection{General results for Schur-convex indices}\label{sec:GeneralTheory}

Although our computational emphasis is on indices that are linear-fractional after sorting, we also present solution-form results for a general Schur-convex inequality index $G_n(\mathbf{y})$. This identifies the structure of sharp optimizers beyond the linear-fractional class while keeping the computational development focused on the measures used in our applications.

\begin{theorem}
\label{thm:Sc2solform3}
Consider a continuous Schur-convex inequality index $G_n(\cdot)$, and suppose the components of $\mathbf y$ are ordered according to \eqref{ordering}. Let the constraints be $C^{(1)}_n\mathbf y=f^{(1)}_n$ and $C^{(2)}_n\mathbf y\le f^{(2)}_n$, where $C^{(1)}_n$ and $C^{(2)}_n$ have $q_1$ and $q_2$ rows, and suppose the feasible region defined by these constraints, \eqref{ordering}, and \eqref{LF3} is nonempty. Write $\widetilde C_n:=\bigl(C_n^{(1)\top},C_n^{(2)\top}\bigr)^\top$. For each group $d$, let $\widetilde C_{n,d}$ be the submatrix obtained by keeping the nonzero columns corresponding to elements of group $d$ and the rows that contain at least one such element. Suppose $\widetilde C_{n,d}$ has block-diagonal form
\[
\widetilde C_{n,d}=\begin{bmatrix}
A^{(d)}_1 & 0 & \cdots & 0 \\
0 & A^{(d)}_2 & \cdots & 0 \\
\vdots & \vdots & \ddots & \vdots \\
0 & 0 & \cdots & A^{(d)}_{k_d}
\end{bmatrix},
\]
where the columns are kept in their original within-group order, each displayed block corresponds to a consecutive constrained subvector of group $d$, and each row in each block $A^{(d)}_j$ consists only of $1$'s or only of $-1$'s.

Then there exist a maximizer $\mathbf y^*_{\max}$ and a minimizer $\mathbf y^*_{\min}$ of $G_n$ subject to \eqref{LF3}, \eqref{ordering}, $C^{(1)}_n\mathbf y=f^{(1)}_n$, and $C^{(2)}_n\mathbf y\le f^{(2)}_n$ such that the following hold.
\begin{itemize}
\item[(a)] If group $d$ has no equality or inequality constraint involving its elements, then $\mathbf y^*_{\max}$ has at most one component in group $d$ that lies strictly inside $\mathcal G_d$, and all components of $\mathbf y^*_{\min}$ in group $d$ are equal.
\item[(b)] If group $d$ has constrained and unconstrained blocks as above, then the components of $\mathbf y^*_{\min}$ in group $d$ take at most $k_d+o_d$ distinct values, provided $k_d+o_d\le n_d$, where $o_d$ is the number of maximal consecutive subvectors of group $d$ that do not appear in any constrained block. The components of $\mathbf y^*_{\max}$ in group $d$ take at most $k_d+o_d+2$ distinct values, provided $k_d+o_d\le n_d-2$.
\end{itemize}
\end{theorem} 
First, Theorem~\ref{thm:Sc2solform3} allows constraints on a subset, rather than all, of the elements of group $d$. Second, conditions on matrix $\widetilde{C}_{n,d}$ allow constraints on several subgroups within group $d$. 
For example, when $d=1$, the restrictions may include an equality or inequality on the sum $y_1+\cdots+y_{[n_1/2]-1}$, a restriction on the median value $y_{[n_1/2]}$ in that group, and a restriction on the sum $y_{[n_1/2]+1}+\cdots+y_{n_1}$. A useful special case of Theorem~\ref{thm:Sc2solform3} occurs when $\widetilde{C}_n$ is invariant to permutations of columns corresponding to elements within each group $d$; equivalently, $\widetilde{C}_n$ is invariant to permutations of columns $\sum_{k=1}^{d-1} n_k+1, \ldots, \sum_{k=1}^{d} n_k$. The condition on matrix $\widetilde{C}_n$ in Theorem~\ref{thm:Sc2solform3} is consistent with many empirically relevant settings because most observed restrictions treat a subset, or all, of the variables within an interval group symmetrically. The constraints considered below, including subgroup means, income ratios, Lorenz-curve points, and overall means, satisfy this condition. This same symmetry is what makes the later measure formulation natural: once the objective and constraints depend only on blockwise masses or moments, permutations within a group are irrelevant, so the explicit ordering constraints used in the computational derivations become bookkeeping devices rather than substantive restrictions.

An important conclusion from Theorem~\ref{thm:Sc2solform3} is that asymptotically we can look at optimizers of $G_n(y)$ as those composed from a finite number of values, both in a sample and asymptotically (as long as the structures of $C^{(1)}_n$ and $C^{(2)}_n$ remain stable as not to lead to the increase of $k_d+o_d$ with $n$).

\section{Computation in Scenario~2}\label{sec:Sc2} 

Scenario~2 allows mixed exact observations and observation-specific intervals that may overlap. Unlike Scenario~1, the data generally cannot be summarized by counts in a fixed set of non-overlapping brackets and therefore cannot be fully ordered ex ante (only partially ordered). As a result, the linear-fractional structure of the inequality measures of interest cannot be exploited or transformed into linear programs, since these measures typically admit known linear-fractional representations only under a full ordering.

We start by splitting observations into two subsets: those with exact values of $y_i$ and those with nondegenerate interval observations. Define the exact-observation set \(P=\{i:\underline a_i=\overline a_i\}\) and the interval set \(Q=\{i:\underline a_i<\overline a_i\}\). Let $\mathcal{B}:=\{\underline{a}_i:i\in Q\}\cup\{\overline{a}_i:i\in Q\}$ and $\mathcal{U}:=\mathcal{B}\cup\{\underline{a}_i:i\in P\}$. Thus $\mathcal{B}$ records boundary values coming only from nondegenerate interval observations in $Q$, while $\mathcal{U}$ augments $\mathcal{B}$ with the realized exact observations from $P$. Exact observations are degenerate intervals; we separate them for computation because their locations are fixed, whereas observations in $Q$ generate unknown allocations.

Unlike Section~\ref{sec:Scenario1}, we first present solution-form results, and only after that discuss computational aspects. In our solution structure results we first consider general Schur-convex inequality indices in Theorem~\ref{th:Sc3}, and then Schur-convex indices with a linear-fractional representation (with a fixed ordering of $y_i$ in the sample) in Theorem~\ref{th:Sc3_LF}.

\begin{theorem}
\label{th:Sc3}
Consider data that comply with Scenario~2, and let $G_n(\mathbf y)$ be a continuous Schur-convex inequality index. There exist a maximizer $\mathbf y^*_{\max}$ and a minimizer $\mathbf y^*_{\min}$ of $G_n(\mathbf y)$ over the Scenario~2 feasible set such that the following hold.
\begin{enumerate}
\item[(a)] Among the interval observations $i\in Q$, at most one component of $\mathbf y^*_{\max}$ takes a value outside $\mathcal B$.
\item[(b)] Among the interval observations $i\in Q$, the set of distinct values taken by $\mathbf y^*_{\min}$ contains at most one value outside $\mathcal B$.
\end{enumerate}
\end{theorem}

The main insight of Theorem~\ref{th:Sc3} is that, at an extremum, almost all interval observations can be placed at the interval boundary values in $\mathcal B$. This identifies the structure of sharp solutions in Scenario~2 beyond the linear-fractional class. Theorem~\ref{th:Sc3_LF} gives the corresponding finite-support statement for linear-fractional indices: the possible exceptional placement for the maximum, and the possible exceptional common value for the minimum, can be chosen from $\mathcal U$.

\begin{theorem}
\label{th:Sc3_LF}
Consider data that comply with Scenario~2. Suppose $G_n(\mathbf y)$ is continuous and Schur-convex, and suppose that on each fixed ordering cone it has a linear-fractional representation with denominator strictly positive on the feasible set. Then the following hold.
\begin{itemize}
\item[(a)] A maximizer $\mathbf y^*_{\max}$ can be chosen such that all its components lie in $\mathcal U$, with at most one component $y^*_{i,\max}$ for $i\in Q$ belonging to $\mathcal U\setminus\mathcal B$.
\item[(b)] A minimizer $\mathbf y^*_{\min}$ can be chosen such that all its components lie in $\mathcal U$, with at most one distinct value in $\{y^*_{i,\min}:i\in Q\}$ belonging to $\mathcal U\setminus\mathcal B$.
\end{itemize}
\end{theorem}

Theorem~\ref{th:Sc3_LF} delivers the key dimensionality reduction. It shows that the search for an extremum can be restricted to allocations on the finite set $\mathcal U$. Exact observations contribute known masses at elements of $\mathcal U$, while the unknown part of the problem is entirely the allocation of interval observations in $Q$ across those same support points subject to feasibility constraints. This is the sense in which the overlapping-interval problem becomes finite-dimensional.

\paragraph*{First step: Ordering} Without loss of generality, arrange all unique elements of $\mathcal{B}$ in increasing order \(b_1 < b_2 < \cdots < b_K\), and order all unique elements in $\mathcal{U}$ increasingly. Let $u(d)$ index the position of $b_d$ in ordered $\mathcal{U}$.

\paragraph*{Second step: Complete set of interval-overlap restrictions.} Let $N_u$ denote the number of interval observations from $Q$ assigned to the support point $u\in\mathcal U$. The feasible counts are characterized by containment lower bounds and overlap upper bounds over consecutive blocks of boundary points.

Take any consecutive block $[b_d,b_{d+k}]=\bigcup_{j=0}^{k-1}[b_{d+j},b_{d+j+1}]$, where $k=1,\ldots,K-1$ and $d=1,\ldots,K-k$. The \textit{containment lower bound} counts intervals $i\in Q$ fully contained in the block: 
 \[
 \sum_{i \in Q} \mathbf{1}(\mathcal{I}_i \subseteq [b_d,b_{d+k}]) 
 \leq \sum_{u=u(d)}^{u(d+k)} N_u.
 \]
The \textit{overlap upper bound} counts intervals $i \in Q$ that overlap the block, even partially: 
 \[
 \sum_{u=u(d)}^{u(d+k)} N_u \leq 
 \sum_{i \in Q} \mathbf{1}(\mathcal{I}_i \cap [b_d,b_{d+k}] \neq \emptyset).
 \]
The case $k=K-1$ yields the equality $\sum_{u=u(1)}^{u(K)} N_u = |Q|$. The formal assignment result behind these rows is Theorem~\ref{th:Sc3_overlap_feasibility} in the online supplement. It shows that, together with the basic nonnegativity, total-mass, and support restrictions below, the consecutive-block lower and upper bounds are necessary and sufficient for assigning every interval observation in $Q$ to an admissible support point in $\mathcal U$. It also implies that non-consecutive unions add no independent restrictions.

\paragraph*{Third step: Optimization.} 
Since it is difficult to optimize subject to integer restrictions on all $N_u$, it is useful to rewrite the constraints obtained in the second step in terms of shares in the overall sample by dividing both the left-hand and the right-hand sides by $n$. Then for any $k=1,\ldots,K-1$ and any $d=1,\ldots,K-k$, letting $\widehat{\phi}_u := \frac{N_u}{n}$ implies that
\begin{equation} \label{Sc3:IneqBounds}
\frac{1}{n}\sum_{i \in Q} \mathbf{1}(\mathcal{I}_i \subseteq [b_d,b_{d+k}])
\;\le\;
\sum_{u=u(d)}^{u(d+k)} \widehat{\phi}_u
\;\le\;
\frac{1}{n}\sum_{i \in Q} \mathbf{1}(\mathcal{I}_i \cap [b_d,b_{d+k}] \neq \emptyset).
\end{equation}
The share vector also satisfies the basic support and mass constraints
\begin{equation}\label{Sc3:massconstraints}
\widehat\phi_u\ge0\quad(u\in\mathcal U),\qquad
\sum_{u\in\mathcal U}\widehat\phi_u=\frac{|Q|}{n},\qquad
\widehat\phi_u=0\quad\text{if }u\notin\mathcal A_Q,
\end{equation}
where \(\mathcal A_Q:=\{u\in\mathcal U:u\in\mathcal I_i\text{ for at least one }i\in Q\}\). The last condition simply removes support points that no interval observation can take.

For Scenario~2 we focus on the Gini coefficient as the benchmark case for computation. Theorem~\ref{th:Sc3_LF} yields a finite-support reduction more generally, but for the Gini this reduction leads to a particularly transparent finite-dimensional program. Define $\widehat{\psi}_u := \frac{1}{n}\sum_{i \in P} \mathbf{1}(\underline a_i = u)$, for $u \in \mathcal{U}$, as the observed share at support point $u$ coming from exact observations. The total share at $u$ can then be written as $\widehat{\psi}_u + \widehat{\phi}_u$, where $\widehat{\psi}_u$ is known and $\widehat{\phi}_u$ is the unknown mass assigned to $u$ by interval observations, subject to (\ref{Sc3:IneqBounds}) and (\ref{Sc3:massconstraints}).

Several equivalent Gini formulas can be written in terms of $\widehat{\psi}_u$ and $\widehat{\phi}_u$. We write the elements of $\mathcal{U}$ in increasing order as $\mathcal{U}=\{u_1<\cdots<u_{|\mathcal{U}|}\}$ and set $\widehat{\phi}=(\widehat{\phi}_{u_1},\ldots,\widehat{\phi}_{u_{|\mathcal{U}|}})$. The form used for computation is
\begin{equation} \label{Giniobjective_Sc3}
G_{2,n}(\widehat{\phi})
=
\frac{\sum_{k=1}^{|\mathcal{U}|} \sum_{j=1}^{|\mathcal{U}|}
(\widehat{\psi}_{u_k}+\widehat{\phi}_{u_k})(\widehat{\psi}_{u_j}+\widehat{\phi}_{u_j})|u_k-u_j|}
{2\sum_{k=1}^{|\mathcal{U}|} (\widehat{\psi}_{u_k}+\widehat{\phi}_{u_k})u_k}.
\end{equation}

Let $\mathcal{S}_{\widehat{\phi}}$ denote the set of all $\widehat{\phi}$ satisfying \eqref{Sc3:IneqBounds} and \eqref{Sc3:massconstraints}. This set is a continuous relaxation of the finite-sample allocation problem. In the finite-sample problem, each admissible support share satisfies $n\widehat{\phi}_{u}\in\mathbb{Z}_{+}$. By contrast, $\mathcal{S}_{\widehat{\phi}}$ may contain fractional share vectors that do not correspond to any allocation of the observed interval-valued units. Consequently, optimization over $\mathcal{S}_{\widehat{\phi}}$ 
 may yield endpoints that differ from the exact finite-sample endpoints (the relaxed maximum is weakly larger than the exact finite-sample maximum, whereas the relaxed minimum is weakly smaller than the exact finite-sample minimum). As established by our asymptotic results, the discrepancy induced by the relaxation is asymptotically negligible.

Let $Q(\widehat{\phi})$ and $M(\widehat{\phi})$ denote, respectively, the numerator and denominator in \eqref{Giniobjective_Sc3}. The upper and lower endpoints of the continuous relaxation can be characterized using the Dinkelbach transformation \citep{dinkelbach67}. 
Focusing on the upper bounds, for a fixed value of $\lambda$, define $f_{2;\max}(\lambda)
=
\max_{\widehat{\phi}\in\mathcal{S}_{\widehat{\phi}}}
\left\{
Q(\widehat{\phi})
-
\lambda M(\widehat{\phi})
\right\}.$ Assuming that all feasible outcomes are nonnegative and that the feasible mean is uniformly positive, $f_{2;\max}(\lambda)$ is continuous, strictly decreasing, and convex on $[0,1]$. Its unique root $\lambda_{\max}^{\star}$ lies in $[0,1]$ and equals  the maximum value of $G_{2,n}(\widehat{\phi})$ over $\mathcal{S}_{\widehat{\phi}}$.  Algorithm \ref{alg:2samplesUP} therefore converges to $\lambda^{\star}_{\max}$ (it is analogous to Scenario 1, using bisection on $\lambda$ with global optimization at each step).  The lower endpoint has the analogous root characterization. In particular, define $f_{2;\min}(\lambda)=\min_{\widehat{\phi}\in\mathcal{S}_{\widehat{\phi}}}\left\{Q(\widehat{\phi})-\lambda M(\widehat{\phi})\right\}$. Its unique root $\lambda_{\min}^{\star}$ equals $\min_{\widehat{\phi}\in\mathcal{S}_{\widehat{\phi}}}G_{2,n}(\widehat{\phi})$. The computational properties of the upper and lower problems, however, differ, as discussed below.

\begin{algorithm}
	\caption{for finding $\max_{\widehat{\phi} \in \mathcal{S}_{\widehat{\phi}}} G_{2,n}(\widehat{\phi})$ with residual tolerance  $\varepsilon$ \label{alg:2samplesUP}}
	\begin{algorithmic}[1]
		\STATE Initiate the algorithm with $\lambda_1= 0.5$. 
		\STATE[Iteration $i$] Take $\widetilde{\lambda}=\lambda_{i}$ and find $f_{2;\max}(\widetilde{\lambda})$ by solving the problem 
     $\max_{{\widehat{\phi}} \in \mathcal{S}_{\widehat{\phi}}} \left(Q(\widehat{\phi})- \widetilde{\lambda} M(\widehat{\phi}) \right).$
		
\IF{$0 \geq f_{2;\max}(\widetilde{\lambda})>-\varepsilon$}
\STATE the algorithm stops and we take $\widetilde{\lambda}$ as $\max_{\widehat{\phi}\in \mathcal{S}_{\widehat{\phi}}} G_{2,n}(\widehat{\phi})$. 
\ELSIF{$f_{2;\max}(\widetilde{\lambda})>0$} 
\STATE take $\lambda_{i+1}=\lambda_i+\frac{1}{2^{i+1}}$ and go back to Step 2 taking $\widetilde{\lambda}=\lambda_{i+1}$
\ELSE 
\STATE (that is, when $f_{2;\max}(\widetilde{\lambda})<-\varepsilon$) take $\lambda_{i+1}=\lambda_i-\frac{1}{2^{i+1}}$ and go back to Step 2 taking $\widetilde{\lambda}=\lambda_{i+1}$. 
\ENDIF
\end{algorithmic}
\end{algorithm}
Note that $\varepsilon$ governs the residual of the Dinkelbach subproblem, which scales with the outcome variable. Since the denominator is bounded away from zero by the positive feasible mean, a tight residual tolerance ensures a tight bound on the computed Gini endpoint. Alternatively, since the algorithm relies on bisection, one can track the width of the $\lambda$ bracket at iteration $i$ (which is $2^{-i}$) to directly bound the scale-free error on the Gini coefficient itself.

Since $Q(\widehat{\phi})$ is concave on the fixed-mass feasible set and $M(\widehat{\phi})$ is linear, $Q(\widehat{\phi})-\lambda M(\widehat{\phi})$ is concave on the fixed-mass feasible set in $\widehat{\phi}$ for every fixed $\lambda$. Each subproblem for finding the maximum value of $G_{2,n}(\widehat{\phi})$ over $\mathcal{S}_{\widehat{\phi}}$ is therefore a concave quadratic maximization problem over a convex polytope and can be solved as a convex-optimization problem.
For minimization the analogous formulation is nonconvex, and, therefore, more demanding in terms of optimization tools. But the specialized finite-sample characterization in Proposition \ref{prop:GiniSc3_original} below provides a more convenient exact computation. By contrast, evaluating $f_{2;\min}(\lambda)$ requires minimizing a concave quadratic function over a convex polytope and is therefore a nonconvex global-optimization problem.

For the Gini index, Proposition \ref{prop:GiniSc3_original} provides a more convenient exact finite-sample computation directly in terms of the original vector $\mathbf{y}$.

\begin{prop}
\label{prop:GiniSc3_original}
Consider data on $\mathbf y$ that comply with Scenario~2. Let $G_n(\mathbf y)$ be the Gini index, and suppose all feasible components are nonnegative and $\sum_{i=1}^n y_i>0$ for every feasible vector.

\begin{itemize}
\item[(a)] There exists a minimizer $\mathbf y^*_{\min}$ satisfying the conclusion of Theorem~\ref{th:Sc3_LF}. In addition, there exists $u_0\in\mathcal U$ such that $y^*_{i,\min}=\overline a_i$ for every $i\in Q$ with $\overline a_i<u_0$, $y^*_{i,\min}=\underline a_i$ for every $i\in Q$ with $\underline a_i>u_0$, and $y^*_{i,\min}=u_0$ for every $i\in Q$ with $u_0\in\mathcal I_i=[\underline a_i,\overline a_i]$.
\item[(b)] There exists a maximizer $\mathbf y^*_{\max}$ satisfying the conclusion of Theorem~\ref{th:Sc3_LF} and the stronger property that $y^*_{i,\max}\in\mathcal B$ for every $i\in Q$. In addition, there exists $u_0\in\mathcal U$ such that $y^*_{i,\max}=\underline a_i$ for every $i\in Q$ with $\overline a_i<u_0$, $y^*_{i,\max}=\overline a_i$ for every $i\in Q$ with $\underline a_i>u_0$, and $y^*_{i,\max}\in\{\underline a_i,\overline a_i\}$ for every $i\in Q$ with $u_0\in\mathcal I_i=[\underline a_i,\overline a_i]$.
\end{itemize}
\end{prop}

Proposition~\ref{prop:GiniSc3_original}(a) establishes that the minimizer compresses unknown values $y_{i}$ for $i \in Q$ toward a common threshold $u_0 \in \mathcal{U}$, pushing bounded intervals to their nearer extreme and aligning overlapping ones at $u_0$. The proposition implies an efficient minimization strategy based on a search over candidate thresholds in $\mathcal{U}$. For each candidate $u_0 \in \mathcal{U}$ construct the assignment for $y_i$, $i \in Q$, 
 according to the pattern in part (a) of the proposition. A single pass over $|\mathcal{U}|\leq 2|Q| + |P|$ candidates, computing the Gini index each time, is guaranteed by the proposition to return the global minimum and a corresponding minimizer $\mathbf{y}^*_{\min}$.

Proposition~\ref{prop:GiniSc3_original}(b) shows that a maximizer stretches the distribution by pushing uncertain values away from a common threshold $u_0 \in \mathcal{U}$. Unlike minimization, it does not uniquely prescribe the boundary assignment for the (typically small) subset $Q(u_0) = \{i \in Q : u_0 \in \mathcal{I}_i\}$. When $|Q(u_0)|$ is small, direct enumeration of the $2^{|Q(u_0)|}$ possible boundary assignments for these intervals is feasible and exact. For larger $|Q(u_0)|$, a binary-integer formulation is preferable. Dominance rules can reduce that problem before it is solved: if $i,i' \in Q(u_0)$ satisfy $\underline{a}_i = \underline{a}_{i'}$ and $\overline{a}_i < \overline{a}_{i'}$, then assigning the narrower interval $i$ to its upper bound $\overline{a}_i$ precludes assigning the wider interval $i'$ to its lower bound $\underline{a}_{i'}$, and symmetrically for the dual case. 
The binary-integer formulation is as follows. Let $ind_i \in \{0,1\}$ indicate whether $y_{i,\max}=\overline a_i$ (as opposed to $\underline a_i$) for $i\in Q$. Given the fixed assignments outside $Q(u_0)$, each realized value is affine in the binary endpoint indicators. The Gini numerator nevertheless contains absolute-difference terms that must be linearized using auxiliary variables and ordering or sign constraints. The resulting mixed-integer linear-fractional problem can then be solved globally using a standard fractional-programming method, with each parametric subproblem solved as a MILP by branch and bound.

The proposition also simplifies computation within the proportion-based optimization framework discussed earlier. All uncertain values lie on the finite boundary set $\mathcal{B}$, so the program does not require continuous proportions over $\mathcal{U} \setminus \mathcal{B}$.

\section{Asymptotics of sharp bounds} \label{sec:asymptotics}
This section establishes asymptotic theory and bootstrap validity for the vector of sharp bound estimators. Our key observation is that each endpoint can be written as the optimal value of a constrained optimization problem over probability measures. Letting the constraint right-hand sides be \(c\), the nuisance parameters indexing the constraint maps (such as bracket boundaries) be \(\theta\), and letting \(\pi\) and \(q\) denote the proportion and distribution of exact observations, we collect these inputs into a vector \(\eta=(c,\theta,\pi,q)\), with population value denoted $\eta_0$. We show below that the population bound endpoints can be written as $V_\infty(\eta_0)$ for a value function $V_\infty: \mathbb{H} \rightarrow \mathbb{R}^2$. Similarly, the estimators computed in Sections~\ref{sec:Scenario1}--\ref{sec:Sc2} can be written as $\hat V^{(n)}(\hat\eta)$\footnote{Up to numerical error.}, where $\hat V^{(n)}$ approximates $V_\infty$ and $\hat\eta$ converges to $\eta_0$.

Under the conditions below, the argument has three parts. First, the solution-form results reduce the endpoint problems to distributions with a uniformly bounded number of support points. Second, the empirical-mass, finite-restriction, and numerical approximations are negligible at the \(n^{-1/2}\) scale, so
\[
\hat V^{(n)}(\hat\eta)-V_\infty(\eta_0)
=
V_\infty(\hat\eta)-V_\infty(\eta_0)+o_p(n^{-1/2}).
\]
Third, \(V_\infty\) is Hadamard directionally differentiable, and the directional delta method of \citet[Theorem~2.1]{FangSantos2019} applies to \(\sqrt n(\hat\eta-\eta_0)\). The endpoint estimator therefore has a root-\(n\) limit under these conditions. Because the derivative need not be linear, the ordinary bootstrap need not reproduce that limit. Our recommended procedure is therefore to use an \(m\)-out-of-\(n\) bootstrap; the ordinary bootstrap is valid under the stronger differentiability and first-stage conditions stated below.

Throughout, hats denote sample analogues and the subscript \(0\) denotes population counterparts. Let \(\YY=[\underline y,\overline y]\subset\mathbb R\) be the support of \(Y\), let \(\mathcal M\) be the set of Borel probability measures on \(\YY\), and let \(\Wone\) denote the Wasserstein-1 distance on \(\mathcal M\). For \(y\in\YY\), let \(\delta_y\) denote the Dirac measure at \(y\), and let \(\mathcal J:\mathcal M\to\mathbb R\) denote the inequality index, viewed as a function of the probability measure of $Y$.

\subsubsection*{Writing the problem using measures}

\paragraph{Population problem.}
Let \(\mu_0\in\mathcal M\) denote the distribution of outcomes for the component over which the endpoint problem optimizes. When exact observations are separated from nondegenerate interval observations, \(\mu_0\) is the conditional distribution for the non-exact component, \(q_0\) is the conditional distribution of exact observations, and \(\pi_0\) is the population share of exact observations. The overall outcome distribution is then $\nu_0=(1-\pi_0)\mu_0+\pi_0q_0$.

In Scenario~1, degenerate brackets can be absorbed into the grouped-data distribution, so we set \(\pi_0=0\) and \(\nu_0=\mu_0\). The restrictions below are written for the optimized component \(\mu_0\), with right-hand sides normalized consistently with this convention. Thus, in Scenario~2, overlap restrictions are restrictions on the conditional distribution of the non-exact observations. If an auxiliary restriction is reported for the full population while exact observations are separated, it is first rewritten as a restriction on \(\mu_0\); for example, a full-population share restriction \(\nu_0(A)=s\) can be written as \(\mu_0(A)=(s-\pi_0q_0(A))/(1-\pi_0)\) when \(\pi_0<1\), or equivalently represented as an affine restriction whose right-hand side depends on \((\pi_0,q_0)\). This convention ensures that exact observations enter the inequality index once, through the mixture in \(F\) below.

The full set of constraints in the population problem may be finite or countably infinite. After row reduction, let \(h(\infty)\) denote the equality rows and \(g(\infty)\) the inequality rows. Write \(h(J_n)\) and \(g(J_n)\) for the finite row sets imposed at sample size \(n\). There are finitely many equality constraints, with \(h(J_n)=h(\infty)\) for all sufficiently large \(n\), while \(g(J_n)\subseteq g(\infty)\) may increase with \(n\). If \(g(\infty)\) is finite, then \(g(J_n)=g(\infty)\) eventually and the omitted-inequality condition below is vacuous. Respondent-specific intervals do not by themselves make the restriction family infinite. If an unfolding-bracket survey can generate only finitely many distinct intervals \([L,U]\), for example, the corresponding overlap restrictions form a finite family even though different respondents may receive different intervals.

Let \(H,G:\mathcal M\times\ell^\infty(\mathbb N)\to\ell^\infty(\mathbb N)\) collect the equality and inequality restrictions for the optimized component, with unused coordinates padded by zeros. Define
\[
\mathcal C_\infty(c,\theta)
:=
\{\mu\in\mathcal M:H(\mu;\theta)=c^1,\ G(\mu;\theta)\le c^2\},
\]
where \(c=(c^1,c^2)\). Scenario~1A uses equality restrictions, Scenario~1B may add equalities or inequalities, and Scenario~2 uses conditional overlap inequalities for the non-exact component. The population information is summarized by the restriction that \(\mu_0\in\mathcal C_\infty(c_0,\theta_0)\).

Define \(F(\mu;\pi,q):=\mathcal J((1-\pi)\mu+\pi q)\). Let \(\mathcal F_0:=\{x\mapsto\mathbf 1(x\le t):t\in\mathbb R\}\), and identify \(q\) with its distribution function when it is viewed as an element of \(\ell^\infty(\mathcal F_0)\). Furthermore, write
\[
\mathbb H
:=
\bigl(\ell^\infty(\mathbb N)\times\ell^\infty(\mathbb N)\bigr)
\times\ell^\infty(\mathbb N)\times\mathbb R\times\ell^\infty(\mathcal F_0),
\]
and equip \(\mathbb H\) with the product sup norm. Finite-dimensional \(\theta\) vectors are embedded by padding with zeros. For a given input vector $\eta = (c, \theta, \pi, q)$, the population lower and upper bounds are therefore given by
\[
V_\infty^{\inf}(\eta):=\inf_{\mu\in\mathcal C_\infty(c,\theta)}F(\mu;\pi,q),
\qquad
V_\infty^{\sup}(\eta):=\sup_{\mu\in\mathcal C_\infty(c,\theta)}F(\mu;\pi,q).
\]
We therefore define the population value function \(V_\infty(\eta):=(V_\infty^{\inf}(\eta),V_\infty^{\sup}(\eta))^\top\), which returns the lower and upper endpoints for the overall distribution \((1-\pi)\mu+\pi q\). Note that via the definition of $\mathcal C_\infty(c,\theta)$, optimization is only taken over probability measures, rather than signed measures. Our proof does not require the Banach space \(\mathbb H\) to be compact, nor for the supremum or infimum values to be attained by measures in the constraint set.

\paragraph{Sample problem.}
Conditional on the observed interval-reporting pattern, the estimators computed in the previous sections have two interpretations. First, they are sharp bounds for the inequality index of the realized sample. Second, under repeated sampling, they are plug-in estimators of the population bounds \(V_\infty(\eta_0)\). The connection is simple: the sample problem has the same form as the population problem, except that the unknown input vector \(\eta_0\) is replaced by its estimate \(\hat\eta\), and the feasible distributions are required to be empirical measures rather than arbitrary probability measures.

Let \(Q_n\) denote the observations whose outcomes are not fixed by the exact-observation component, and let \(N_n:=|Q_n|\). In Scenario~1 without a separated exact component, \(Q_n=\{1,\ldots,n\}\) and \(N_n=n\). In Scenario~2, \(Q_n\) is the interval-observation set \(Q\) defined in Section~\ref{sec:Sc2}. The sample constraints for the optimized component depend on the unobserved vector only through
\[
\mu_{y,Q}:=N_n^{-1}\sum_{i\in Q_n}\delta_{y_i},
\]
while exact observations enter through \(\hat\pi\) and \(\hat q\). The overall empirical distribution represented by a feasible conditional measure is \((1-\hat\pi)\mu_{y,Q}+\hat\pi\hat q\). Equivalently, the Scenario~2 computation in Section~\ref{sec:Sc2} uses overall interval shares \(\widehat\phi_u=(1-\hat\pi)\mu_{y,Q}(\{u\})\) together with exact shares \(\widehat\psi_u=\hat\pi\hat q(\{u\})\).\footnote{The measure formulation does not require the components of \(y\) to be ordered.}

Write the estimated input vector as \(\hat\eta=(\hat c,\hat\theta,\hat\pi,\hat q)\). For example, when the population mean appears as a component of \(c_0\), the corresponding component of \(\hat c\) is normalized consistently with whether the restriction is imposed on the optimized component or on the full mixture. For a generic input value \(\eta=(c,\theta,\pi,q)\), let \(\mathcal C_J(c,\theta)\) denote the set of probability measures satisfying all equality rows and the inequalities in \(g(J)\). For integers \(a,b\), let \(\mathcal C_J^{(a)}(c,\theta)\) be the subset of \(\mathcal C_J(c,\theta)\) consisting of measures supported on at most \(a\) points, and let \(\mathcal C_J^{(a,b)}(c,\theta)\) be the further subset whose masses are all multiples of \(1/b\). Lemma~\ref{lem:empirical} shows that an \(N_2\)-observation empirical measure with at most \(N_1\) distinct support points is exactly an element of \(\mathcal C_J^{(N_1,N_2)}(c,\theta)\). Thus the finite-sample restrictions for the optimized component are equivalent to requiring
\(\mu_{y,Q}\in\mathcal C_{J_n}^{(N_n,N_n)}(\hat c,\hat\theta)\).

For any generic \(\eta=(c,\theta,\pi,q)\), define the exact finite-sample value function, conditional on the reporting pattern and hence on \(N_n\), by
\[
V_{J_n}^{N_n,N_n}(\eta)
:=
\begin{pmatrix}
\inf_{\mu\in\mathcal C_{J_n}^{(N_n,N_n)}(c,\theta)}F(\mu;\pi,q)\\
\sup_{\mu\in\mathcal C_{J_n}^{(N_n,N_n)}(c,\theta)}F(\mu;\pi,q)
\end{pmatrix}.
\]
The reported sample bounds can therefore be written as
\(\hat V^{(n)}(\hat\eta):=V_{J_n}^{N_n,N_n}(\hat\eta)+\varepsilon_n\),
where \(\varepsilon_n\) is potential numerical optimization error. This notation keeps the exact finite-sample value \(V_{J_n}^{N_n,N_n}(\hat\eta)\) separate from its computed analogue \(\hat V^{(n)}(\hat\eta)\). To avoid clutter, later approximation displays write \(n\) and \(m\) for the relevant empirical grid sizes. In mixed exact/non-exact data these are the corresponding non-exact subsample sizes; under Assumption~\ref{ass:input_process} with \(\pi_0<1\), they are of the same order as the total sample size, so the \(O(1/n)\) grid approximations and the \(\sqrt n\) sampling normalization are unchanged.

\subsubsection*{Regularity conditions}\label{subsec:asymp_assumptions}
The regularity conditions below are organized around how they are verified in empirical work. Assumptions~\ref{ass:support} and~\ref{ass:valid_coarsening} formalize the population object and the scenarios discussed earlier. Assumption~\ref{ass:restriction_geometry} then specifies the allowed constraints, and Assumption~\ref{ass:objective} specifies the allowed inequality indices. Assumptions~\ref{ass:finite_complexity} and~\ref{ass:tail_existence} control the remaining terms in the asymptotic expansion, and refer to the number and complexity of constraints. Assumptions~\ref{ass:input_process} and~\ref{ass:implementation} further give the assumptions on sampling and on numerical error. Finally, we require two stability checks of the optimization problem, which limit certain knife-edge solutions, and hold fairly generically.

For any \(J\), write \(V_J\) for the value function obtained from \(V_\infty\) by replacing \(\mathcal C_\infty\) with \(\mathcal C_J\). Define \(V_J^a\) and \(V_J^{a,b}\) analogously, using \(\mathcal C_J^{(a)}\) and \(\mathcal C_J^{(a,b)}\). Thus the population, finite-\(J\), and empirical value functions differ only in the input value at which they are evaluated and in the class of probability measures allowed by the constraints. Let \(\mathbb D_0\subseteq\mathbb H\) collect the admissible first-stage directions used for the directional delta method. This tangent set enforces probability-mass preservation for the exact-observation distribution, treats that coordinate as inactive when \(\pi_0=0\), and includes the objective-specific continuity restrictions needed for indices such as the Gini, Hoover, and quantile ratios.\footnote{Formally, \(\mathbb D_0\) is the set of \(h\in\mathbb H\) for which there are admissible local inputs \(\eta_r\) in the maintained extension and numbers \(t_r\downarrow0\) such that \(t_r^{-1}(\eta_r-\eta_0)\to h\). Because \(q\) is a distribution function, admissible \(q\)-directions preserve total probability mass. When \(\pi_0=0\), we fix \(q_0=q^\dagger\), set \(\hat q=q^\dagger\) whenever \(\hat\pi=0\), and set the \(q\)-coordinate of \(\mathbb D_0\) to zero. For inequality indices that depend on the exact-observation distribution function, the admissible \(q\)-perturbations also satisfy the continuity requirements verified in Appendix~\ref{sec:appendix_objective_verification}. For the local approximation, the value functions are evaluated on a small enlargement of the \((\pi,q)\) coordinates: \(\pi\) may move in a real neighborhood of \(\pi_0\), and \(q\) is interpreted through the index-specific formulas in Appendix~\ref{sec:appendix_objective_verification}. This enlargement agrees with the original definitions on valid probability inputs and ensures that \(\eta_0+t h\) is defined for every \(h\in\mathbb D_0\) and all sufficiently small \(|t|\). It affects only how the index is evaluated; the underlying optimization remains over probability measures in \(\mathcal M\).} All conditions are local on a product neighborhood \(\mathcal N_\eta\) of \(\eta_0\), with projections \(\mathcal N_\theta\) and \(\mathcal N_{\pi q}\). Our first assumption is the following.

\begin{assumption}\label{ass:support}
\(P_0(Y\in\YY)=1\), where \(\YY=[\underline y,\overline y]\subset\mathbb R\) is compact.
\end{assumption}

\assinterp{This condition fixes the variable whose inequality is being bounded. With an open-ended top bracket, changing the imposed upper bound can change the estimand. Without an upper support restriction, the upper bound on many inequality indices is uninformative.}

We also formalize our two observation scenarios. In Scenario~1, the common brackets form a partition \(I_1=[a_0,a_1]\) and \(I_b=(a_{b-1},a_b]\), \(b=2,\ldots,B\). In Scenario~2, write the sampling unit as \(W=(D,Y,L,U,X)\), with \(D=1\) denoting an exact observation.

\begin{assumption}\label{ass:valid_coarsening}
In Scenario~1, \(\YY=\bigcup_{b=1}^BI_b\) and the brackets are disjoint. In Scenario~2, \(P_0(L\le Y\le U)=1\) and \(P_0(D=1,L=U=Y)=P_0(D=1)\).
\end{assumption}

\assinterp{This condition rules out interval miscoding and states that exact observations are correctly recorded. In a grouped table, the bracket partition supplies the restrictions. In an unfolding-bracket survey, each respondent's answers determine the interval \([L,U]\).}

The next condition restricts the functional form of the constraints in the
optimization problem. It does not limit the number of released inequalities.
Instead, after the support is partitioned at the cutoffs relevant for the
restrictions, each restriction must be affine in the variable of interest within
each cell. Write \(\theta=(\tau,\gamma)\), where \(\tau\) collects the cutoff values at
which a restriction may change form and \(\gamma\) collects the remaining nuisance
parameters. Our assumptions require that for each \(n\), there exists a partition $\underline y=t_{0,n}(\tau)<\cdots<t_{s_n,n}(\tau)=\overline y$ of \(\YY\) into adjacent cells
\(I_{d,n}(\tau)\). Each equality or inequality restriction has the form
\[
\int f_{u,n}(y;\theta)\,d\mu(y)=c_u
\qquad\text{or}\qquad
\int f_{u,n}(y;\theta)\,d\mu(y)\le c_u,
\]
where, within each cell,
\[
f_{u,n}(y;\theta)=a_{u,d,n}(\gamma)y+b_{u,d,n}(\gamma)
\qquad\text{for }y\in I_{d,n}(\tau).
\]
Thus each restriction value is determined by the probability mass and first moment of
\(\mu\) in each cell. We call these step restrictions when $a_{u,d,n}(\gamma)\equiv0$ on every relevant cell, because their left-hand sides are constant within cells. The formal condition is as follows.

\begin{assumption}\label{ass:restriction_geometry}
The following conditions hold uniformly over all sufficiently large \(n\), all
rows \(u\) covered below, all cells \(d\), and all
\(\theta=(\tau,\gamma)\in\mathcal N_\theta\). The reduced equality system has
uniformly bounded dimension; in particular, the number of retained equality rows
and the number of retained non-step equality rows are bounded uniformly in \(n\).

\begin{enumerate}

\item[(i)] Every retained equality row and every inequality row in \(g(\infty)\)
has the affine-within-cell representation above. A step row means a row for
which \(a_{u,d,n}(\gamma)\equiv0\) for every relevant cell \(d\) and every
\(\gamma\in\mathcal N_\gamma\).

\item[(ii)] The cutoff maps \(t_{d,n}(\tau)\) are continuously differentiable in
\(\tau\). The coefficient maps \(a_{u,d,n}(\gamma)\) and
\(b_{u,d,n}(\gamma)\) are continuously differentiable in \(\gamma\). All these
derivatives are uniformly bounded and have a common local modulus of continuity.

\item[(iii)] The \(f_{u,n}(y;\theta)\) have uniformly bounded sup norms and uniformly bounded total variation. Their within-cell slopes \(|a_{u,d,n}(\gamma)|\) are uniformly bounded.

\item[(iv)] Boundary atoms are handled by the cutoff convention in
Appendix~\ref{sec:appendix_regularities}. A support point at a cutoff is
evaluated using the one-sided affine or step value associated with its chosen
cell-assigned branch; step rows are not required to take the same value on the
two adjacent branches. For retained step equality rows, the boundary
compatibility condition \eqref{eq:boundary_consistency} holds uniformly: at
every cutoff where a retained step equality changes value, the cumulative mass
on either side of the cutoff is determined by the retained step equality rows
with integer coefficients whose absolute sum is bounded uniformly in \(n\).
\end{enumerate}

Finally, the baseline feasible set is nonempty: $\mathcal C_\infty(c_0,\theta_0)\ne\varnothing$.
\end{assumption}

The standard grouped-data and survey restrictions considered in this paper
satisfy this structure. Subgroup probabilities, adjacent-subgroup restrictions,
and cumulative-share restrictions are step restrictions, while overall means, subgroup
means, income shares, and points on the Lorenz curve are affine restrictions. Linear aggregate
ratios are also included after cross-multiplication when the denominator has a
known sign, and overlap restrictions from a finite set of possible survey intervals are covered by treating
the corresponding interval endpoints as cell boundaries. Appendix~\ref{sec:appendix_regularities} verifies
the functional form, smoothness, and cutoff-boundary convention for these
cases.

Our next assumption places restrictions on the kinds of inequality indices we can apply our statistical theory to. It requires the index to be stable when the optimizing distribution is moved slightly, to be continuous in the exact-observation inputs \((\pi,q)\), and to have a uniform first-order expansion once the endpoint program has been reduced to finitely many support points.

More formally, let \(\mathcal O\) be a fixed product neighborhood, using \(\Wone\) in the measure coordinate, of the closure of all triples \((\mu,\pi,q)\) with \(\mu\in\mathcal C_{J_n}(c,\theta)\) or \(\mu\in\mathcal C_\infty(c,\theta)\), \(\eta\in\mathcal N_\eta\), and \(n\) sufficiently large. For objectives whose regularity is verified on generated finite branches, such as quantile ratios, \(\mathcal O\) is understood to be the corresponding generated local domain containing the endpoint branches and the support-reduction, mass-rounding, equality-repair, and Slater-mixture paths used in the approximation and regularity arguments below. On the local extension of the \((\pi,q)\) coordinates, write
\[
d_{\pi q}((\pi,q),(\tilde\pi,\tilde q))
:=
|\pi-\tilde\pi|+\|q-\tilde q\|_{\ell^\infty(\mathcal F_0)}.
\]
For a finite-support candidate \(x=(p,z)\), write \(\mu_x=\sum_jp_j\delta_{z_j}\) and \(\Phi_k(x;\pi,q)=F(\mu_x;\pi,q)\). The finite-dimensional expansion below is required only on fixed neighborhoods of the reduced baseline optimizer sets, but the remainder must be uniform across those neighborhoods, all sufficiently large \(n\), and directions in compact subsets of \(\mathbb D_0\).

\begin{assumption}\label{ass:objective}
The inequality measure \(\mathcal J\), equivalently the objective \(F(\mu;\pi,q)=\mathcal J((1-\pi)\mu+\pi q)\), is admissible on the local objective domain \(\mathcal O\) in the following sense.

\begin{enumerate}
\item[(i)] Every denominator that appears in the objective is uniformly bounded away from zero on \(\mathcal O\).

\item[(ii)] There is \(L<\infty\) such that $|F(\mu;\pi,q)-F(\tilde\mu;\pi,q)|
\le L\Wone(\mu,\tilde\mu)$ whenever \((\mu,\pi,q)\) and \((\tilde\mu,\pi,q)\) belong to \(\mathcal O\). Furthermore, $|F(\mu;\pi,q)-F(\mu;\tilde\pi,\tilde q)|
\le
\omega_F(d_{\pi q}((\pi,q),(\tilde\pi,\tilde q)))$ for a common modulus \(\omega_F(a)\to0\).

\item[(iii)] On the retained optimizer neighborhoods, the support-coordinate gradients of \(\Phi_k\) are uniformly bounded and have a common modulus of continuity. Moreover, \(\Phi_k\) has the common finite-dimensional value and gradient expansion stated in Appendix~\ref{sec:appendix_regularities}.
\end{enumerate}
\end{assumption}

\noindent Appendix~\ref{sec:appendix_objective_verification} proves the following coverage results.
\begin{enumerate}
\item[(a)] Smooth functions of finitely many \(C^2\) moments are allowed when their moment vector stays in the interior of the domain of the formula and all denominators are locally bounded away from zero. This covers the mean log deviation, the Theil index, \(\mathrm{GE}_\alpha\), the Atkinson class, the Kolm class, and linear moment functionals, subject to their usual domain restrictions such as positive support for logarithms or negative powers and a positive mean when the mean is a denominator.
\item[(b)] The Gini index is allowed when the feasible mean is locally bounded away from zero, repeated optimizer atoms are merged, distinct optimizer support points remain locally separated, and the exact-observation distribution function \(q\), together with the admissible \(q\)-direction functions, is locally continuous at those optimizer support points.
\item[(c)] The Hoover index is allowed under the Gini-type requirements and, additionally, the optimizer mean must stay locally separated from the optimizer support points.
\item[(d)] Quantile ratios \(Q_{\tau_2}/Q_{\tau_1}\) are allowed only when the lower quantile is positive and the endpoint program has a stable finite quantile branch: the relevant quantile atoms have positive mass, are separated from neighboring support points, have cumulative-probability gaps around \(\tau_1\) and \(\tau_2\), and remain the same matched atoms through the support-reduction, mass-rounding, equality-repair, and Slater-mixture steps used in the proof.
\end{enumerate}
Thus the objective condition is broad but not automatic. For smooth moment-based indices, verification is an ordinary formula-domain check: the support must lie where the index is defined, the relevant moment vector must stay in the interior of the formula's domain, and denominators must remain bounded away from zero. For Gini, Hoover, and quantile ratios, verification is a finite endpoint diagnostic: after solving the reduced programs, merge repeated atoms and check the separation, no-kink, or quantile-branch conditions listed above.

Assumption~\ref{ass:finite_complexity} states that, for \(\sqrt n\) asymptotics, the endpoint programs can be asymptotically reduced to measures with a bounded number of support points.

\begin{assumption}\label{ass:finite_complexity}
There are integers \(k_n\) with \(\sup_n k_n<\infty\) and deterministic \(r_n=o(n^{-1/2})\) such that, uniformly over \(\eta=(c,\theta,\pi,q)\in\mathcal N_\eta\),
\[
\sup_{N\ge k_n:\,\mathcal C_{J_n}^{(N,N)}(c,\theta)\ne\varnothing}
\bigl\|V_{J_n}^{N,N}(\eta)-V_{J_n}^{k_n,N}(\eta)\bigr\|_2\le r_n.
\]
Moreover, whenever \(\mathcal C_{J_n}^{(N,N)}(c,\theta)\ne\varnothing\), we also have \(\mathcal C_{J_n}^{(k_n,N)}(c,\theta)\ne\varnothing\).
\end{assumption}

Assumption~\ref{ass:finite_complexity} requires the at-most-\(k_n\)-support restriction to change the endpoint vector by only \(o(n^{-1/2})\), uniformly over local perturbations of \(\eta\). In our main cases of interest, this holds exactly: one may take \(k_n\le 2D\) in Scenario~1A, \(k_n\le 2D+q_1+q_2\) for the linear-fractional Scenario~1B problem with a fixed number of auxiliary rows, and \(k_n\le |\mathcal B|+1\) in Scenario~2 when the set \(\mathcal B\) of possible interval endpoints has bounded cardinality; see Propositions~\ref{prop:base_cornersol} and~\ref{prop:Sc1BsolLF} and Theorems~\ref{th:Sc3}--\ref{th:Sc3_LF}. In all of these cases \(r_n=0\) exactly. Thus Scenario~2 is not excluded by the fixed-\(D\) language used for Scenario~1; it enters through the finite endpoint set \(\mathcal B\), or through the more general bounded-effective-support condition stated here. The empirical content of this assumption is that effective support size is controlled by dimension of the constraints, not by sample size. An infinite family of inequalities may still satisfy the assumption when its endpoint solutions use only a fixed number of atoms.

The full population problem can impose more inequalities than the \(J_n\)-program, in order to allow for Scenario~2. As mentioned above, equality rows are eventually all included and always imposed exactly. The next condition also imposes that we include all necessary inequalities fast enough that we do not create an extra term in the asymptotic distribution. In effect, this says that the omitted inequalities affect the constraint set by $o(n^{-1/2})$.

\begin{assumption}\label{ass:tail_existence}
For all large \(n\), \(h(J_n)=h(\infty)\), and there is \(\kappa_{J_n}=o(n^{-1/2})\) such that
\[
\sup_{\eta\in\mathcal N_\eta}
\sup_{\mu\in\mathcal C_{J_n}(c,\theta)}
\sup_{u\in g(\infty)\setminus g(J_n)}
[G_u(\mu;\theta)-c_u^2]_+
\le\kappa_{J_n}.
\]
\end{assumption}

\assinterp{The innermost supremum is defined as zero when no inequality is omitted. If \(g(\infty)\) is finite, all inequalities are eventually included and \(\kappa_{J_n}=0\). This includes an unfolding-bracket survey with finitely many possible \([L,U]\) intervals. If \(g(\infty)\) is infinite, the condition allows \(J_n\) to grow but requires every distribution feasible for the retained restrictions to violate each omitted restriction by at most \(o(n^{-1/2})\). Appendix~\ref{subsec:appendix_approximation_primitives} gives two sufficient cases: a compactly indexed, uniformly Lipschitz set of constraints with grid mesh of size \(o(n^{-1/2})\), and the moving interval inequalities in Scenario~2, where monotonicity applies when every omitted interval is bracketed by retained inner and outer grid intervals and the released containment lower-bound and overlap upper-bound probabilities change by \(o(n^{-1/2})\) over that bracket.}

The next condition is the only sampling-based assumption. It requires a central
limit theorem for the first-stage input vector \(\hat\eta\), together with the
corresponding conditional central limit theorem for the resampled input used
for inference. Let \(\eta_m^*\) be the input vector recomputed by the
resampling procedure, and let \(\eta_n^*\) denote its ordinary-bootstrap
analogue when that procedure is used.

\begin{assumption}\label{ass:input_process}
The first-stage inputs satisfy $\sqrt n(\hat\eta-\eta_0)\Rightarrow Z
 \, \text{in }\mathbb H$, where \(Z\) is a tight, mean-zero Gaussian element and
\(Z\in\mathbb D_0\) almost surely. For every \(m\to\infty\) with
\(m/n\to0\), $\sqrt m(\eta_m^*-\hat\eta)\Rightarrow Z$ conditionally in probability in \(\mathbb H\). If ordinary-bootstrap inference
is used, then $\sqrt n(\eta_n^*-\hat\eta)\Rightarrow Z$ conditionally in probability in \(\mathbb H\).
\end{assumption}

\assinterp{The condition follows under standard assumptions on how $\hat{\eta}$ is
constructed. In the simplest case, for example, \(\hat c\) may contain a sample
mean and \(c_0\) the corresponding population mean; then an i.i.d.\ sample
suffices for the usual central limit theorem, with the required moment
conditions supplied by Assumption~\ref{ass:support}. More generally, the
finite-dimensional coordinates of \((c,\gamma,\pi)\) may be smooth functions
of empirical moments with finite second moments, cutoff coordinates may be
fixed or empirical quantiles with a density continuous and bounded above and
away from zero near the target, and \(q\) may be the empirical distribution
among exact observations. With a growing family of released coordinates, the
corresponding indexed empirical process must satisfy a uniform central limit
theorem. In Scenario~2, the unconditional containment and overlap counts are averages of
bounded threshold indicators in \((D,L,U)\), and the conditional shares used for the optimized non-exact component are smooth transformations of these averages and \(1-\hat\pi\) when \(\pi_0<1\). These indicators form bounded VC
classes and are therefore \(P_0\)-Donsker, so the required growing-coordinate
central limit theorem applies to every finite or increasing retained
subcollection. In survey, administrative, or tabulation settings, the same
condition may instead follow from a design central limit theorem and from the
replication, resampling, or simulation method supplied with the release.
Appendix~\ref{subsec:appendix_approximation_primitives} gives formal i.i.d.\ sufficient
conditions.}

Furthermore, we assume that numerical approximation is asymptotically
negligible. There are two possible implementation errors. First, the endpoint
programs may be solved only up to a tolerance. Second, empirical-mass programs
can use only grid-valued block masses: the relevant masses must lie on the
\(1/n\) grid for the sample program and on the \(1/m\) grid for an
\(m\)-out-of-\(n\) resample. Unrestricted-mass programs have no such grid
requirement. The grid requirement is automatic when the step-equality block masses are
empirical shares; otherwise, we may project the released inputs onto the
appropriate grid by an asymptotically negligible adjustment of the
step-equality right-hand sides. Below, the same symbols \(\hat\eta\),
\(\eta_m^*\), and \(\eta_n^*\) denote the possibly projected inputs. Let
\(\varepsilon_n\), \(\varepsilon_m^*\), and \(\varepsilon_n^*\) denote the
corresponding sample, \(m\)-out-of-\(n\), and ordinary-bootstrap optimization
errors.

\begin{assumption}\label{ass:implementation}
The optimization errors are first-order negligible: \(\|\varepsilon_n\|_2=o_p(n^{-1/2})\), \(\|\varepsilon_m^*\|_2=o_{P^*}(m^{-1/2})\), and, when the ordinary bootstrap is used, \(\|\varepsilon_n^*\|_2=o_{P^*}(n^{-1/2})\). Any projection used to make the block masses implied by the step equalities grid-valued is \(o_p(n^{-1/2})\) for the sample input, \(o_{P^*}(m^{-1/2})\) for the \(m\)-out-of-\(n\) input, and \(o_{P^*}(n^{-1/2})\) for the ordinary-bootstrap input.
\end{assumption}

\assinterp{The assumption permits solver tolerances and, for empirical-mass
programs, small mass-grid adjustments. It requires both to be smaller than the
sampling error, so neither affects the first-order limit. Because any grid
projection is negligible in the input norm, it does not change the central
limit theorem in Assumption~\ref{ass:input_process}. Appendix~\ref{subsec:appendix_approximation_primitives}
states when the grid requirement holds exactly and how a projection enters the
approximation.}

The final two assumptions are stability checks for the finite programs that
compute the bounds. Assumption~\ref{ass:feasibility_rank} concerns the feasible
set before optimizing the index. Assumption~\ref{ass:endpoint_rank} concerns the
lower- and upper-bound solutions after the endpoint programs have been solved. We state the checks in cell-moment form because this is how they are verified
numerically. Let \(M=(m,s)\), where \(m_d\) is the probability of cell \(d\) and
\(s_d\) is its first moment. For a cutoff vector \(\tau\), define
\[
\mathcal P_n(\tau)
:=
\{(m,s):m_d\ge0,\ \sum_dm_d=1,\ t_{d-1,n}(\tau)m_d\le s_d\le t_{d,n}(\tau)m_d\}.
\]
Let \(R_n^H(M;\theta)\) denote the row-reduced equalities and let
\(R_{\infty,n}^G(M;\theta)\) denote the full inequality family. Set
\[
\mathcal P_n^H(c,\theta)
:=
\{M\in\mathcal P_n(\tau):R_n^H(M;\theta)=c^1\}.
\]
Thus \(\mathcal P_n^H(c,\theta)\) is the cell-moment feasible set after imposing
the equalities.\footnote{Interiority is always relative to this equality face:
cells or directions forced by the equalities do not count as failures of
interiority. A finite-support representation may also include zero masses,
repeated locations, and the fixed padding coordinates already counted in
\(k_n\). Appendix~\ref{subsec:appendix_regularity} records the corresponding
equality Jacobian.}

\begin{assumption}\label{ass:feasibility_rank}
The following hold uniformly over \(\eta\in\mathcal N_\eta\).

\begin{enumerate}
\item[(i)] After redundant equality rows are removed, the smallest singular value
of the cell-moment equality matrix is bounded away from zero.

\item[(ii)] There exist \(\underline s>0\) and a measure
\(\mu_n^{\mathrm{sl},\infty}(\eta)\) with cell-moment vector
\(M_n^{\mathrm{sl}}(\eta)\in\operatorname{relint}\mathcal P_n^H(c,\theta)\) such
that
\[
R_{\infty,n}^G(M_n^{\mathrm{sl}};\theta)\le c^2-\underline s\mathbf 1.
\]

\item[(iii)] For every \(N\ge k_n\), each feasible measure with at most \(N\)
support points admits an augmented \(N\)-coordinate representation and a
same-support probability vector satisfying the equalities and all inequalities
in \(g(J_n)\) with slack at least \(\underline s\). For the mass and
interior-location coordinates selected for equality repair, the step-mass block
and the unweighted affine-slope block in
Appendix~\ref{subsec:appendix_regularity} have uniformly bounded inverses. Under
the strictly feasible repair weights, the selected affine-adjustment masses are
bounded away from zero and the selected locations remain a positive distance
from their cell boundaries.
\end{enumerate}
\end{assumption}

Assumption~\ref{ass:feasibility_rank} rules out knife-edge feasibility. It is
checked in three steps. First, remove redundant equality rows and compute the
smallest singular value of the resulting equality matrix. Second, maximize the
minimum inequality slack over \(\mathcal P_n^H(c,\theta)\); a positive optimum
verifies the strict-feasibility part of the assumption.\footnote{With finitely
many inequalities this is an ordinary finite-dimensional auxiliary program.
With an infinite family, the same positive margin must hold uniformly over the
normalized rows. In the settings considered here this can be checked by the
finite reduction or by the Lipschitz/grid and interval-monotonicity
arguments in Appendix~\ref{subsec:appendix_approximation_primitives}.} Third, check that the finite-support representation
has enough local freedom to repair small equality perturbations without leaving
the feasible set. Numerically, this can be implemented via the nonsingularity check for the
step-mass and affine-slope blocks displayed in
Appendix~\ref{subsec:appendix_regularity}.

The intuition is simple: the equalities should define a well-behaved face, the
inequalities should leave room inside that face, and the fixed padding
coordinates should absorb the small perturbations created by replacing
population inputs with sample inputs. These requirements are generic. Once
redundant equalities have been removed, and inequalities that bind for every
feasible distribution have been treated as equalities, failure requires an exact
degeneracy: a rank loss, the disappearance of strict slack, or a chosen repair
direction becoming tangent to the boundary. Such events are not stable under
small perturbations of right-hand sides or nuisance parameters. In
Scenario~1A, the bracket-share block is independent after the simplex redundancy
and equality-implied zero cells are removed; additional mean or Lorenz
restrictions require the selected within-cell slope matrix to have full rank. In
the baseline Scenario~2 formulation there are no nontrivial equalities beyond
total mass, so the overlap restrictions are checked through the strict-feasibility
and endpoint active-set conditions.

The final regularity condition is the finite-dimensional analogue of a standard regularity condition for constrained optimization. It does not require the lower or upper endpoint optimizer to be unique. It requires only that the solved endpoint program not sit at a knife-edge: optimizer atoms should not vanish or collide, inactive inequalities should not be arbitrarily close to binding, and the active constraints should have full rank. The condition is checked after solving the lower- and upper-bound programs. For each endpoint, start from a cleaned optimizer representation: merge repeated atoms, remove zero-mass atoms, and delete duplicate or locally implied active rows.\footnote{If an optimizer atom lies exactly at a cutoff, retain each adjacent-cell representation that is consistent with the step equalities and can arise from a local feasible perturbation. Away from cutoffs there is only one relevant local representation.} Let \(\mathcal A_n(x^*;\eta)\) be the derivative matrix formed from the simplex row, the retained equalities, the binding inequalities, and the binding support-position bounds. For \(\circ\in\{\inf,\sup\}\), set \(\sigma_{\inf}=1\) and \(\sigma_{\sup}=-1\), and let \(\mathcal S_n^\circ(\eta)\) be the set of optimizer measures for \(V_{J_n}^{k_n,\circ}(\eta)\).

\begin{assumption}\label{ass:endpoint_rank}
On a neighborhood of \(\eta_0\), both reduced endpoint values are attained for
all sufficiently large \(n\). For each such \(n\), each endpoint, and each local
input, the compatible reduced optimizer representations form a nonempty compact
set. Uniformly over these representations, positive optimizer masses are bounded
away from zero, distinct optimizer support points remain separated, and
nonboundary atoms remain a positive distance from cell boundaries. The
active-gradient matrices are uniformly full rank, in the sense that $\sigma_{\min}(\mathcal A_n(x^*;\eta)\mathcal A_n(x^*;\eta)^\top)$ is bounded away from zero. Inactive inequalities and inactive support bounds
have common positive slack. Finally, for each \(\circ\in\{\inf,\sup\}\) and
every \(\varepsilon>0\), there is \(\gamma_\varepsilon>0\) such that, uniformly
over all sufficiently large \(n\) and every
\(\mu\in\mathcal C_{J_n}^{(k_n)}(c_0,\theta_0)\),
\[
 \inf_{\nu\in\mathcal S_n^\circ(\eta_0)}\Wone(\mu,\nu)\ge\varepsilon
 \quad\Longrightarrow\quad
 \sigma_\circ F(\mu;\pi_0,q_0)
 \ge\sigma_\circ V_{J_n}^{k_n,\circ}(\eta_0)+\gamma_\varepsilon.
\]
\end{assumption}

Assumption~\ref{ass:endpoint_rank} formalizes this stability requirement. Several regular optimizers may coexist, which is one reason the value function is generally directionally rather than fully differentiable. The assumption rules out fragile optima, such as atoms with vanishing mass, colliding support points, deficient active-constraint rank, inactive constraints with vanishing slack, or feasible distributions away from the optimizer set with objective values arbitrarily close to the endpoint.

The practical diagnostics are the quantities produced by the endpoint programs.
For each lower and upper endpoint, and for each compatible optimizer
representation, compute the minimum positive atom mass, the minimum distance
between distinct support points, the minimum distance of nonboundary atoms from
cell boundaries, the minimum inactive slack, and $\sigma_{\min}(\mathcal A_n(x^*;\eta)\mathcal A_n(x^*;\eta)^\top)$. These quantities should be bounded away from zero uniformly over the local
neighborhood. The final objective-gap condition can be checked by resolving the
finite endpoint problem after excluding an \(\varepsilon\)-neighborhood of the
baseline optimizer set in \(\Wone\). Equivalently, one can use the finite
active-set sufficient condition in Appendix~\ref{subsec:appendix_regularity}:
optimal active sets must pass the rank and slack checks, while nonoptimal active
sets must have objective values separated from the endpoint by a common margin.

These endpoint conditions are also generic. Strict positivity of masses,
separations, slacks, singular values, and objective gaps persists under small
perturbations. Failure therefore corresponds to a lower-dimensional coincidence:
an atom disappears, two atoms merge, an inactive constraint becomes active
without a gap, or the active-gradient matrix loses rank. When such a degeneracy
is intrinsic to the model, the sharp bounds are still well-defined, but the
local expansion must be written on the appropriate lower-dimensional face or
treated as a more general nonsmooth problem. In applications, the sample
analogues of the diagnostics above are useful checks of whether the population
regularity conditions are plausible.

With these conditions in place, we can state the main statistical results.

\subsubsection*{Main results}\label{subsec:asymp_results}

The results have three steps. First, we study the deterministic value function \(V_\infty\), which sends the inputs into the sharp lower and upper endpoints. Second, we obtain the limiting distribution of the computed endpoints via the CLT and the directional delta method. Third, we show how that distribution can be estimated using the $m$-out-of-$n$ bootstrap.

The main point of the first step is that the value function is generally directionally, rather than fully, differentiable. This is because an endpoint may be attained by more than one optimizer. The derivative therefore allows the active optimizer to change with the perturbation direction: it takes the relevant envelope selection over the baseline optimizers. When each endpoint has a unique locally stable optimizer, and all compatible representations of any boundary atom give the same linear envelope derivative, this switching disappears and the derivative is linear.

\begin{prop}\label{prop:hdd_limit_phi_n}
Under Assumptions~\ref{ass:support},~\ref{ass:restriction_geometry},~\ref{ass:objective},~\ref{ass:finite_complexity},~\ref{ass:tail_existence},~\ref{ass:feasibility_rank}, and~\ref{ass:endpoint_rank}, \(V_\infty\) is Hadamard directionally differentiable at \(\eta_0\) tangentially to \(\mathbb D_0\), and its derivative \(V'_{\eta_0}:\mathbb D_0\to\mathbb R^2\) is Lipschitz continuous. If, in addition, for all sufficiently large reductions each endpoint has a unique optimizer measure and all compatible local representations of that optimizer have the same linear envelope derivative, then \(V_\infty\) is Hadamard differentiable at \(\eta_0\) tangentially to \(\mathbb D_0\).
\end{prop}

The derivative in Proposition~\ref{prop:hdd_limit_phi_n} is the many-constraints limit of the finite-program envelope derivatives constructed in Appendix~\ref{subsec:appendix_local_sensitivity}. For a retained local parameterization, it is obtained by differentiating the Lagrangian with respect to the inputs and then taking the appropriate envelope selection over the baseline optimizers, with opposite sign conventions for the lower and upper endpoints. This derivative is not needed to compute the sharp endpoints themselves; it is only used as a proof device to show the estimators' asymptotic distribution and to justify the bootstrap.

The second step adds the sample law. The finite-support reduction, mass discretization, omitted-inequality approximation, and numerical error are assumed to be smaller than the \(n^{-1/2}\) sampling scale. Consequently, the first-order distribution of the estimator is obtained by passing the limiting input process through the derivative \(V'_{\eta_0}\).

\begin{prop}\label{prop:asymptotics1}
Under the assumptions of Proposition~\ref{prop:hdd_limit_phi_n} and Assumptions~\ref{ass:valid_coarsening},~\ref{ass:input_process}, and~\ref{ass:implementation},
\[
\sqrt{n}\,[\hat V^{(n)}(\hat\eta)-V_\infty(\eta_0)]
\Rightarrow
V'_{\eta_0}(Z)
\]
in \(\mathbb R^2\).
\end{prop}

The limit in Proposition~\ref{prop:asymptotics1} is Gaussian when \(V'_{\eta_0}\) is linear, but it can be non-Gaussian when the value function is only directionally differentiable. For that reason, the bootstrap is stated in terms of a centered and scaled bootstrap statistic. In the \(m\)-out-of-\(n\) bootstrap, recompute \(\eta_m^*\) on an \(m\)-out-of-\(n\) resample and set \(\hat V_m^*:=V_{J_n}^{m,m}(\eta_m^*)+\varepsilon_m^*\), where, in mixed exact/non-exact data, the superscripts use the non-exact subsample size in the resample. The set of constraints \(J_n\) is kept fixed across resamples, so the bootstrap approximates the sampling law of the inputs to the same \(n\)-th endpoint problem. For the ordinary bootstrap, set \(\hat V_n^*:=V_{J_n}^{n,n}(\eta_n^*)+\varepsilon_n^*\), with the same grid-size convention. Define the centered bootstrap statistic
\[
R_m^*:=\sqrt{m}\,[\hat V_m^*-\hat V^{(n)}(\hat\eta)].
\]

\begin{prop}\label{prop:bootstrap}
Suppose the assumptions of Proposition~\ref{prop:asymptotics1} hold. If the \(m\)-out-of-\(n\) bootstrap satisfies Assumptions~\ref{ass:input_process} and~\ref{ass:implementation}, with \(m\to\infty\) and \(m/n\to0\), then \(R_m^*\Rightarrow V'_{\eta_0}(Z)\) conditionally in probability in \(\mathbb R^2\). If, instead, the ordinary bootstrap satisfies Assumptions~\ref{ass:input_process} and~\ref{ass:implementation}, and if \(V_\infty\) is Hadamard differentiable at \(\eta_0\) tangentially to \(\mathbb D_0\), then \(R_n^*:=\sqrt{n}\,[\hat V_n^*-\hat V^{(n)}(\hat\eta)]\Rightarrow V'_{\eta_0}(Z)\) conditionally in probability in \(\mathbb R^2\).
\end{prop}

The estimator and the bootstrap use different normalizations for different
purposes. The sampling object of interest is the statistic $\sqrt{n}\,[\hat V^{(n)}(\hat\eta)-V_\infty(\eta_0)]$. The \(m\)-out-of-\(n\) bootstrap uses \(\sqrt m\) to estimate the law of this
root, not to set the length of the final confidence interval. Thus, if
\(q_\alpha^{*,\circ}\) is the conditional \(\alpha\)-quantile of the bootstrap
root \(R_m^{*,\circ}\), the corresponding displacement on the endpoint scale is $\Delta_{\alpha,n}^{*,\circ}
:=
q_\alpha^{*,\circ}/\sqrt n$, rather than \(q_\alpha^{*,\circ}/\sqrt m\) and rather than a raw percentile of the bootstrap
endpoint. For endpoint coordinate \(\circ\in\{\inf,\sup\}\), write
\(\hat V^{(n),\circ}(\hat\eta)\) for the corresponding component of
\(\hat V^{(n)}(\hat\eta)\). If the limiting coordinate law is continuous at its
\(\alpha/2\) and \(1-\alpha/2\) quantiles, a two-sided root-based interval with
asymptotic coverage \(1-\alpha\) is therefore
\[
\left[
\hat V^{(n),\circ}(\hat\eta)-\frac{q_{1-\alpha/2}^{*,\circ}}{\sqrt n},
\quad
\hat V^{(n),\circ}(\hat\eta)-\frac{q_{\alpha/2}^{*,\circ}}{\sqrt n}
\right].
\]

The same bootstrap roots can cover the entire identified set, which is an interval here. Let
\(c_{1-\alpha}^*\) be the conditional \((1-\alpha)\)-quantile of
\(\max(R_m^{*,\inf},-R_m^{*,\sup})\). Equivalently, the endpoint-scale
expansion is \(c_{1-\alpha}^*/\sqrt n\). If the limiting law is continuous at
this quantile, then
\[
\left[
\hat V^{(n),\inf}(\hat\eta)-\frac{c_{1-\alpha}^*}{\sqrt n},
\quad
\hat V^{(n),\sup}(\hat\eta)+\frac{c_{1-\alpha}^*}{\sqrt n}
\right]
\]
contains the population identified set
\([V_\infty^{\inf}(\eta_0),V_\infty^{\sup}(\eta_0)]\) with asymptotic
probability at least \(1-\alpha\). The analogous max-absolute root gives
simultaneous two-sided endpoint intervals.

Two implementation cautions are important. First, when \(m/n\to0\), raw
percentiles of \(\hat V_m^*\) are not the intervals justified by
Proposition~\ref{prop:bootstrap}. The justified intervals are obtained from
bootstrap root quantiles and then put on the endpoint scale by dividing by
\(\sqrt n\), as in the displays above. Second, Gaussian critical values are
justified by the general theory only when the derivative is linear. Empirical
work should therefore report \(m\), the number of resamples, and the root
transformation used to form the interval. The asymptotic theory alone does not select a unique finite-sample value of
\(m\), but it requires an intermediate sequence with \(m\to\infty\) and
\(m/n\to0\) \citep{Shao1994}. A common implementation is
\(m=\lfloor n^\gamma\rfloor\) for \(\gamma\in(1/2,1)\). If the ordinary bootstrap is used instead, the full Hadamard
differentiability and the ordinary-bootstrap version of
Assumptions~\ref{ass:input_process} and~\ref{ass:implementation} should be
verified.

Finally, the bootstrap must resample the objects that generate the estimated
inputs. In Scenarios~1A and~2, the inputs are empirical averages over sampling
units, so one can resample those units and recompute
\((\hat c,\hat\theta,\hat\pi,\hat q)\). In Scenario~1B, the same procedure is
valid only if all extra components of $\hat{\eta}$---such as subgroup means---are recoverable from the resampled
microdata, or when they can be resampled jointly using replicate estimates or a
covariance model. Otherwise the sharp endpoints remain well defined, but
inference without further assumptions applies to the outer interval obtained
after omitting the non-resampleable auxiliary constraints.

\section{Applications} 
\label{sec:applications}
This section illustrates the framework with two applications, one for each observational scenario. The first uses the English Longitudinal Study of Ageing (ELSA), where many households report exact values while others provide respondent-specific intervals generated by unfolding brackets (Scenario~2). The second uses published U.S.\ income distribution tables from the mid-twentieth century, which report frequencies in non-overlapping income brackets and, in some years, additional aggregates such as subgroup means or selected quantiles (Scenario~1). In the first application we report conventional point estimates based on common missing-data treatments alongside our sharp identified sets. This comparison shows how much of the precision reported by conventional imputation-based estimates is attributable to the imputation assumption itself rather than to information contained in the data.

\subsection{Household level wealth data}
\label{app:wealth}

Household wealth surveys routinely face item nonresponse: respondents may be unwilling to report exact amounts or may not know them precisely. To reduce missingness while limiting respondent burden, many wealth surveys use unfolding brackets: if a respondent does not report an exact value, the survey asks a short sequence of Yes/No threshold questions (e.g.\ ``Is it more or less than \(X\)?'') that places the value in an interval. The thresholds are typically randomized to mitigate anchoring and response-order effects. \citet{JusterSmith1997} and \citet{JusterSmithStafford1999} discuss the design and performance of this approach, which is now used in a range of surveys including HRS, PSID, ELSA and the Survey of Health, Ageing and Retirement in Europe.

This design naturally produces exact-value and interval-valued observations with respondent-specific intervals (Scenario~2). We use the 2018/19 wave of ELSA (see \citet{ELSA_IJE2013}), focusing on households whose financial respondent is aged 50--74. This yields \(n=4{,}422\) observations. We study two measures of liquid savings: (i) a narrow single-question measure with a simple interval data structure and (ii) a broader composite measure that aggregates three components and therefore exhibits more interval complexity. For each measure we compare sharp bounds with commonly used imputation-based point estimates.

The most straightforward, and narrowest definition of liquid precautionary savings is balances in savings and checking accounts at banks or building societies (a mutual financial institution in the UK that serves a similar role as community banks or credit unions in the US). ELSA financial respondents are asked to give the value of their household's total current balances in bank and building society accounts in a single question response, returning a value of 0 if the household does not have such accounts. Of 4,422 observations in our sample, respondents were able and willing to give exact values for this variable in 3,827 (86\%) of the cases and sample statistics for this exact-value subsample are given in panel A of Table~\ref{table:WealthSAV}. Of the remaining 595 observations with respondent-specific intervals, 233 had some kind of bounded interval data generated from the unfolding bracket procedure, and a further 362 had interval data generated from unfolding brackets that was unbounded at the top. For simplicity here we have taken twice the final cutoff of the open-ended top bracket as the upper interval limit for all unbounded cases. Specifically, for this variable in ELSA, the final bracket cutoff is \textsterling 150,000, so we have set the value for an open-ended top bracket to be \textsterling 300,000.\footnote{Alternative choices are straightforward to implement.} Overall, the 595 observations generate only 11 distinct intervals.\footnote{These unique intervals are 
$[0,999]$, $[0,4999]$, $[0,19999]$, $[0,300000]$, $[1001,4999]$, $[1001,300000]$, 
$[5001,19999]$, $[5001,300000]$, $[20001,149999]$, $[20001,300000]$, $[150001,300000]$.
}

\begin{table}[tbp]
	\caption{Distribution of Wealth in Liquid Savings: Summary statistics}
	\label{table:WealthSAV}
	\begin{center}
		{Summary statistics computed for the subset of $N_p$ households with exact values}
	\end{center}
	\par
\small
	\begin{tabular}{ccccccccc}
		\hline
		Mean & St.\ dev. & 1st decile & 1st quartile & Median & 3rd quartile & 9th decile & Min & Max \\ \hline
\multicolumn{9}{l}{\textit{A. Narrow definition ($N_p=3,827$)}} \\
		29,247& 76,813 & 100 & 1000 & 7,000 & 26,000 & 70,000 & 0 & 1,300,000\\ 
\multicolumn{9}{l}{\textit{B. Broad definition ($N_p=3,708$)}} \\
 43,201& 95,775 & 150 & 1,731 & 11,500 & 45,000 & 110,000 & 0 & 1,450,000\\ 
 \hline	
 \multicolumn{9}{l}{Note: Values in \textsterling; Sample size is 4,422. Remaining observations have interval data. } 
	\end{tabular}
\end{table}

Other financial assets are functionally close to balances in bank and building society accounts because they offer similar liquidity, risk, and interest-rate characteristics. In the UK savings landscape, National Savings products and tax-advantaged Individual Savings Accounts held as cash, rather than stocks and shares, fall into this category. A broader measure of liquid savings therefore includes these balances. Constructing this broader measure requires aggregating three interview questions, each with its own missingness pattern. A respondent has an exact value for the broader variable if and only if all three components have exact values. If at least one component has an interval-valued response, the lower endpoint of the broader interval is the sum of the component lower endpoints and the upper endpoint is the sum of the component upper endpoints. A component reported as an exact value is treated as having identical lower and upper endpoints.

Aggregating variables, each with its own missingness pattern, generates more interval observations and a more complex interval structure. Out of the 4,422 observations in our sample, 3,708 have exact values for the broader measure of liquid savings and 714 respondents have interval-valued observations. Panel B of Table~\ref{table:WealthSAV} gives summary statistics for the subset of individuals with exact values. For the interval observations for this variable, there are 199 unique interval types; the five most frequent are $[0, 300000]$ (157 times), $[0, 340000]$ (109 times), $[0, 999]$ (39 times), 
 $[1001, 4999]$ (34 times),
 $[5001, 19999]$ (34 times). 

To place the bounds in context, we begin by computing values for the Gini coefficient in this household savings data using commonly-used approaches to deal with missing data. The most simplistic of these is simply to drop cases where exact values are missing and calculate the Gini solely on the basis of the continuous part of the sample. Researchers concerned with whether the data are missing at random, however, typically want to include information from the whole sample and therefore use a variety of imputation based methods. 

The simplest imputation approaches assign either the midpoint of the relevant bracket or the mean value computed from the exact-value subsample lying within the bracket to every observation whose exact value is missing. These procedures use the non-missing data, but they are not appropriate for inequality analysis because they mechanically reduce dispersion in the completed sample. Hot-deck imputation, which assigns each missing exact value a random donor from the exact observations lying within the relevant bracket, does not mechanically collapse within-bracket dispersion. We report a single hot-deck imputation estimate and a multiple hot-deck imputation estimate based on ten imputations, reducing dependence on any single draw (see \citet{Rubin1987} for an overview). Additional parametric imputation exercises, including shifted-lognormal and generalized Pareto specifications, are reported in Appendix \ref{app:additional_wealth_imputation}.

\begin{table}[tbp]
	\caption{Inequality in Liquid Savings}
	\label{table:GiniSAV}
	\begin{center}
		{Gini coefficients under different methodologies}
	\end{center}
	\par
    \begin{center}
	\begin{tabular}{lcccc}
		\hline
    & \multicolumn{2}{c}{A: Narrow definition} & \multicolumn{2}{c}{B: Broad definition} \\ 
		Method & Gini & se & Gini & se \\ \hline
    A. Continuous Data only & 0.7617 & 0.007396 & 0.7388 & 0.006685\\ 
    B. Mean imputation & 0.7358 & 0.007866 & 0.7094 & 0.007066\\
    C. Midpoint imputation & 0.7339 & 0.005792 & 0.7016 & 0.005653 \\ 
    D. Single hot-deck & 0.7567 & 0.006718 & 0.7288 & 0.006050 \\
    E. Multiple hot-deck & 0.7568 & 0.006717 & 0.7288 & 0.006050 \\
    \hline
\multicolumn{5}{l}{Note: Rows D and E present the mean value from 1000 repetitions.}\\ 
\multicolumn{5}{l}{Bootstrap standard errors are from 1000 bootstrap draws.}\\ 
	\end{tabular}
    \end{center}
\end{table}

The results from these exercises for both the narrow and broad definitions of liquid savings are presented in Table~\ref{table:GiniSAV}. The level of inequality is high compared to that typically seen in income distributions but this is entirely to be expected, and particularly so for households at the older end of the cycle where differences in income and expenditure will have accumulated up over time. Looking across the various measures, mean and midpoint imputation methods yield a lower measure of inequality as expected, while the inequality in the continuous (non-missing) sample is higher than that measured by the hot-deck imputation methods that take into account the unfolding-bracket information for the missing cases. The Gini coefficients (and their s.e.) for the broader savings definition presented in Panel B of the table are lower than for the narrow single-variable case, which is consistent with some substitutability between the three different types of assets in the class.

For each of the hot-deck imputation methods we ran the procedure 1000 times, reporting the mean values and bootstrap confidence intervals for the Gini coefficient in the table. For every hot-deck run of 1000 repetitions the distribution is more compressed in the multiple imputation case, as expected, but the aggregated values across 1000 repetitions have similar bootstrap standard errors. 

\begin{figure}[h]

  \begin{minipage}{0.48\textwidth}
    \centering
    \includegraphics[width=\textwidth]{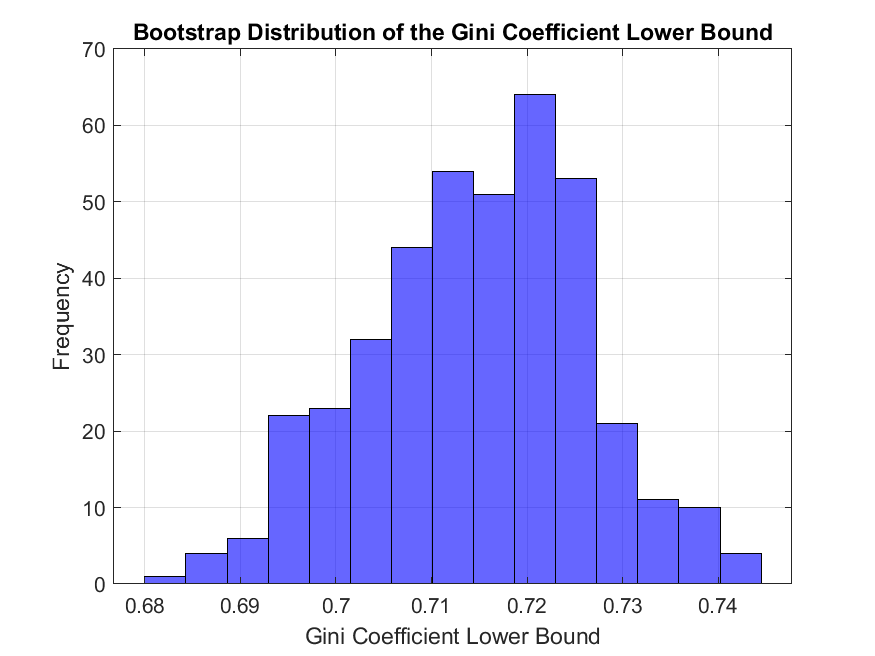}
  \end{minipage}
  \hfill 
  \begin{minipage}{0.48\textwidth}
    \centering
    \includegraphics[width=\textwidth]{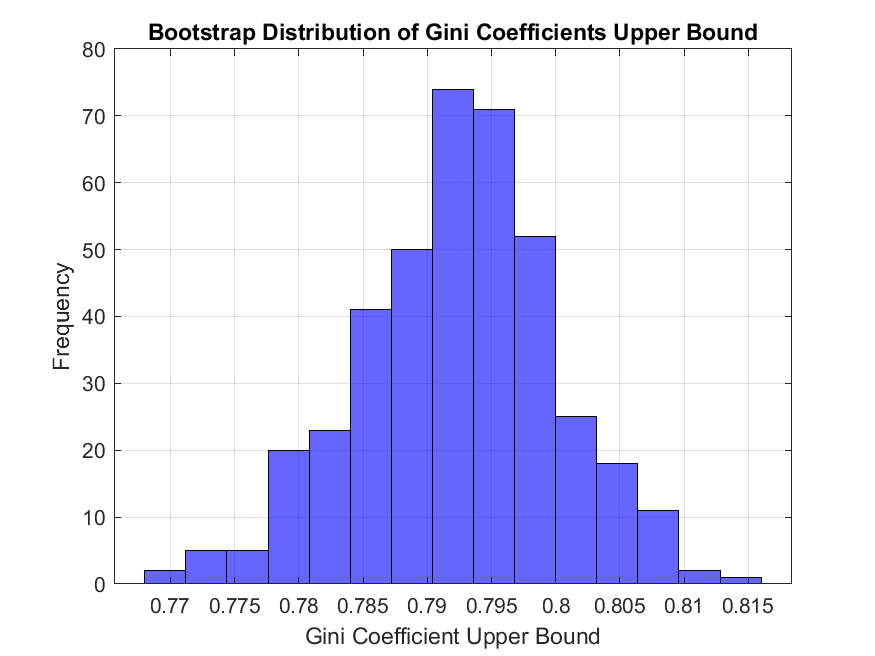}
  \end{minipage}

  \vspace{0.25cm}

  \centering
  \includegraphics[width=0.5\textwidth]{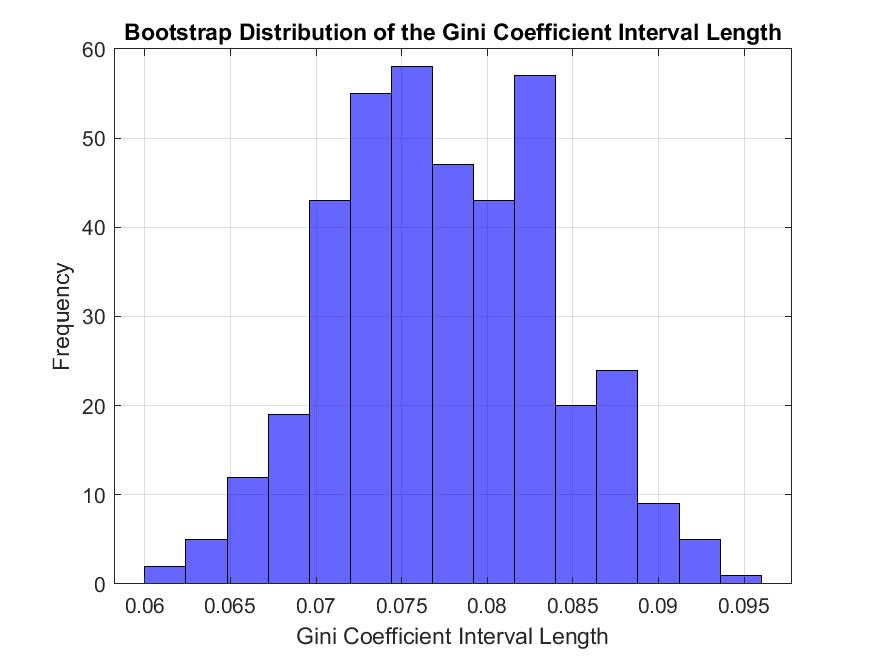}

 \caption{$m$-out-of-$n$ bootstrap distributions for the Gini inequality index (narrow  savings definition). $m \approx n^{0.9}=1,910$ as $n=4,422$.}
 \label{fig:sav_boothist} 
\end{figure}

The sharp bounds on the Gini coefficient computed using our methodology for Scenario~2 are given in Table~\ref{table:BoundSAVCombined}\footnote{As expected, the sharp upper bound on the Gini is sensitive to our treatment of the unbounded top interval. When we set the upper interval limit for entries with the open right bracket to 1,500,000 (instead of 300,000), the sharp upper bound on the Gini for the narrow definition of liquid savings changes to 0.8829.}. The bounds are considerably wider than either the confidence intervals on Gini coefficients calculated on the imputed data or on the range of values for the Gini covered by the 1000 draws of the hot-deck imputation samples. For $m \approx n^{0.9}=1,910$,  the $m$-out-of-$n$  bootstrap standard error for the length of the sharp Gini interval in the narrow case is 0.0042 and the 95\% confidence interval
for that length constructed from the 2.5th and 97.5th percentiles of the centered $m$-out-of-$n$ bootstrap deviations after rescaling them to the full-sample size, is $[0.0731,0.0885]$. The $m$-out-of-$n$ bootstrap distributions of lower and upper sharp bounds for the narrow savings definition, as well as the sharp Gini interval length, are given in Figure~\ref{fig:sav_boothist}. 

The role of the top interval is key. Whether a parametric assumption would tighten or loosen identification relative to the sharp bounds depends on what auxiliary information is available to constrain the upper tail. Appendix~\ref{app:additional_wealth_imputation} shows that when the top interval for this application is treated as genuinely open-ended, the fitted shifted-lognormal and generalized Pareto specifications produce heavy-tailed, highly dispersed imputed Gini coefficients since nothing in the data disciplines the tail. In this regime parametric imputation does not deliver tight, assumption-free precision and it is the shape of the parametric family that alone is doing the work, and different reasonable choices of family disagree sharply. Only once the top interval is closed, giving the fit something to anchor to, do the imputed values cluster tightly. We therefore treat our own top-coding choice, and its documented sensitivity, in the same spirit  as an assumption whose consequences we report, not as a device for narrowing the bounds artificially.

\begin{table}[tbp]
	\caption{Sharp Bounds on the Gini Coefficient for Liquid Savings}
	\label{table:BoundSAVCombined}
	\par
	\begin{tabular}{lcccc}
		\hline
        & \multicolumn{2}{c}{A: Narrow Definition} & \multicolumn{2}{c}{B: Broad Definition} \\
        \cline{2-3} \cline{4-5}
        & Lower Bound & Upper Bound & Lower Bound & Upper Bound \\ \hline
	  Gini 
      & 0.7144 & 0.7918 
      & 0.6857 & 0.7666 \\ 
      
    Bootstrap s.e. 
      & 0.0075 & 0.0051 
      & 0.0064 & 0.005 \\ 
      
      Conf interval: $\pm 1.96\cdot se$ 
      & $[0.6996,0.7292]$ & $[0.7819,0.8017]$ 
      & $[0.6731,0.6984]$ & $[0.7568,0.7763]$ \\

      Conf interval:  Bootstrap 
      & $[0.6995,0.7290]$ & $[0.7817,0.8019]$ 
      & $[0.6732,0.6977]$ & $[0.7569,0.7760]$ \\
		\hline
\multicolumn{5}{l}{\footnotesize Note: All bootstrap quantities are based on 400 \(m\)-out-of-\(n\) resamples, with \(m=\lfloor n^{0.9}\rfloor=1{,}910\) and \(n=4{,}422\).}\\
\multicolumn{5}{l}{\footnotesize Bootstrap confidence intervals use the 2.5th and 97.5th percentiles of \(\sqrt m(\widehat\theta_m^*-\widehat\theta_n)\), rescaled by \(1/\sqrt n\).}\\
\multicolumn{5}{l}{\footnotesize Confidence intervals based on \(\pm1.96\cdot\mathrm{s.e.}\) are reported as descriptive normal approximations.}
	\end{tabular}
\end{table}

Note that while the imputation based methods in Table~\ref{table:GiniSAV} indicate a lower Gini and smaller standard error in the broad savings case than for narrow savings measure, comparing Panel A and Panel B of Table~\ref{table:BoundSAVCombined} shows that the bounds on the Gini for the broad definition are actually wider, reflecting the increasing missingness and more complex interval nature of the broader data --- the width of the bounds is 0.0809 in comparison to 0.0774 in the narrow case. 

The results presented here carry a sharp practical message for applied researchers studying wealth inequality with survey data. The imputation-based estimates in Table~\ref{table:GiniSAV}  range from 0.734 to 0.757 for the narrow savings definition, a spread of less than 0.03 that a researcher might attribute to differences in imputation methods. Our sharp bounds show that this apparent precision is misleading: the data are consistent with population Gini values anywhere in an identified set of width 0.077, more than twice the full range of the imputation estimates. Therefore, conventional practice of reporting imputed point estimates with standard errors substantially understates identification uncertainty and may lead to overly confident conclusions about the level of wealth inequality or its evolution. Our approach provides a transparent alternative. Rather than relying on point estimates driven by untestable assumptions about within-bracket distributions, researchers can report the full range of inequality values consistent with the observed data.

From the computational perspective we maximize the Gini index represented as $G_{2,n}(\widehat{\phi})$ in (\ref{Giniobjective_Sc3}) subject to the complete set of interval-overlap restrictions in (\ref{Sc3:IneqBounds}), together with the nonnegativity, zero-mass support, and total-mass constraints in (\ref{Sc3:massconstraints}). In the broad savings definition case $|\mathcal{U}|=961$ (in contrast, in the narrow definition $|\mathcal{U}|=521$), so our optimization is over 961 variables and, since $|\mathcal{B}|=240$ (in the narrow definition $|\mathcal{B}|=10$), we have 57,358 interval-overlap inequality constraints, compared with 88 such constraints in the narrow-definition case. 

This illustrates how the computational problem in Section~\ref{sec:Sc2} grows as additional variables contain interval-valued or missing components. The number of constraints increases with the number of distinct aggregate intervals among observations in $Q$, and the number of unknown support shares increases with $|\mathcal{U}|$, although Theorem~\ref{th:Sc3_LF} can reduce that dimension. A measure of \textit{total} net financial wealth may combine ten to fifteen subcomponents, so the missing-data structure can be intricate. The algorithm is unchanged; the effect is computational rather than conceptual. 

\subsection{Income inequality in the U.S. from historical data}

In our second example we consider a different type of interval data corresponding to Scenarios~1A and~1B in the theoretical analysis above. This is the scenario where data are given in non-overlapping intervals, as is often the case when information on how a distribution breaks down into summary intervals is presented in distributional tables typically presenting the number of observations falling into different ranges, perhaps with additional subgroup means or medians. For our empirical example we consider the case of historical data on US income inequality in the early to mid twentieth century and use data from the distributional tables covering various years from 1929--1971 that were produced by government agencies prior to their public release of microdata for analysis by researchers. 

Early descriptions of the income distribution in America were developed by Selma Goldsmith from the US Department of Commerce Office of Business Economics, who produced various summary distributional tables for select years between 1929 and 1950 (see \cite{Goldsmith1954} and \cite{Goldsmith1958}). Subsequently, the Office of Business Economics (OBE) produced distributional tables as part of their regular outputs in many years from 1958 onwards (\cite{OBE1958USIncomeOutput}). Inequality statistics were not produced, and in addition the distributional tables were in different formats for different years. The 1929 income distribution was presented in 8 income brackets, with additional aggregate income ratios presented for the bottom 40\%, 40-60\%, 60-80\%, 80-95\% and top 5\%. Data covering select years 1935 to 1950 (in \cite{Goldsmith1958}) used the same 8 income brackets but instead presented subgroup mean incomes for the bottom 40\%, 40-60\%, 60-80\%, 80-95\% and top 5\% groups. In contrast to this, data from the OBE (\cite{OBE1958USIncomeOutput}) presented the distribution in 13 brackets along with subgroup means until 1954, then with the addition of the median, 20th, 40th, 60th, 80th and 95th quantiles for the years 1955--1962, before switching to 25 brackets along with subgroup means within each bracket from 1964 onwards. 

These data provided a wealth of information for scholars of historical inequality in the US (see, for example, \cite{Budd1970} or \cite{Lindert2000}) but did not provide statistics, such as the Gini, which could be combined with calculations from modern microdata to produce long run series. While some individual studies have used assumptions on the underlying distribution to allow an estimate of the Gini from subsets of the available distributional statistics, our method allows for the computation of sharp bounds on Gini coefficients or quantile ratios for the US income distribution incorporating all the information from the published statistics, regardless of the fact that the nature of the information available changes over time.

\begin{figure}[tbp]
\centering
\includegraphics[width=12cm]{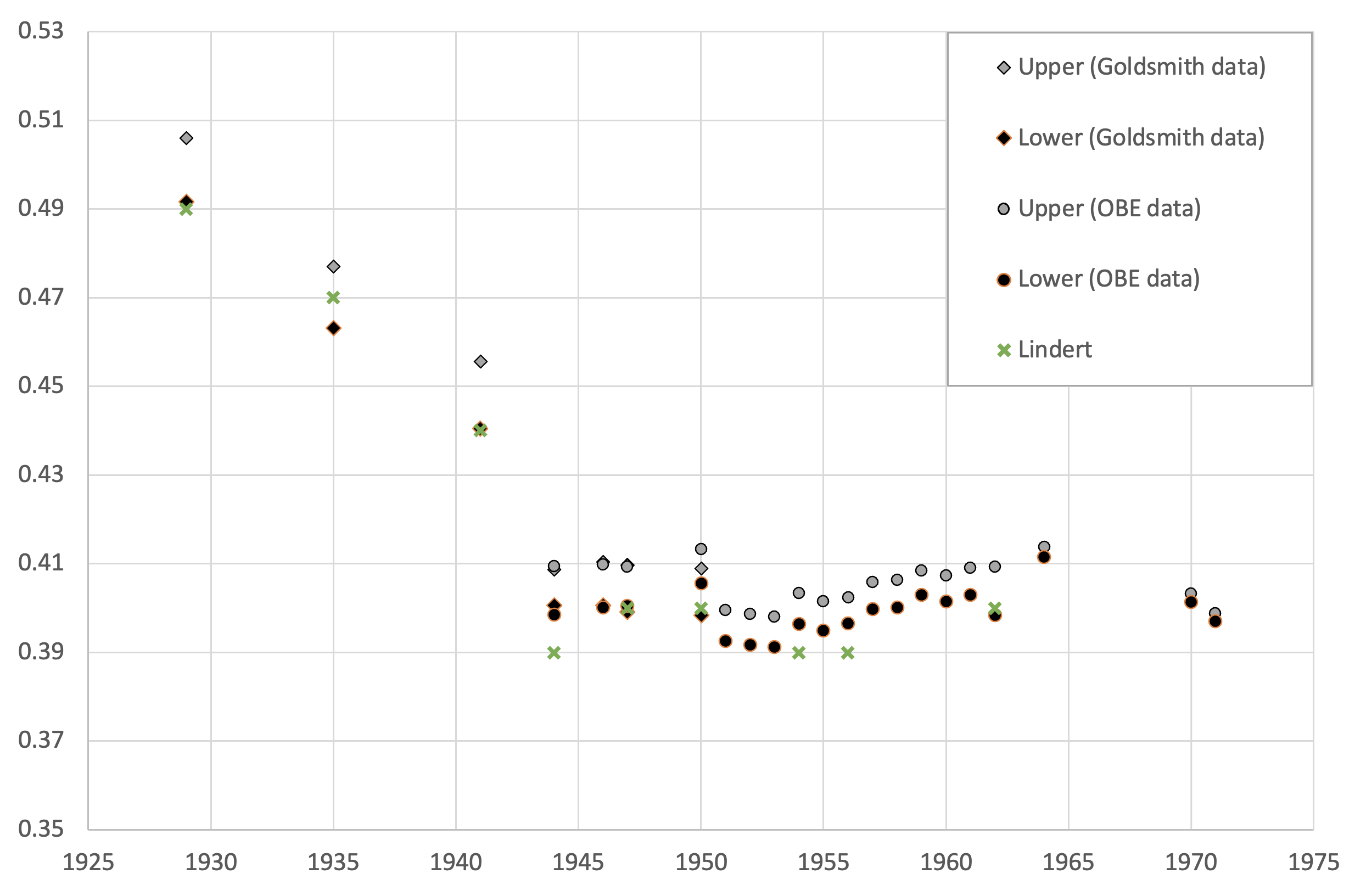}
\caption{\protect\small Sharp bounds on Gini inequality for U.S. historical data.}
\label{fig:App2_Gini}
\end{figure}

Sharp Gini bounds computed from these historical US Income data are given in Figure~\ref{fig:App2_Gini}. The figure presents our calculations for two time-series of upper and lower bounds, depending on whether the source data come from \citet{Goldsmith1958} covering years from 1929 to 1950 or the various OBE sources covering for the period from 1944 onwards. In addition, and as a comparison, we present the point estimates for the Gini over this time period that have been reported by \cite{Lindert2000}.\footnote{Gini coefficients reported by \cite{Lindert2000} should be interpreted as a compiled secondary series rather than as estimates constructed by Lindert from grouped observations. The mid-twentieth-century entries in \cite{Lindert2000}'s Table 4 are labelled as coming from the OBE--Goldsmith consumer-unit series. Tracing this source shows that Goldsmith’s chapter in \cite{Budd1967Book} documents the underlying OBE family/consumer-unit income-distribution series and reports mainly income shares, especially top shares, rather than Gini coefficients. The Gini, or “concentration ratio,” figures appear instead in Budd’s introductory update to the same volume. Budd  does not document the computational procedure used to obtain the concentration ratios from the grouped OBE--Goldsmith data. Moreover, these figures are not identical to the Gini coefficients later reported by \citet{Budd1970}, where Budd explicitly constructs smoothed Lorenz curves using polynomial and upper-tail interpolation. We therefore treat Lindert’s U.S. Gini series as an externally reported OBE--Goldsmith/Budd concentration-ratio series, not as an imputation-based estimate constructed from Lindert’s grouped data.} 

The Gini coefficient falls substantially from the late 1920s to the beginning of the Second World War, even though the bounds are relatively wide in the early years because the distributional tables are coarse. The lower bound on the Gini for 1929 is also higher than current estimates, despite other measures of inequality such as the top 1\% share of income having already returned to, or surpassed, their 1920s \textit{Gilded Age} levels (see, for example, \cite{Saez2023}). 
By the mid-1940s, the upper bound on the Gini falls below the lower bound for 1929, so the data establish, without any parametric within-bracket assumption, that inequality was strictly lower in the mid-1940s than in 1929. This is a more credible statement than one based on point estimates, which require assumptions that cannot be tested from the tabulations alone. 

The figure also shows that the bounds narrow considerably from 1964 onwards, when the OBE tables begin reporting a much finer bracket structure alongside subgroup means.

\begin{figure}[tbp]
\centering
\includegraphics[width=12cm]{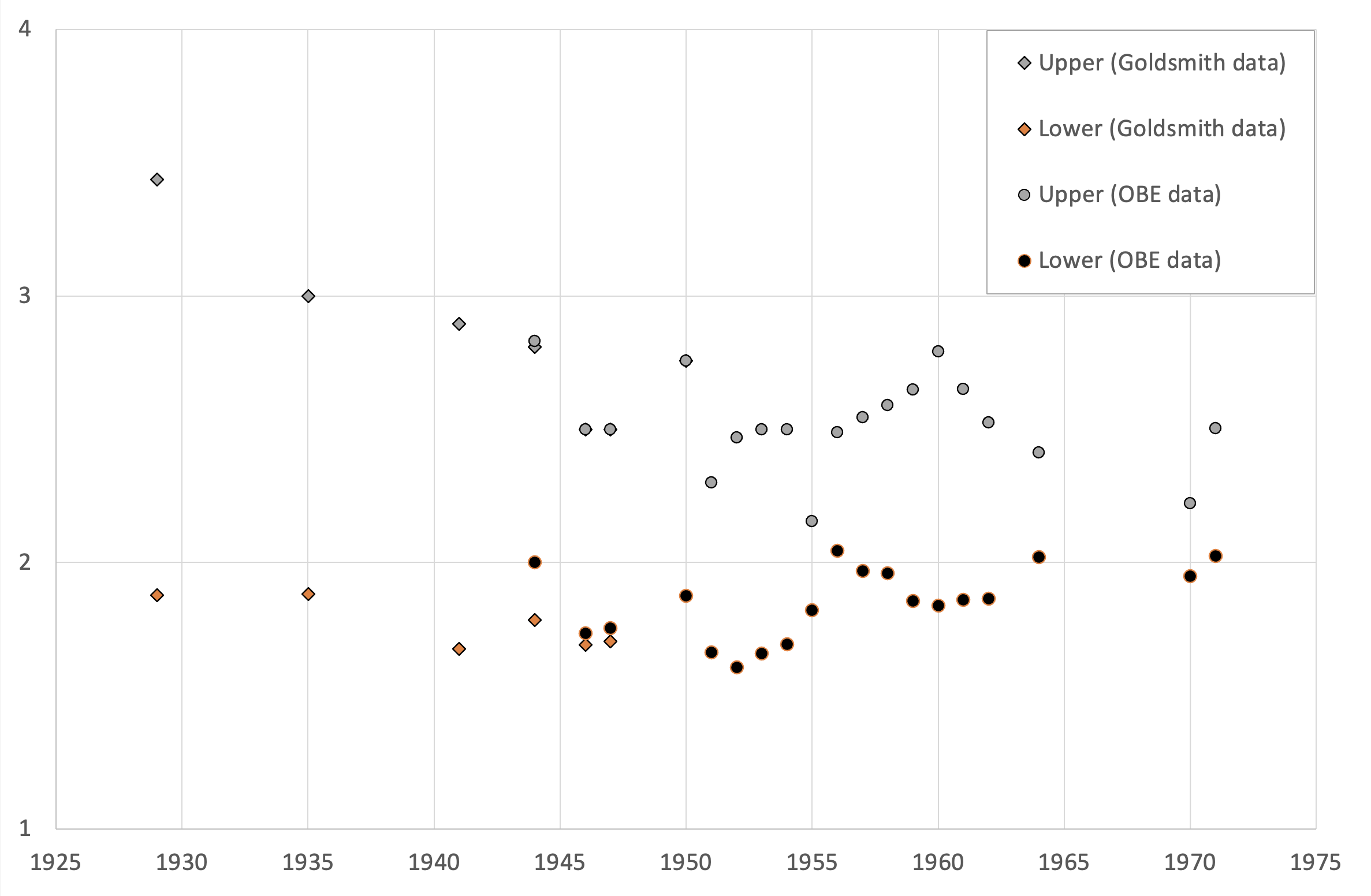}
\caption{\protect\small Sharp bounds on the 90/50 percentile ratio for U.S. historical data.}
\label{fig:App2_9050}
\end{figure}

As discussed previously, our procedures can also deliver bounds for percentile ratios of the distribution. In the case when the frequencies in various intervals are all that is known then such bounds would be straightforward to compute but when additional information on, for example, subgroup means or income shares needs to be taken into account then it becomes a non-trivial operation. We apply our algorithm to produce sharp bounds for the 90/50 percentile ratio from these same historical US Income data, presented in Figure~\ref{fig:App2_9050} in order to examine the role of upper middle inequality in driving these changes in the Gini.\footnote{It is of course possible to compute bounds on any quantile ratios. We choose to present the 90/50 since the upper bound for the other commonly used 90/10 and 50/10 ratios is extremely large or even infinite in the earliest years of the data where published distributional information shows that it is possible that considerable numbers of families at the bottom of the distribution had extremely low income or even zero income.} Unlike the bounds for the Gini, the bounds on the 90/50 percentile ratio do not narrow so much with time when the published distributional tables begin to be produced using more brackets. 

\section{Conclusion}
\label{sec:conclusion}

We develop a framework for sharp partial identification and inference on inequality indices when outcomes are observed through grouped data or observation-specific intervals, possibly combined with auxiliary linear restrictions. For many commonly used indices, a linear-fractional representation after sorting reduces bound computation to tractable linear (or closely related) programs. Our solution-form characterizations also support the asymptotic analysis in Section~\ref{sec:asymptotics}: under the environment stated there, the value function is directionally differentiable and the estimated endpoints have a \(\sqrt n\) asymptotic distribution, with a generally valid \(m\)-out-of-\(n\) bootstrap implementation and ordinary bootstrap validity under full Hadamard differentiability when the standard bootstrap consistently estimates the first-stage distribution.

The applications illustrate two practical messages. First, the gap between conventional point estimates and the sharp bounds gives a clear indication of how much identifying content any given imputation assumption or procedure is  contributing. When that gap is large, conclusions rest more heavily on the specific assumption used than on the data. Second, incorporating credible auxiliary information (e.g.\ subgroup means or Lorenz-curve restrictions) can shrink this gap directly, without recourse to distributional assumptions, sometimes turning wide bounds into informative intervals on their own.


Extending these computational ideas to additional Schur-convex inequality indices is one promising direction for future work. More broadly, the linear constraint framework developed here is not specific to scalar inequality indices: the same feasibility characterization, finite-dimensional reduction, and inferential approach may be applicable to a wider class of distributional functionals (some of which could be linear combinations of inequality functionals) in settings where outcomes are only partially observed, and we view this as a fruitful avenue for future research.

{\normalsize 
\bibliographystyle{econometrica}
\bibliography{overallineq}
}

\clearpage
\renewcommand{\thepage}{S.\arabic{page}}
\setcounter{page}{1}

\begin{center}
  {\LARGE \textbf{Online Supplement}}
\end{center}

\appendix

\section{Additional figures}

\raggedbottom

\renewcommand\thefigure{\thesection.\arabic{figure}}  
\setcounter{figure}{0}  

\begin{figure}[htbp]
  \centering
  \includegraphics[width=0.99\linewidth]{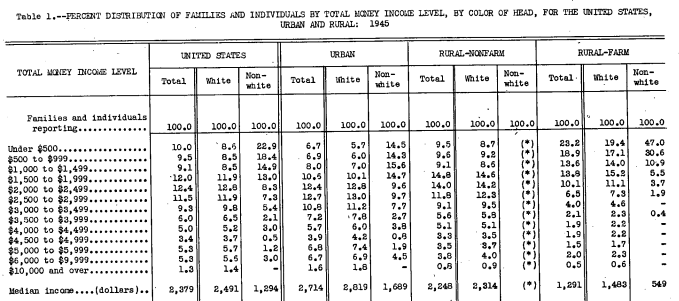}
  \caption{\small Example for Scenario~1A: 1945 income distribution tables from \citet{Census1948P60-02}.}
  \label{fig:Sc1_example1945}
\end{figure}

\begin{figure}[htbp]
  \centering
  \includegraphics[width=0.99\linewidth]{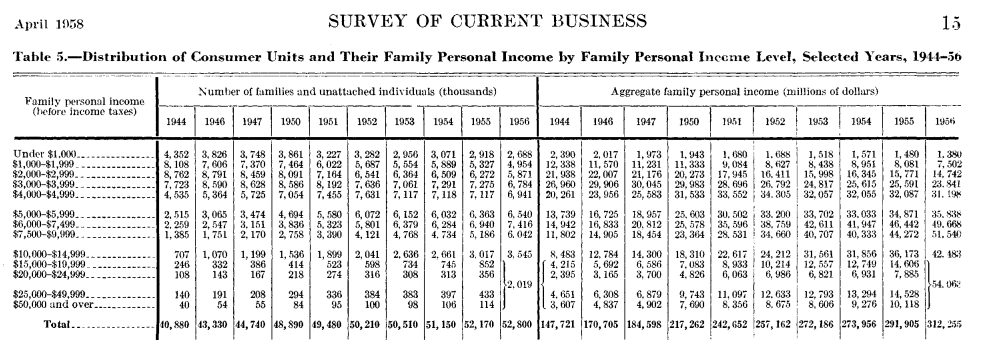}
  \caption{\small Example for Scenario~1B: income tables from the U.S. Department of Commerce, Office of Business Economics, \textit{U.S. Income and Output} (1958) \citep{OBE1958USIncomeOutput}.}
  \label{fig:Sc1B_example1958}
\end{figure}

\section{Proofs of computational results} 

\begin{proof}[Proof of Proposition~\ref{prop:base_cornersol}]
Fix $n$ and write $\mathcal Y_0$ for the set of vectors satisfying \eqref{ordering} and \eqref{LF3}. This set is a nonempty compact polytope, and the maintained positivity of $r_2(n)^\top\mathbf y$ implies that $G_n$ is continuous on $\mathcal Y_0$. Hence both endpoint problems attain their values.

We first reduce each endpoint problem to a linear program. Let $\mathbf y^+$ maximize $G_n$ on $\mathcal Y_0$ and set $\lambda^+=G_n(\mathbf y^+)$. For every $\mathbf y\in\mathcal Y_0$, $r_1(n)^\top\mathbf y-\lambda^+ r_2(n)^\top\mathbf y\le 0=r_1(n)^\top\mathbf y^+-\lambda^+ r_2(n)^\top\mathbf y^+$. Thus $\mathbf y^+$ maximizes the linear form $r_1(n)^\top\mathbf y-\lambda^+r_2(n)^\top\mathbf y$ on $\mathcal Y_0$. The same argument with the reverse inequality applies to a minimizer. A linear form attains its optimum over a nonempty compact polytope at an extreme point, so each endpoint has an optimizer that is an extreme point of $\mathcal Y_0$.

It remains to identify these extreme points. Scenario~1 brackets are non-overlapping and ordered so that $\overline a_d\le\underline a_{d+1}$. Hence cross-group ordering is implied by the box restrictions; when two adjacent brackets meet, any cross-group tie occurs only at their common boundary, which is an endpoint for both groups. Within group $d$, the feasible coordinates form the monotone box $\underline a_d\le y_{m_d+1}\le\cdots\le y_{m_d+n_d}\le\overline a_d$, where $m_d=\sum_{\ell<d}n_\ell$. If a maximal block of tied coordinates in this group has a common value strictly between $\underline a_d$ and $\overline a_d$, then the whole block can be shifted up and down by a sufficiently small common amount while preserving the ordering, including at the two neighboring group boundaries, and the box constraints. The original point is then the midpoint of two distinct feasible points and is not extreme. If an interior coordinate is not tied to a neighbor, the same conclusion follows by shifting that coordinate alone by a sufficiently small amount. Hence every group coordinate of an extreme point is at $\underline a_d$ or $\overline a_d$. This proves the first claim.

Now suppose $G_n$ is strictly Schur-convex. All majorization comparisons below are between vectors with the same sum. Choose a boundary-valued minimizer, whose existence follows from the first part. If a nondegenerate group $d$ contains both $\underline a_d$ and $\overline a_d$, then, because coordinates are ordered within the group, there are adjacent entries taking these two values. Replacing them by $\underline a_d+\delta$ and $\overline a_d-\delta$, with $0<\delta\le(\overline a_d-\underline a_d)/2$, preserves feasibility and the total sum. The new vector is strictly majorized by the original vector, so strict Schur-convexity gives a strictly smaller value, contradicting minimality. Thus each nondegenerate group is assigned entirely to one endpoint; degenerate groups may be assigned to either side of a switch because their endpoints coincide.

If there are nondegenerate groups $d<e$ such that group $d$ is assigned to its lower endpoint and group $e$ is assigned to its upper endpoint, choose the last coordinate in group $d$ and the first coordinate in group $e$. Increasing the former by a small $\delta>0$ and decreasing the latter by the same $\delta$ preserves all box constraints, the ordering, and the total sum. The resulting vector is strictly majorized by the original vector, again contradicting minimality. Therefore the endpoint assignment for a minimizer cannot switch from a lower endpoint in an earlier group to an upper endpoint in a later group. After assigning degenerate groups arbitrarily when necessary, there is $d_0\in\{0,\ldots,D\}$ such that groups weakly before $d_0$ are at upper endpoints and groups after $d_0$ are at lower endpoints.

For maximization, choose a boundary-valued maximizer. If there are nondegenerate groups $d<e$ such that group $d$ contains an upper-endpoint coordinate and group $e$ contains a lower-endpoint coordinate, decrease the first upper-endpoint coordinate in group $d$ by a small $\delta>0$ and increase the last lower-endpoint coordinate in group $e$ by the same $\delta$. Feasibility, ordering, and the total sum are preserved. This is a mean-preserving spread, so the new vector strictly majorizes the original vector and strict Schur-convexity gives a strictly larger value, contradicting maximality. Hence an upper endpoint cannot occur in an earlier nondegenerate group while a lower endpoint occurs in a later nondegenerate group. Since each group is ordered and boundary-valued, this implies that all groups before some $d_0$ are at their lower endpoints, all groups after $d_0$ are at their upper endpoints, and only group $d_0$ can contain both endpoint values. Degenerate groups can be assigned consistently with this switch without changing the vector.
\end{proof}

\begin{proof}[Proof of Proposition~\ref{prop:Sc1BsolLF}]
Let $\mathcal Y_1$ denote the feasible set after adding $C^{(1)}_n\mathbf y=f^{(1)}_n$ and $C^{(2)}_n\mathbf y\le f^{(2)}_n$ to \eqref{ordering} and \eqref{LF3}. This is a nonempty compact polytope. By the same linear-fractional argument used in the proof of Proposition~\ref{prop:base_cornersol}, each endpoint problem has an optimizer that is an extreme point of $\mathcal Y_1$. Fix such an optimizer $\mathbf y^*$, for either the upper or lower endpoint.

We use a finite-dimensional extreme-point count. Consider first a fixed group $d$ and hold all coordinates outside group $d$ at their values in $\mathbf y^*$. The remaining feasible set for the group-$d$ subvector is an ordered box intersected with the equalities inherited from $C^{(1)}_n$ and with those inequalities from $C^{(2)}_n$ that bind at $\mathbf y^*$. Rows that do not involve group $d$ disappear after the other coordinates are fixed. Therefore the rank of the inherited linear system is no larger than $q_1(d)+q_2(d)$. If the group-$d$ subvector were not an extreme point of this sliced polytope, then the resulting local perturbation could be taken small enough that every inactive inequality row remains slack. The vector $\mathbf y^*$ would then be the midpoint of two distinct feasible points of $\mathcal Y_1$, contradicting extremality.

Let $A_d x=b_d$ denote the inherited active linear system in group $d$, with $\operatorname{rank}(A_d)\le q_1(d)+q_2(d)$. Partition the coordinates of the group-$d$ subvector into maximal constant blocks whose common value lies strictly inside $(\underline a_d,\overline a_d)$. If there are $R_d$ such interior blocks and $R_d>\operatorname{rank}(A_d)$, then the $R_d$ block-indicator vectors are linearly dependent after applying $A_d$. Hence there is a nonzero perturbation, constant on each interior block and zero elsewhere, that leaves $A_d x$ unchanged. For sufficiently small positive and negative multiples, the perturbation preserves the ordering and box constraints because the blocks are interior and separated from neighboring distinct values. This contradicts the extreme-point property of the slice. Thus $R_d\le\operatorname{rank}(A_d)\le q_1(d)+q_2(d)$. The only additional values available in group $d$ are the two endpoints $\underline a_d$ and $\overline a_d$, so group $d$ contains at most $q_1(d)+q_2(d)+2$ distinct values.

For the global bound, apply the same argument without fixing other groups. Let $A$ collect the equality rows and the inequalities binding at $\mathbf y^*$, excluding ordering and box constraints. Its rank is at most $q_1+q_2$. Consider all maximal constant blocks whose common value is not an endpoint of the group to which the block belongs. If the number of such interior blocks exceeded $\operatorname{rank}(A)$, a nonzero linear combination of their indicator vectors would preserve $A\mathbf y$ and, for sufficiently small positive and negative multiples, preserve all ordering and box constraints. Because the inequalities omitted from $A$ are inactive at $\mathbf y^*$, the same small multiples also preserve those inequalities. This would contradict that $\mathbf y^*$ is an extreme point. Hence there are at most $q_1+q_2$ interior block values globally. Adding the at most two endpoint values in each of the $D$ groups gives at most $q_1+q_2+2D$ distinct values. The argument does not depend on whether $\mathbf y^*$ solves the upper or lower endpoint problem.
\end{proof}

\begin{proof}[Proof of Theorem~\ref{thm:Sc2solform3}]
The feasible set is a nonempty compact subset of $\mathbb R^n$, and $G_n$ is continuous, so both endpoint problems attain their values. The proof repeatedly uses the following implication of Schur-convexity. If two feasible vectors have the same sum and one is obtained from the other by averaging two coordinates, then the averaged vector is majorized by the original one and has weakly smaller value of $G_n$. If one vector is obtained by moving two ordered coordinates farther apart while preserving their sum, then it majorizes the original vector and has weakly larger value of $G_n$.

Consider first a group $d$ with no constraint involving its elements. For minimization, replace the coordinates in group $d$ by their within-group average. The average lies in $[\underline a_d,\overline a_d]$, the group remains ordered, no constraint is affected, and the new vector is majorized by the original vector. Starting from any minimizer and applying this operation group by group gives a minimizer whose unconstrained groups are constant. For maximization, suppose a maximizer has at least two coordinates in group $d$ strictly inside $\mathcal G_d$. Choose the first coordinate among those attaining the smallest interior value and the last coordinate among those attaining the largest interior value. Move the former down and the latter up by the largest common amount that keeps both coordinates in $[\underline a_d,\overline a_d]$. Ordering and the group sum are preserved, at least one selected coordinate reaches a boundary, and Schur-convexity weakly increases the objective. Iterating the operation yields a maximizer with at most one strictly interior coordinate in the unconstrained group. This proves part (a).

Now fix a group $d$ satisfying the block condition in part (b). The rows in a constrained block are all sums, or all negative sums, over that block. Therefore any transformation that preserves the sum within each constrained block leaves every equality and inequality row involving that block unchanged. Transformations inside an unconstrained block do not affect any row by definition.

For minimization, take any minimizer. In each constrained block $A^{(d)}_j$, replace all coordinates in the block by their block average. The replacement preserves the block sum, and hence all rows in the block, and it is feasible because the average lies between the smallest and largest coordinates of the block. It also preserves the ordering relative to neighboring coordinates. Schur-convexity weakly lowers the objective. Apply the same averaging operation to each consecutive unconstrained block. After all such operations, the resulting vector is still a minimizer, each of the $k_d$ constrained blocks contributes at most one value, and each of the $o_d$ unconstrained blocks contributes at most one value. Thus the group contributes at most $k_d+o_d$ distinct values in a minimizer.

For maximization, take any maximizer. Treat each constrained block and each maximal consecutive unconstrained block as one consecutive block. Within such a block, suppose there are two coordinates whose values are strictly between the adjacent feasible boundary values for that block, where these boundaries are the neighboring coordinates in group $d$ or, at the ends of the group, $\underline a_d$ and $\overline a_d$. Choose the first coordinate attaining the smaller of two such values and the last coordinate attaining the larger value. Move the former down and the latter up by the largest common amount that preserves the block's adjacent boundary values, the group box, and the within-group ordering. The block sum is unchanged, so every constrained row in that block is unchanged; unconstrained blocks affect no row. The new vector majorizes the old one and therefore weakly increases $G_n$. Repeating this spreading operation gives a maximizer in which each constrained or unconstrained block has at most one value strictly between its two adjacent boundary values. Reading the blocks from left to right, each block can introduce at most one new value beyond the value already present at its left boundary, and the group contributes the two possible endpoints $\underline a_d$ and $\overline a_d$. Hence the number of distinct values in group $d$ is at most $k_d+o_d+2$.
\end{proof}

\begin{proof}[Proof of Theorem~\ref{th:Sc3}]
The Scenario~2 feasible set is the product of the compact intervals $\mathcal I_i=[\underline a_i,\overline a_i]$, with exact observations corresponding to degenerate intervals. Since $G_n$ is continuous, maximizers and minimizers exist. Exact observations are fixed, so only coordinates in $Q$ need to be considered.

For the maximum, start from any maximizer. If two coordinates $i,j\in Q$ take values outside $\mathcal B$, then both values are interior to their own intervals. Relabel them so that $y_i\le y_j$. Decrease $y_i$ and increase $y_j$ by the largest common amount $\delta>0$ that keeps both coordinates in their intervals. Feasibility and the total sum are preserved, at least one selected coordinate reaches an endpoint in $\mathcal B$, and the transformed vector is a mean-preserving spread of the original vector. Schur-convexity therefore weakly increases the objective. Repeating this operation reduces the number of interval coordinates outside $\mathcal B$ until at most one remains. Hence some maximizer has at most one coordinate indexed by $Q$ outside $\mathcal B$.

For the minimum, start from any minimizer. If the coordinates in $Q$ take two distinct values outside $\mathcal B$, choose one at the smaller value and one at the larger value. Increase the smaller value and decrease the larger value by a common amount, stopping when the two selected values coincide or one reaches an endpoint of its interval. This transformation preserves feasibility and the total sum, and the transformed vector is majorized by the original vector. Schur-convexity therefore weakly lowers the objective. Iterating over distinct outside-$\mathcal B$ values gives a minimizer for which all coordinates in $Q$ that remain outside $\mathcal B$, if any, share one common value. This is exactly the claim in part (b).
\end{proof}

\begin{proof}[Proof of Theorem~\ref{th:Sc3_LF}]
We prove the claim for both endpoints at once. By Theorem~\ref{th:Sc3}, there is an endpoint optimizer with the stated outside-$\mathcal B$ structure: for the maximum there is at most one interval coordinate outside $\mathcal B$, and for the minimum there is at most one distinct outside-$\mathcal B$ value among interval coordinates. If no such outside value is present, then every interval coordinate lies in $\mathcal B$ and every exact observation lies in $\mathcal U$ by definition, so the conclusion follows.

It remains to handle the single exceptional outside-$\mathcal B$ value. For the maximum, let $I$ contain the one interval observation taking this value. For the minimum, let $I$ contain all interval observations in $Q$ that take the common exceptional value. Hold all coordinates outside $I$ fixed and move the coordinates in $I$ together to a common value $x$ in the compact interval $J=\cap_{i\in I}\mathcal I_i$. The current optimizer corresponds to some $x\in J$. The endpoints of $J$ are endpoints of intervals belonging to observations in $I$, and therefore lie in $\mathcal B$.

As $x$ varies over $J$, the ordering of the full vector changes only when $x$ reaches the value of a fixed coordinate. All fixed interval coordinates have values in $\mathcal B$, and all fixed exact observations have values in $\mathcal U$, so every such breakpoint lies in $\mathcal U$. On each open subinterval between consecutive breakpoints, the ordering is fixed. By the maintained linear-fractional representation, the objective along that subinterval has the form $(a x+b)/(c x+d)$ with $c x+d>0$. Its derivative has constant sign, $(ad-bc)/(c x+d)^2$, so the function is monotone on the subinterval unless it is constant. Consequently an endpoint optimum over the closure of that subinterval is attained at one of its endpoints, and if the function is constant an endpoint is again optimal.

Taking the best among the finitely many subinterval endpoints gives an endpoint optimizer for which the exceptional value belongs to $\mathcal U$. The outside-$\mathcal B$ structure from Theorem~\ref{th:Sc3} is preserved because the maximum case moves only one coordinate and the minimum case moves all coordinates with the common exceptional value together. This proves both parts.
\end{proof}

The next theorem records the finite assignment fact used in the Scenario~2 computation. Let $b_1<\cdots<b_K$ be the ordered elements of $\mathcal B$, and, for $1\le d<e\le K$, write $T_{d,e}:=\{u\in\mathcal U:b_d\le u\le b_e\}$.

\begin{theorem}
\label{th:Sc3_overlap_feasibility}
Suppose the data comply with Scenario~2, and let $A_i:=\mathcal I_i\cap\mathcal U$ for $i\in Q$. Let $(N_u)_{u\in\mathcal U}$ be an integer vector satisfying $N_u\ge0$, $\sum_{u\in\mathcal U}N_u=|Q|$, and $N_u=0$ whenever $u\notin \bigcup_{i\in Q}A_i$. Then there exists an assignment $a:Q\to\mathcal U$ such that $a(i)\in A_i$ for every $i\in Q$ and $\#\{i\in Q:a(i)=u\}=N_u$ for every $u\in\mathcal U$ if and only if, for every $1\le d<e\le K$,
\[
\#\{i\in Q:A_i\subseteq T_{d,e}\}
\le
\sum_{u\in T_{d,e}}N_u
\le
\#\{i\in Q:A_i\cap T_{d,e}\ne\emptyset\}.
\]
The same conditions imply the analogous lower and upper inequalities for finite unions of disjoint boundary blocks.
\end{theorem}

\begin{proof}[Proof of Theorem~\ref{th:Sc3_overlap_feasibility}]
For each $i\in Q$, the set $A_i$ is nonempty because it contains the two endpoints of $\mathcal I_i$. Since these endpoints belong to $\mathcal B$ and $\mathcal U$ is ordered on the real line, $A_i$ is a consecutive block of support points. We first prove necessity. If an assignment $a$ exists and $T_{d,e}$ is a boundary block, then every observation with $A_i\subseteq T_{d,e}$ must be assigned to a support point in $T_{d,e}$. This gives the lower inequality. Conversely, every observation assigned to a support point in $T_{d,e}$ must have $A_i\cap T_{d,e}\ne\emptyset$, which gives the upper inequality. Nonnegativity, total mass, and the zero restrictions outside $\bigcup_{i\in Q}A_i$ are also necessary.

We now prove sufficiency. Replace each support point $u\in\mathcal U$ by $N_u$ identical copies, and let $\mathcal V_N$ denote the resulting multiset. Connect interval observation $i$ to a copy of $u$ whenever $u\in A_i$. By Hall's theorem, it is enough to show that, for every subset $R\subseteq Q$, the number of copies adjacent to $R$ is at least $|R|$. Let $\Gamma(R):=\bigcup_{i\in R}A_i$. If $R$ is empty there is nothing to show. Otherwise, because each $A_i$ is a boundary block, $\Gamma(R)$ is a finite union of disjoint maximal boundary blocks, say $T_{d_1,e_1},\ldots,T_{d_m,e_m}$, ordered from left to right. The lower inequalities in the theorem give
\[
\sum_{u\in\Gamma(R)}N_u
=
\sum_{r=1}^m\sum_{u\in T_{d_r,e_r}}N_u
\ge
\sum_{r=1}^m\#\{i\in Q:A_i\subseteq T_{d_r,e_r}\}.
\]
A consecutive set $A_i$ that is contained in the disjoint union $\Gamma(R)$ must be contained in one of its maximal components; otherwise it would cross a gap between two components and would include support points outside $\Gamma(R)$. Hence the last sum equals $\#\{i\in Q:A_i\subseteq\Gamma(R)\}$. Every $i\in R$ satisfies $A_i\subseteq\Gamma(R)$, so
\[
\sum_{u\in\Gamma(R)}N_u
\ge
\#\{i\in Q:A_i\subseteq\Gamma(R)\}
\ge |R|.
\]
The left-hand side is exactly the number of copies in $\mathcal V_N$ adjacent to $R$, so Hall's condition holds. There is therefore a matching that assigns every observation in $Q$ to a distinct adjacent copy. Since $\sum_{u\in\mathcal U}N_u=|Q|$, all copies are used, and the number of observations assigned to each support point $u$ is exactly $N_u$. This gives the required assignment.

It remains only to justify the final sentence of the theorem. The preceding argument used only the lower inequalities for the maximal boundary components of $\Gamma(R)$, and therefore shows that the corresponding lower inequality holds for any finite union of disjoint boundary blocks by summing over its components. The upper inequality for such a union follows from the assignment just constructed and the necessity argument in the first paragraph. Thus non-consecutive unions do not add independent feasibility restrictions.
\end{proof}

\begin{proof}[Proof of Proposition~\ref{prop:GiniSc3_original}]
Let $\mathbf z=(z_1,\ldots,z_n)$ denote the nondecreasing rearrangement of a feasible vector, and write $S(\mathbf z)=\sum_{j=1}^n z_j$ and $T(\mathbf z)=\sum_{j=1}^n jz_j$. The maintained nonnegativity and positive-sum conditions give $S(\mathbf z)>0$. The ordered-sample Gini coefficient satisfies $G_n(\mathbf z)=2T(\mathbf z)/(nS(\mathbf z))-(n+1)/n$. Thus optimizing $G_n$ is equivalent to optimizing $R(\mathbf z):=T(\mathbf z)/S(\mathbf z)$. Because the weights $z_j/S(\mathbf z)$ are nonnegative and sum to one, $R(\mathbf z)\in[1,n]$.

We first record the one-coordinate calculation used throughout. Hold all coordinates except the coordinate occupying rank $k$ fixed, and write $R(x)=(kx+w)/(x+q)$, where $w$ and $q$ are the corresponding sums over the other coordinates. On any region where the ordering is fixed, $x+q>0$ and $R'(x)=(kq-w)/(x+q)^2=(k-R(x))/(x+q)$. Hence increasing a rank-$k$ coordinate lowers $R$ when $k<R(x)$ and raises $R$ when $k>R(x)$; decreasing the coordinate has the opposite effect. If the moving coordinate crosses a fixed coordinate with value $v$, its rank rises by one and that fixed coordinate's rank falls by one. The numerator $kq-w$ then increases by $q+v$, which is the total sum at the crossing point and is strictly positive. Therefore, as a single coordinate is moved upward with the vector resorted after crossings, the sign of the derivative can change only from negative to positive, never from positive to negative. In particular, the Gini objective has no strict interior local maximum as a function of one coordinate over its feasible interval.

For the minimum, choose a minimizer satisfying Theorem~\ref{th:Sc3_LF}, and let $\mathbf z^*$ be its ordered version. Set $R^*=R(\mathbf z^*)$, $k^*=\lceil R^*\rceil$, and $u_0=z^*_{k^*}$. Theorem~\ref{th:Sc3_LF} gives $u_0\in\mathcal U$. Consider any interval observation $i\in Q$. If $\overline a_i<u_0$ and $y^*_{i,\min}<\overline a_i$, then a sufficiently small increase of $y^*_{i,\min}$ keeps the observation below $u_0$ after resorting. The moved observation occupies a rank $k<k^*$, hence $k<R^*$, so the preceding derivative calculation strictly lowers $R$ and therefore $G_n$, contradicting minimality. Thus $y^*_{i,\min}=\overline a_i$ whenever $\overline a_i<u_0$. If $\underline a_i>u_0$ and $y^*_{i,\min}>\underline a_i$, then a sufficiently small decrease keeps the observation above $u_0$ and at a rank $k>k^*\ge R^*$, which again strictly lowers $R$ and contradicts minimality. Hence $y^*_{i,\min}=\underline a_i$ whenever $\underline a_i>u_0$. Finally suppose $u_0\in\mathcal I_i$. If $y^*_{i,\min}<u_0$, the same increasing perturbation lowers the objective; if $y^*_{i,\min}>u_0$, the same decreasing perturbation lowers the objective. Therefore $y^*_{i,\min}=u_0$. This proves part (a).

For the maximum, first choose a maximizer satisfying Theorem~\ref{th:Sc3_LF}. If every interval observation in $Q$ is already assigned to $\mathcal B$, keep this maximizer. Otherwise, Theorem~\ref{th:Sc3_LF} leaves at most one interval observation $d\in Q$ assigned to a value in $\mathcal U\setminus\mathcal B$, which is strictly inside $\mathcal I_d$. Holding all other coordinates fixed, view $R$ as a function of $y_d$ over $\mathcal I_d$, with the vector resorted after each crossing. The one-coordinate calculation above shows that this function is monotone or decreases and then increases; it has no strict interior local maximum. Hence the maximum over $\mathcal I_d$ is attained at an endpoint, and replacing $y_d$ by an endpoint of $\mathcal I_d$ does not lower $G_n$. We may therefore take a maximizer with $y^*_{i,\max}\in\mathcal B$ for every $i\in Q$.

Let $\mathbf z^*$ be the ordered version of such a maximizer, set $R^*=R(\mathbf z^*)$, $k^*=\lceil R^*\rceil$, and $u_0=z^*_{k^*}\in\mathcal U$. If $\overline a_i<u_0$ and $y^*_{i,\max}>\underline a_i$, then a sufficiently small decrease of $y^*_{i,\max}$ keeps the observation below $u_0$ and at a rank $k<k^*$, hence $k<R^*$. Since increasing such a coordinate would lower $R$, decreasing it strictly raises $R$ and therefore raises the Gini coefficient, contradicting maximality. Thus $y^*_{i,\max}=\underline a_i$ whenever $\overline a_i<u_0$. If $\underline a_i>u_0$ and $y^*_{i,\max}<\overline a_i$, then a sufficiently small increase keeps the observation above $u_0$ and at a rank $k>k^*\ge R^*$, so the derivative calculation strictly raises $R$ and contradicts maximality. Hence $y^*_{i,\max}=\overline a_i$ whenever $\underline a_i>u_0$. If $u_0\in\mathcal I_i$, the already established boundary-support property gives $y^*_{i,\max}\in\{\underline a_i,\overline a_i\}$. This proves part (b).
\end{proof}

\section{Proofs of results in Section~\ref{sec:asymptotics}}\label{app:asymptotic_proofs}

This appendix proves the asymptotic results by separating primitive verification, deterministic approximation, finite-dimensional regularity, and local sensitivity. The first subsections record the reduced-row notation, boundary convention, objective expansions, and sampling conditions used throughout. The finite-rank and slack arguments are stated before the approximation lemmas because they supply the equality-repair constants used to compare the unrestricted, finite-support, and empirical-mass programs. The final subsections prove local sensitivity of the reduced endpoint maps and then apply the directional delta method and bootstrap arguments to the full value function.

\subsection{Notation and input-level implications}\label{sec:appendix_regularities}
For the proofs, let \(h(J)\) and \(g(J)\) index the equality and inequality rows imposed in the \(J\)-th problem after row reduction. The equality set is finite and satisfies \(h(J_n)=h(\infty)\) for all sufficiently large \(n\); the inequality sets satisfy \(g(J_n)\subseteq g(\infty)\) and may increase with \(n\). Duplicate or uniformly locally implied inequalities are removed, and inequalities that bind throughout the feasible set are treated as equalities. All nonzero rows use the fixed normalization described in Section~\ref{sec:asymptotics}. The finite-restriction feasible set is
\[
\mathcal C_J(c,\theta)
:=
\{\mu\in\mathcal M:H_j(\mu;\theta)=c_j^1\ \text{for }j\in h(J),\quad
G_u(\mu;\theta)\le c_u^2\ \text{for }u\in g(J)\}.
\]
For \(a,b\in\mathbb N\), write \(\Delta^{a-1}:=\{p\in[0,1]^a:\sum_jp_j=1\}\) and \(b^{-1}\mathbb Z_+^a:=\{p:bp\in\mathbb Z_+^a\}\). The set \(\mathcal C_J^{(a)}(c,\theta)\) contains measures \(\sum_{j=1}^ap_j\delta_{z_j}\) with \(p\in\Delta^{a-1}\) and \(z\in\YY^a\); \(\mathcal C_J^{(a,b)}(c,\theta)\) additionally requires \(p\in b^{-1}\mathbb Z_+^a\). The associated lower and upper value vectors are denoted by \(V_J^a\), \(V_J^{a,b}\), and \(V_J\). Throughout this appendix, the objective and value functions use the local extension described in Section~\ref{sec:asymptotics}. Thus every straight path \(\eta_0+t h\), \(h\in\mathbb D_0\), is defined for sufficiently small \(|t|\), while the value on every valid probability input is unchanged.

Let \(\mathcal L_{\mathrm{step},n}\) collect rows whose within-cell slope is zero, let \(\mathcal L_{\mathrm{step,eq},n}\) be its equality subset, and let \(\mathcal L_{\mathrm{aff,eq},n}\) collect the remaining equality rows. Write \(m_{\mathrm{aff},n}:=|\mathcal L_{\mathrm{aff,eq},n}|\) and let \(m_{\mathrm{aff}}\) be a finite uniform upper bound. Assumption~\ref{ass:restriction_geometry} gives constants \(\bar f\), \(\bar f_\partial\), and \(\bar V_f\) that bound, respectively, the row sup norms, the within-cell slopes, and total variation, uniformly over all rows in the full family, all large \(n\), and local nuisance values.

Write \(\bar k:=\sup_n k_n<\infty\). For a step row \(u\), let \(v_{u,n}\in\mathbb R^{s_n}\) collect its cell values. Let \(\mathcal D_n\) be the set of boundary indices at which at least one retained step equality changes value, and write \(e_{1:d}:=\sum_{\ell\le d}e_\ell\). The boundary compatibility condition used in Assumption~\ref{ass:restriction_geometry} is
\begin{equation}\label{eq:boundary_consistency}
 e_{1:d}=\sum_{u\in\mathcal L_{\mathrm{step,eq},n}}b_{d,u,n}v_{u,n},
 \qquad
 \sum_{u\in\mathcal L_{\mathrm{step,eq},n}}|b_{d,u,n}|\le\bar b,
 \qquad d\in\mathcal D_n,
\end{equation}
for integers \(b_{d,u,n}\) and a constant \(\bar b\) independent of \(n\). Thus the cumulative mass to either side of a relevant cutoff is determined by the retained step equalities. The condition is used only when a boundary atom is assigned to an adjacent cell or when empirical masses are rounded.

\begin{lemma}\label{lem:boundary_consistency_standard}
Suppose Assumption~\ref{ass:valid_coarsening} gives a finite ordered bracket partition and the retained step-equality basis is chosen after deleting only simplex-redundant rows. In the standard nonoverlapping bracket-share system, \eqref{eq:boundary_consistency} holds with a constant independent of \(n\). The same conclusion holds for cumulative bracket-share rows and for equality rows formed from finite unions of adjacent brackets whenever the row-reduced system spans the elementary bracket-share rows with uniformly bounded integer coefficients.
\end{lemma}

\begin{proof}[Proof of Lemma~\ref{lem:boundary_consistency_standard}]
Fix \(n\) and write the elementary bracket cells as \(I_{1,n},\ldots,I_{s_n,n}\). Let \(e_b\in\mathbb R^{s_n}\) denote the vector that is one on cell \(b\) and zero elsewhere, so the population mass of bracket \(b\) is \(e_b^\top m\) when \(m\) is the vector of cell probabilities. A relevant boundary \(d\) is a boundary at which some retained step equality changes value, and the cumulative mass on its left is \(e_{1:d}^\top m\), where \(e_{1:d}:=\sum_{b\le d}e_b\).

For elementary nonoverlapping bracket shares, choose the row-reduced basis by retaining the elementary share rows needed to span the proper cumulative vectors and deleting only a terminal share that is redundant with the simplex row. Then, for every proper relevant boundary, \(e_{1:d}=\sum_{b\le d}e_b\). The coefficients are integers and their absolute sum is at most the number of retained elementary bracket-share equalities. Assumption~\ref{ass:restriction_geometry} imposes a uniform bound on that equality dimension, so the coefficient sum is uniformly bounded.

If the retained equalities are cumulative shares, the row vector for the cumulative share at boundary \(d\) is exactly \(e_{1:d}\). Hence \eqref{eq:boundary_consistency} holds with one coefficient equal to one whenever the boundary is relevant. Finally, suppose the released rows are finite unions of adjacent brackets and the row-reduced equality matrix spans the elementary bracket-share rows with a uniformly bounded integer right inverse. Let \(A_n\) be the integer matrix whose rows are the retained union rows and let \(B_n\) be such an integer right inverse on the span relevant for proper cumulative masses. Since each \(e_{1:d}\) is an integer sum of elementary rows, \(e_{1:d}=b_{d,n}^\top A_n\) for an integer vector \(b_{d,n}\) whose \(\ell_1\)-norm is bounded by the product of the uniform inverse bound and the uniform equality dimension. This is exactly \eqref{eq:boundary_consistency}. The argument concerns only proper cutoffs; the endpoints have cumulative masses zero and one and do not require adjacent-cell compatibility.
\end{proof}

For \(x=(p,z)\in\Delta^{k-1}\times\YY^k\), set \(\mu_x:=\sum_{j=1}^kp_j\delta_{z_j}\), \(H(x;\theta):=H(\mu_x;\theta)\), \(G(x;\theta):=G(\mu_x;\theta)\), and \(\Phi_k(x;\pi,q):=F(\mu_x;\pi,q)\). Let \(\mathfrak X_n\) be the finite collection of compact, cell-specific neighborhoods of the baseline endpoint optimizer representations obtained by fixing an endpoint, a support size no larger than \(k_n\), and a compatible cell assignment for each support point. The neighborhoods retain every compatible assignment of a boundary atom and are chosen within the positive mass, support-separation, inactive-slack, and cell-boundary margins in Assumption~\ref{ass:endpoint_rank}. The compactness clause of Assumption~\ref{ass:endpoint_rank} permits a finite subcover for each \(n\). The common-gradient clause in Assumption~\ref{ass:objective} gives a modulus \(\bar\omega^\Phi(a)\to0\) such that \(\|\nabla_x\Phi_k(x;\pi,q)-\nabla_x\Phi_k(\tilde x;\pi,q)\|\le\bar\omega^\Phi(\|x-\tilde x\|)\) uniformly over \(X\in\mathfrak X_n\), \(x,\tilde x\in X\), and \((\pi,q)\) in the maintained local extension \(\mathcal N_{\pi q}\). The uniform expansion in Assumption~\ref{ass:objective} means that, for every compact \(K\subset\mathbb D_0\) and every \(M<\infty\), there are moduli \(\omega_{K,M}^\Phi(a)\to0\) and \(\rho_{K,M}^\Phi(t)\to0\) such that, uniformly over all large \(n\), \(X\in\mathfrak X_n\), \(x\in X\), \(\|\Delta_t\|\le M\) with \(x_t=x+t\Delta_t\in X\), and paths \(\eta_t=\eta_0+th_t\) with \(h_t\in K\) and \(h_t\to h\),
\begin{equation}\label{eq:objective_uniform_expansion}
\begin{aligned}
&\big|\Phi_k(x_t;\pi_t,q_t)-\Phi_k(x;\pi_0,q_0)
-D_x\Phi_k(x;\pi_0,q_0)[x_t-x]-t\dot\Phi_{k,x}(h_t)\big|
\le t\rho_{K,M}^\Phi(t),\\
&\big\|\nabla_x\Phi_k(x_t;\pi_t,q_t)-\nabla_x\Phi_k(x;\pi_0,q_0)
-t\dot G_{k,x}(h_t)\big\|
\le\omega_{K,M}^\Phi(\|x_t-x\|)+t\rho_{K,M}^\Phi(t),
\end{aligned}
\end{equation}
where the linear maps \(\dot\Phi_{k,x}\) and \(\dot G_{k,x}\) have uniformly bounded norms. For every compact \(K\subset\mathbb D_0\), the evaluated maps \(x\mapsto\dot\Phi_{k,x}(h)\) and \(x\mapsto\dot G_{k,x}(h)\), \(h\in K\), are uniformly bounded and have a common modulus of continuity on the retained neighborhoods. This is the common remainder bound used in the envelope argument.

The restriction smoothness in Assumption~\ref{ass:restriction_geometry} gives an analogous common expansion on every compatible cell-assigned endpoint representation with at most \(\bar k\) coordinates, including the augmented endpoint representations and optimizer neighborhoods used below. For any row or support-position bound, write its left-hand side as \(R_{u,n}(x;\eta)\). For every compact \(K\subset\mathbb D_0\) and every \(M<\infty\), there are moduli \(\omega_{K,M}^R(a)\to0\) and \(\rho_{K,M}^R(t)\to0\) such that, uniformly over those representations and paths with \(\|\Delta_t\|\le M\),
\begin{equation}\label{eq:restriction_uniform_expansion}
\begin{aligned}
&|R_{u,n}(x_t;\eta_t)-R_{u,n}(x;\eta_0)-D_xR_{u,n}(x;\eta_0)[x_t-x]-t\dot R_{u,n,x}(h_t)|\le t\rho_{K,M}^R(t),\\
&\|\nabla_xR_{u,n}(x_t;\eta_t)-\nabla_xR_{u,n}(x;\eta_0)-t\dot G^R_{u,n,x}(h_t)\|\le\omega_{K,M}^R(\|x_t-x\|)+t\rho_{K,M}^R(t).
\end{aligned}
\end{equation}
The derivative maps in \eqref{eq:restriction_uniform_expansion} are uniformly bounded. For every compact \(K\subset\mathbb D_0\), the evaluated maps \(x\mapsto\dot R_{u,n,x}(h)\) and \(x\mapsto\dot G^R_{u,n,x}(h)\), \(h\in K\), have a common modulus of continuity on the retained representations, uniformly over included rows and large \(n\). The same smoothness gives a common modulus for \(\nabla_xR_{u,n}(\cdot;\eta)\) on these representations, uniformly over local inputs. Step restrictions are locally constant within a retained cell assignment; the expansion for a moving support bound follows from the cutoff map.

Row reduction is used throughout. Within a retained endpoint representation, repeated optimizer atoms are merged and zero-mass atoms are deleted. An active inequality is removed as uniformly locally implied only when, on a common neighborhood of the input and coordinate system, its residual is a nonnegative linear combination of the retained inequality residuals plus a linear combination of the equality residuals, with uniformly bounded coefficients. The same relation must hold for the perturbed right-hand sides. Removing such a row leaves the local feasible set and its directional derivative unchanged; no row is deleted merely because its gradient is dependent at the baseline point.

The row-form clauses of Assumption~\ref{ass:restriction_geometry} are stated in terms of the standard restriction classes used in the paper. The next result records this verification separately from the finite-rank and slack calculations below.

\begin{prop}\label{prop:restriction_rows_primitives}
Suppose Assumptions~\ref{ass:support} and~\ref{ass:valid_coarsening} hold, and the population right-hand sides are the corresponding released shares, moments, or overlap probabilities. For the standard nonoverlapping bracket-share rows, finitely many cumulative or adjacent-bracket share rows satisfying Lemma~\ref{lem:boundary_consistency_standard}, finitely many overall or subgroup moment rows with \(C^1\) nuisance coefficients, finitely many income-share or Lorenz-ordinate rows, and finitely many linear ratio restrictions after cross-multiplication by a denominator with known sign, the restriction-form clauses of Assumption~\ref{ass:restriction_geometry} hold. In Scenario~2, the same conclusion holds for any finite set of containment lower-bound and overlap upper-bound interval rows. If the set of possible interval rows is countable, the conclusion holds row by row under the same uniform coefficient, cutoff, and total-variation bounds.
\end{prop}

\begin{proof}[Proof of Proposition~\ref{prop:restriction_rows_primitives}]
Fix a row after the positive normalization used in Section~\ref{sec:asymptotics}. Bracket-share, cumulative-share, and adjacent-bracket-share restrictions have integrands that are step functions on the bracket partition. Their step vectors are elementary cell indicators or finite integer sums of adjacent indicators, and Lemma~\ref{lem:boundary_consistency_standard} gives the required convention for a mass placed at a reported cutoff. Because the number of equality rows retained after row reduction is uniformly bounded, the row-reduction coefficients for these standard share systems remain uniformly bounded under the hypotheses of the proposition.

A mean row has integrand \(y\). More generally, once the released normalizing constants and any subgroup indicators are fixed, subgroup means, income shares, and Lorenz ordinates are linear combinations of cell probabilities and cell first moments. On each cell \(I_{d,n}(\tau)\), each such row therefore has the form \(a_{u,d,n}(\gamma)y+b_{u,d,n}(\gamma)\). The same representation applies to linear aggregate restrictions. If a ratio restriction is written as \(A/B\le r\) and the denominator has a known positive sign on the local domain, it is equivalent to \(A-rB\le0\); if the sign is negative, multiplying by minus one gives the normalized inequality. Thus no division by a local quantity is used in the row map, and the denominator condition is only needed to fix the direction of the inequality.

The compact support \(\YY\) converts uniform bounds on the coefficients into uniform sup-norm and total-variation bounds for the integrands. The assumed \(C^1\) nuisance coefficient and cutoff maps, together with their common derivative modulus, give the local expansion and derivative-modulus parts of Assumption~\ref{ass:restriction_geometry}. For step restrictions, the only discontinuities occur at retained cutoffs. For affine restrictions, the within-cell slope is the coefficient \(a_{u,d,n}(\gamma)\), which is uniformly bounded. Row reduction preserves these properties because it uses only finitely many uniformly bounded linear combinations in the finite-row cases, and the countable case assumes the same coefficient, cutoff, and total-variation bounds uniformly row by row.

In Scenario~2, the containment lower-bound and overlap upper-bound rows for an interval \([a,b]\) use the step integrand \(\mathbf 1\{a\le y\le b\}\), with the one-sided value determined by the partition generated by the retained interval endpoints. A finite set of possible survey intervals gives a finite set of normalized rows. A countable set is covered one row at a time, and uniformly over the retained sequence, under the stated common bounds. Finally, Assumption~\ref{ass:valid_coarsening} gives an unobserved population distribution that is consistent with the released shares, moments, and overlap probabilities. Since the proposition takes the population right-hand sides to be those released quantities, that distribution belongs to \(\mathcal C_\infty(c_0,\theta_0)\). The nonemptiness clause of Assumption~\ref{ass:restriction_geometry} follows.
\end{proof}

\subsection{Objective verification}\label{sec:appendix_objective_verification}

This subsection verifies Assumption~\ref{ass:objective} for the objective classes used in the main text. The distinction mirrors implementation. Smooth functions of finitely many moments require only formula-domain checks and denominator bounds. Gini and Hoover require separation from the nonsmooth kink locations on the optimizer neighborhoods. Quantile ratios require a stable finite quantile branch, with atom and cumulative-probability gaps that survive the support-reduction, mass-rounding, equality-repair, and Slater-mixture steps used below.

For moment and fixed-kink functionals, the local extension of the distribution-function coordinate can be written explicitly. This extension is only a device for deterministic directional derivatives; on valid probability inputs the objective is unchanged. Let \(Q_0=F_{q_0}\), and let \(Q\) be a bounded local coordinate with \(Q(\overline y)=1\). For \(f\in C^1(\YY)\), define \(\mathcal I_f(Q):=\int f\,dq_0-\int_{\underline y}^{\overline y}[Q(t)-Q_0(t)]f'(t)\,dt\) and, for fixed \(z\in\YY\), define
\[
\mathcal A_z(Q)
:=\int |y-z|\,dq_0(y)
 +\int_{\underline y}^{z}\bigl(Q(t)-Q_0(t)\bigr)\,dt
 -\int_{z}^{\overline y}\bigl(Q(t)-Q_0(t)\bigr)\,dt.
\]
Integration by parts shows that \(\mathcal I_f(F_q)=\int f\,dq\) and \(\mathcal A_z(F_q)=\int|y-z|\,dq(y)\) for every probability distribution \(q\) supported on \(\YY\). These affine formulas therefore agree with the original objective on valid inputs and define it on a sup-norm neighborhood of \(Q_0\). Quantile functionals require a separate finite-branch local extension, stated before the quantile-ratio verification below.

The following integration-by-parts bound is used repeatedly. For a mass-preserving \(q\)-tangent \(h_q\in\ell^\infty(\mathcal F_0)\), write \(H_{h_q}\) for \(t\mapsto h_q((-\infty,t])\), with \(H_{h_q}(\overline y)=0\) and zero extension outside \(\YY\). If \(f\in C^1(\YY)\), let \(\Lambda_f(h_q):=-\int_{\underline y}^{\overline y}H_{h_q}(t)f'(t)\,dt\), so \(|\Lambda_f(h_q)|\le \diam(\YY)\|f'\|_\infty\|h_q\|_{\ell^\infty(\mathcal F_0)}\). For the fixed-kink map \(y\mapsto |y-z|\), let \(\Lambda_z^{\mathrm{abs}}(h_q):=\int_{\underline y}^{z}H_{h_q}(t)\,dt-\int_{z}^{\overline y}H_{h_q}(t)\,dt\), so \(|\Lambda_z^{\mathrm{abs}}(h_q)|\le \diam(\YY)\|h_q\|_{\ell^\infty(\mathcal F_0)}\). If \(q_t\) is either a path of distributions or an admissible local-extension path with \(t^{-1}(F_{q_t}-F_q)\to H_{h_q}\) uniformly, the affine formulas give \(\int f\,dq_t-\int f\,dq=t\Lambda_f(h_q)+o(t)\). They also give \(\int |y-z|\,dq_t(y)-\int |y-z|\,dq(y)=t\Lambda_z^{\mathrm{abs}}(h_q)+o(t)\), uniformly for \(z\in\YY\).

\begin{prop}\label{prop:objective_examples}
Let \(g=(g_1,\ldots,g_m)\), with each \(g_r\in C^2(\YY)\) and bounded together with its first two derivatives, and let \(\mathcal J(\nu)=\Psi(\int g\,d\nu)\). Suppose the extended moment vectors generated by the local objective domain \(\mathcal O\) lie in a compact set contained in the open domain of a \(C^2\) map \(\Psi\), and that \(D\Psi\) and \(D^2\Psi\) are bounded on a neighborhood of that compact set. Then the objective satisfies the objective-specific parts of Assumption~\ref{ass:objective}.
\end{prop}

When \(\underline y>0\), Proposition~\ref{prop:objective_examples} covers the mean log deviation, the Theil index, \(\mathrm{GE}_\alpha\) for \(\alpha\neq0,1\), and the Atkinson class. On compact \(\YY\) it also covers the Kolm class \(K_\alpha(\nu)=\int y\,d\nu+\alpha^{-1}\log\int e^{-\alpha y}\,d\nu\) for \(\alpha>0\), as well as any linear functional \(\nu\mapsto\int\varphi\,d\nu\) with \(\varphi\in C^2(\YY)\).

\begin{proof}[Proof of Proposition~\ref{prop:objective_examples}]
Let \(T(\nu):=\int g\,d\nu\). Since every \(g_r\) is continuously differentiable on the compact interval \(\YY\), \(\|T(\nu)-T(\nu')\|_\infty\le (\max_r\|g_r'\|_\infty)\Wone(\nu,\nu')\). When only the optimization measure changes, the exact-observation component cancels and the mixture moment changes by \((1-\pi)\bigl(T(\mu)-T(\nu)\bigr)\). The local objective domain keeps \(|1-\pi|\) bounded and keeps the extended moment vector in a compact subset of the open domain of \(\Psi\). The mean-value theorem and boundedness of \(D\Psi\) therefore give \(|F(\mu;\pi,q)-F(\nu;\pi,q)|\le C\Wone(\mu,\nu)\), uniformly on \(\mathcal O\).

Now keep \(\mu\) fixed. For valid probability inputs, the integration-by-parts formula gives \(|\int g_r\,dq-\int g_r\,d\tilde q|\le C\|q-\tilde q\|_{\ell^\infty(\mathcal F_0)}\) for each coordinate. The same inequality holds on the maintained local extension because \(\mathcal I_{g_r}(Q)\) is affine and continuous in the sup norm. Boundedness of \(g\) controls the term generated by changing \(\pi\). Applying the mean-value theorem to \(\Psi\) again gives the input-continuity clause of Assumption~\ref{ass:objective}, with a linear modulus in \(|\pi-\tilde\pi|+\|q-\tilde q\|_{\ell^\infty(\mathcal F_0)}\).

For the local finite-dimensional expansion, write \(T_k(x;\pi,q)=(1-\pi)\sum_{j=1}^k p_jg(z_j)+\pi\int g\,dq\) and \(\Phi_k(x;\pi,q)=\Psi(T_k(x;\pi,q))\). The support size is bounded by \(\bar k\). The first support-coordinate derivatives of \(T_k\) are \((1-\pi)g(z_j)\) with respect to the mass coordinates and \((1-\pi)p_jg'(z_j)\) with respect to the location coordinates. The bounded first and second derivatives of \(g\), together with bounded \(\pi\) and bounded support size, give uniform bounds and a common modulus for these derivatives and for the second support-coordinate derivatives of \(T_k\).

Let \(\eta_t=\eta_0+t h_t\), with \(h_t\) in a compact subset of \(\mathbb D_0\), and let \(x_t=x+t\Delta_t\), with \(\|\Delta_t\|\le M\). The auxiliary bound gives \(\int g\,dq_t-\int g\,dq_0=t\Lambda_g(h_{q,t})+o(t)\) uniformly over the compact direction set, where \(\Lambda_g\) is applied coordinatewise. Hence \(T_k(x_t;\pi_t,q_t)\), \(D_xT_k(x_t;\pi_t,q_t)\), and their input derivatives admit first-order expansions with a common \(o(t)\) remainder, uniformly over all retained parameterizations and all large \(n\). The input derivative of the moment vector at \((x,\eta_0)\) is the linear map \(\dot T_{k,x}(h)=h_\pi[\int g\,dq_0-\sum_{j=1}^kp_jg(z_j)]+\pi_0\Lambda_g(h_q)\), with the second term omitted when the \(q\)-coordinate is fixed.

The compactness hypothesis keeps all intermediate moment vectors in a compact subset on which \(D\Psi\) and \(D^2\Psi\) are bounded and \(D^2\Psi\) is uniformly continuous. Taylor's theorem applied to \(\Psi\) gives the value expansion in \eqref{eq:objective_uniform_expansion}, with a remainder bounded by a common \(O(t^2)+t o(1)\) term. Applying the same Taylor argument to \(D_x\Phi_k=D\Psi(T_k)D_xT_k\) gives the gradient expansion. The displayed derivative formulas and the common moduli for \(g'\), \(D\Psi\), and \(D^2\Psi\) also give the common modulus for the evaluated maps \(x\mapsto\dot\Phi_{k,x}(h)\) and \(x\mapsto\dot G_{k,x}(h)\), uniformly over \(h\) in compact subsets of \(\mathbb D_0\). These are exactly the objective-specific clauses of Assumption~\ref{ass:objective}.
\end{proof}

\begin{corollary}\label{cor:smooth_objective_domains}
Suppose the conditions of Proposition~\ref{prop:objective_examples} hold. If every denominator entering \(\Psi\) is bounded below by a positive constant on the compact extended moment range, then all objective-specific clauses of Assumption~\ref{ass:objective} hold for \(\mathcal J(\nu)=\Psi(\int g\,d\nu)\). A positive lower support bound, a released positive mean equality, or a finite endpoint check over the local feasible moment range is sufficient for this denominator condition, depending on the index.
\end{corollary}

\begin{proof}[Proof of Corollary~\ref{cor:smooth_objective_domains}]
Proposition~\ref{prop:objective_examples} proves the \(\Wone\)-Lipschitz, input-continuity, and local value-and-gradient expansion clauses. It remains only to verify the denominator part of Assumption~\ref{ass:objective}. If the lower support point is positive and the denominator is a positive continuous moment on that support, compactness of \(\YY\) gives a positive lower bound. If the denominator is the mean and a released equality fixes the mean at a positive value, then continuity of the local right-hand side keeps the mean bounded below on a sufficiently small neighborhood of \(\eta_0\). In the remaining finite-dimensional cases, the denominator is a continuous function of the compact extended moment vector; a finite endpoint or local feasible-range check showing that it is positive on that compact set implies a positive minimum by Weierstrass' theorem. Under any of these sufficient conditions, every denominator entering \(\Psi\) is uniformly bounded away from zero on the maintained compact range, so all objective-specific clauses of Assumption~\ref{ass:objective} hold.
\end{proof}

Quantile ratios require stability of the quantile branch used by the finite endpoint programs. A density condition for a released cutoff is not enough, because here the quantile is evaluated at distributions chosen by the endpoint program and those distributions may be discrete. We therefore verify the objective condition through a finite-branch atom-gap condition on the measures actually generated by the endpoint, reduction, rounding, repair, and Slater-mixture steps. Fix \(0<\tau_1<\tau_2<1\), let \(Q_\tau(\nu):=\inf\{y:F_\nu(y)\ge\tau\}\), and set \(\mathcal J(\nu)=Q_{\tau_2}(\nu)/Q_{\tau_1}(\nu)\).

For this objective, let \(\mathcal O_Q\) denote the part of the local objective domain generated by the finite endpoint parameterizations, the bounded-support reductions, the mass-grid rounding and equality-repair operations, and the Slater mixtures used to pass from \(J_n\) to the full inequality family. The following condition is imposed on that generated domain. There are constants \(\underline Q>0\), \(\underline p_Q>0\), \(\delta_Q>0\), \(\Delta_Q>0\), and \(\xi_Q>0\) such that, for all sufficiently large \(n\), every local input, every represented optimization measure \(\mu=\sum_jp_j\delta_{z_j}\) in \(\mathcal O_Q\), and \(r=1,2\), the quantile \(Q_{\tau_r}((1-\pi)\mu+\pi q)\) is a support coordinate \(z_{j_r}\) satisfying \(z_{j_1}\ge\underline Q\), optimization-measure atom mass \(p_{j_r}\ge\underline p_Q\), and support separation at least \(\delta_Q\). Its cumulative probability in the combined distribution \((1-\pi)\mu+\pi q\) is at most \(\tau_r-\Delta_Q\) below \(z_{j_r}\) and at least \(\tau_r+\Delta_Q\) at or below \(z_{j_r}\). The same atom branch is retained after every support reduction, rounding, equality repair, and Slater mixture whose \(\Wone\) distance from the original optimization measure is below \(\xi_Q\). If two represented optimization measures in \(\mathcal O_Q\) are closer than \(\xi_Q\) in \(\Wone\), their \(\tau_r\)-quantile atoms are paired by the natural atom matching. If they are not closer than \(\xi_Q\), the bounded support of \(\YY\) controls the quantile difference. Finally, when the optimization measure is fixed, a change in \((\pi,q)\) by \(d_{\pi q}\) changes each cumulative probability in the preceding inequalities by at most \(C_Qd_{\pi q}\).

\begin{prop}\label{prop:quantile_ratio_examples}
Suppose the finite-branch quantile stability condition in the preceding paragraph holds on \(\mathcal O_Q\), and the maintained local extension fixes each quantile on its retained atom branch. Then the quantile-ratio objective satisfies the objective-specific parts of Assumption~\ref{ass:objective} on \(\mathcal O_Q\). In particular, the conclusion does not require a separate global \(\Wone\)-Lipschitz assumption for the quantile map.
\end{prop}

\begin{proof}[Proof of Proposition~\ref{prop:quantile_ratio_examples}]
We first prove the \(\Wone\)-Lipschitz bound for each quantile branch on the generated domain. Fix \(r\in\{1,2\}\) and two represented optimization measures \(\mu\) and \(\tilde\mu\) in \(\mathcal O_Q\), with the same local input \((\pi,q)\). If \(\Wone(\mu,\tilde\mu)\ge\xi_Q\), compact support gives \(|Q_{\tau_r}((1-\pi)\mu+\pi q)-Q_{\tau_r}((1-\pi)\tilde\mu+\pi q)|\le\diam(\YY)\le\diam(\YY)\Wone(\mu,\tilde\mu)/\xi_Q\). Now suppose \(\Wone(\mu,\tilde\mu)<\xi_Q\), reducing \(\xi_Q\) if necessary so that \(\xi_Q<\underline p_Q\delta_Q/12\). Let \(z_{j_r}\) be the retained \(\tau_r\)-quantile atom of \(\mu\). If \(\tilde\mu\) placed no support point within \(\delta_Q/3\) of \(z_{j_r}\), then at least \(\underline p_Q\) mass initially located at \(z_{j_r}\) would have to move a distance at least \(\delta_Q/3\) under every coupling, contradicting the choice of \(\xi_Q\). Hence a matched support point \(\tilde z_{j_r}\) exists. It is unique because the branch condition gives support separation at least \(\delta_Q\), and the branch-pairing clause identifies it with the retained \(\tau_r\)-quantile atom of \(\tilde\mu\). Thus both matched atoms have mass at least \(\underline p_Q\). Because \(\YY\) is compact, an optimal coupling \(\Gamma^\star\) exists. If \(\Gamma^\star\) transported less than \(\underline p_Q/2\) mass from \(z_{j_r}\) to \(\tilde z_{j_r}\), then at least \(\underline p_Q/2\) units of mass would either leave \(z_{j_r}\) for an unmatched atom of \(\tilde\mu\) or arrive at \(\tilde z_{j_r}\) from an unmatched atom of \(\mu\). Every unmatched atom on either side is at distance at least \(2\delta_Q/3\) from the opposite matched atom, so this would force transportation cost at least \(\underline p_Q\delta_Q/3\), contradicting \(\Wone(\mu,\tilde\mu)<\xi_Q\). Therefore \(\Gamma^\star\) transports at least \(\underline p_Q/2\) mass from \(z_{j_r}\) to \(\tilde z_{j_r}\), and its cost on that flow is at least \((\underline p_Q/2)|z_{j_r}-\tilde z_{j_r}|\). Consequently \(|z_{j_r}-\tilde z_{j_r}|\le2\Wone(\mu,\tilde\mu)/\underline p_Q\). The finite-branch condition states that, for such close measures, the paired atoms remain the \(\tau_r\)-quantile atoms of the combined distributions after mixing with the same \((\pi,q)\). Thus this is the desired quantile bound. The large-distance and small-distance cases together give a uniform \(\Wone\)-Lipschitz constant for both quantiles on \(\mathcal O_Q\).

The input-continuity clause uses only the cumulative-probability gap. Fix \(\mu\) and change \((\pi,q)\) to \((\tilde\pi,\tilde q)\). By assumption, every cumulative probability entering the lower and upper gap inequalities changes by at most \(C_Qd_{\pi q}((\pi,q),(\tilde\pi,\tilde q))\). If this quantity is at most \(\Delta_Q/2\), the same atom still has cumulative mass below it at most \(\tau_r-\Delta_Q/2\) and cumulative mass at or below it at least \(\tau_r+\Delta_Q/2\), so the quantile is unchanged. If the quantity exceeds \(\Delta_Q/2\), compact support gives the bound \(\diam(\YY)\), which is dominated by \(2\diam(\YY)C_Qd_{\pi q}/\Delta_Q\). This gives a common modulus that tends to zero at the origin. Since \(Q_{\tau_1}\ge\underline Q\), the elementary quotient inequality transfers both continuity bounds to \(Q_{\tau_2}/Q_{\tau_1}\).

It remains to verify the local value-and-gradient expansion. Fix a compact \(K\subset\mathbb D_0\), a retained optimizer parameterization, and paths \(x_t=x+t\Delta_t\) and \(\eta_t=\eta_0+t h_t\), with \(h_t\in K\) and \(\|\Delta_t\|\) bounded. The atom masses, support separations, branch pairings, denominator lower bound, and cumulative-probability gaps all have uniform positive margins on \(\mathcal O_Q\). Hence, for all sufficiently small \(t\), the same coordinates \(j_1\) and \(j_2\) remain the retained quantile branches. The maintained local extension fixes each quantile on that branch, so \(Q_{\tau_r}((1-\pi_t)\mu_{x_t}+\pi_tq_t)=z_{j_r,t}\) for \(r=1,2\). Thus \(D_pQ_{\tau_r}=0\), \(\partial_{z_{j_r}}Q_{\tau_r}=1\), \(\partial_{z_j}Q_{\tau_r}=0\) for \(j\ne j_r\), and the input derivative with respect to \((\pi,q)\) is zero on the retained branch. These identities are exact on the local branch, so the value and support-gradient expansions in \eqref{eq:objective_uniform_expansion} have zero branch remainder. Applying the quotient rule to \(Q_{\tau_2}/Q_{\tau_1}\), using \(Q_{\tau_1}\ge\underline Q\), gives the required expansion and the common modulus for the evaluated input-derivative maps. No global Lipschitz property of the quantile map is used.
\end{proof}

The final verification covers the Gini and Hoover indices under local separation conditions at endpoint optimizers. These conditions can be checked once the feasible mean is bounded away from zero, distinct optimizer support points are uniformly separated, and in the Hoover case the optimizer mean is uniformly separated from those support points. The condition on the exact-observation distribution is local: it concerns only the support points and, for Hoover, the endpoint mean that appear in the finite optimizer representations. It can fail in applications with exact-observation mass at those exact locations, in which case the sharp bounds remain valid but the smooth local expansion below should be replaced by a kink-specific formulation. The local objective domain \(\mathcal O\) includes the local extension of the distribution-function coordinate used for deterministic derivative paths. Empirical and bootstrap paths need only belong to the domain of the value function; their stochastic effect is handled later through Assumption~\ref{ass:input_process} and the objective continuity bounds.

Let \(\nu=(1-\pi)\mu+\pi q\) and \(m(\nu):=\int y\,d\nu(y)\). For both indices, suppose there is \(\underline m>0\) such that \(m((1-\pi)\mu+\pi q)\ge\underline m\) on \(\mathcal O\). For Gini, also impose condition~\textup{(a)} below; for Hoover, impose condition~\textup{(b)}.
\begin{enumerate}
\item[(a)] For every compact \(K\subset\mathbb D_0\), there exist \(N_K\in\mathbb N\), \(\delta_K>0\), and a modulus \(\omega_K^G\), with \(\omega_K^G(a)\to0\) as \(a\downarrow0\), such that, for all \(n\ge N_K\), both endpoints, and every retained optimizer local parameterization after repeated atoms are merged, distinct support points remain separated on the associated cell-specific neighborhood, \(\min_{i<j}|z_i-z_j|\ge\delta_K\). For every support point \(z_j\), the baseline CDF \(F_q\) and the direction CDF perturbations \(H_{h_q}\), \(h\in K\), are continuous on \([z_j-\delta_K,z_j+\delta_K]\) and satisfy a common modulus there.

\item[(b)] For every compact \(K\subset\mathbb D_0\), there exist \(N_K\in\mathbb N\), \(\delta_K>0\), and a modulus \(\omega_K^H\), with \(\omega_K^H(a)\to0\) as \(a\downarrow0\), such that, for all \(n\ge N_K\), both endpoints, and every retained optimizer local parameterization, the mean \(m(x;\pi,q)\) stays at least \(\delta_K\) away from every support point. On the \(\delta_K\)-neighborhood of that mean, the baseline CDF \(F_q\) and the direction CDF perturbations \(H_{h_q}\), \(h\in K\), are continuous and satisfy a common modulus.
\end{enumerate}

\begin{prop}\label{prop:gini_hoover_examples}
Suppose \(m((1-\pi)\mu+\pi q)\ge\underline m\) on \(\mathcal O\). Under condition~\textup{(a)}, the Gini objective satisfies the objective-specific parts of Assumption~\ref{ass:objective}; under condition~\textup{(b)}, the Hoover objective does so. In each case the conclusion includes the two continuity clauses and the expansion in \eqref{eq:objective_uniform_expansion}.
\end{prop}

\begin{proof}[Proof of Proposition~\ref{prop:gini_hoover_examples}]
First consider the Gini index. Define \(m(\nu):=\int y\,d\nu(y)\), \(a_\nu(y):=\int |y-y'|\,d\nu(y')\), and \(B(\nu):=\int a_\nu(y)\,d\nu(y)\), so \(\mathrm{Gi}(\nu)=B(\nu)/(2m(\nu))\). The map \(y\mapsto y\) is \(1\)-Lipschitz, and \(a_\nu\) is \(1\)-Lipschitz with \(\|a_\nu-a_{\nu'}\|_\infty\le\Wone(\nu,\nu')\). Therefore \(|m(\nu)-m(\nu')|\le\Wone(\nu,\nu')\) and \(|B(\nu)-B(\nu')|\le2\Wone(\nu,\nu')\). Since \(0\le B(\nu)\le\diam(\YY)\) and the mean is bounded below by \(\underline m\), the quotient inequality gives a Lipschitz bound for \(\mathrm{Gi}\) in the distribution argument. For \(\nu=(1-\pi)\mu+\pi q\) and \(\tilde\nu=(1-\pi)\tilde\mu+\pi q\), \(\Wone(\nu,\tilde\nu)\le |1-\pi|\Wone(\mu,\tilde\mu)\); the local domain bounds \(|1-\pi|\), so the \(\Wone\)-Lipschitz clause follows on valid probability inputs. On the local CDF extension, the formula for \(\mathcal A_z(Q)\) gives \(|\mathcal A_z(Q)-\mathcal A_{z'}(Q)|\le[1+2\|Q-Q_0\|_\infty]|z-z'|\). The term depending only on \(q\) cancels when only \(\mu\) changes, and the displayed bound controls the cross term uniformly. Thus the same Lipschitz conclusion holds throughout the maintained local extension.

For Hoover, let \(L(\nu):=\int |y-m(\nu)|\,d\nu(y)\) and \(\mathrm{Ho}(\nu)=L(\nu)/(2m(\nu))\). For any \(\nu\) and \(\nu'\), \(|L(\nu)-L(\nu')|\le|\int |y-m(\nu)|\,d(\nu-\nu')(y)|+|m(\nu)-m(\nu')|\le2\Wone(\nu,\nu')\), because \(y\mapsto |y-m(\nu)|\) is \(1\)-Lipschitz. Also \(0\le L(\nu)\le\diam(\YY)\). The same quotient argument gives the \(\Wone\)-Lipschitz bound on valid probability inputs. On the local extension, write \(L((1-\pi)\mu+\pi q)=(1-\pi)\int|y-m|\,d\mu(y)+\pi\ell_q(m)\), where \(m=(1-\pi)\int y\,d\mu+\pi\int y\,dq\) and \(\ell_q(m)=\int|y-m|\,dq(y)\). Changing \(\mu\) to \(\tilde\mu\) changes \(m\) by \(O(\Wone(\mu,\tilde\mu))\), changes the first term by the same order, and changes \(\ell_q(m)\) by the same order because the integration-by-parts extension makes \(m\mapsto\ell_q(m)\) uniformly Lipschitz on the local domain. The Hoover \(\Wone\)-Lipschitz clause follows.

Now fix \(\mu\) and vary \((\pi,q)\). For valid probability inputs, the induced CDFs of \((1-\pi)\mu+\pi q\) and \((1-\tilde\pi)\mu+\tilde\pi\tilde q\) differ in sup norm by at most a constant times \(d_{\pi q}((\pi,q),(\tilde\pi,\tilde q))\). The identity \(\Wone(\nu,\tilde\nu)=\int_{\underline y}^{\overline y}|F_\nu(s)-F_{\tilde\nu}(s)|\,ds\), followed by the Lipschitz bounds just proved, yields a common linear input-continuity modulus. On the maintained local extension, the affine integration-by-parts formulas for the mean and fixed-kink terms, the identity \(B(q)=2\int_{\underline y}^{\overline y}F_q(s)(1-F_q(s))\,ds\), and boundedness of the local CDF coordinates give the same modulus directly. This proves the two continuity clauses for both indices.

It remains to prove the local value-and-gradient expansion. Fix a compact \(K\subset\mathbb D_0\), an endpoint, and a retained local parameterization around an optimizer. For Gini, condition~\textup{(a)} gives a neighborhood on which repeated atoms have been merged, distinct support points remain separated, and the signs of \(z_i-z_j\) are fixed. For Hoover, condition~\textup{(b)} gives a neighborhood on which the signs of \(z_j-m(x;\pi,q)\) are fixed. Write \(\nu_x=(1-\pi)\mu_x+\pi q\).

For Gini,
\[
B(\nu_x)=(1-\pi)^2\sum_{i,j}p_ip_j|z_i-z_j|+2\pi(1-\pi)\sum_jp_j a_q(z_j)+\pi^2B(q),
\]
where \(a_q(z)=\int|z-y|\,dq(y)\). The first term is \(C^1\) in \(x\) on the retained neighborhood because its sign pattern is fixed. By condition~\textup{(a)}, \(F_q\) is continuous at each support point with a common modulus, and therefore \(a_q'(z)=2F_q(z)-1\) locally, with \(|a_q(z+\Delta)-a_q(z)-\Delta(2F_q(z)-1)|\le2|\Delta|\omega_K^G(|\Delta|)\). For a path with \(t^{-1}(F_{q_t}-F_q)\to H_{h_q}\) uniformly, the auxiliary integration-by-parts formula gives \(a_{q_t}(z)-a_q(z)=t\Lambda_z^{\mathrm{abs}}(h_q)+o(t)\), uniformly over the support points in the retained neighborhoods. The compact-set modulus for \(H_{h_q}\), \(h\in K\), gives the corresponding uniform expansion of the \(z\)-gradient, with perturbation \(2tH_{h_q}(z)+o(t)\). Finally, \(B(q)=2\int F_q(1-F_q)\) implies \(B(q_t)-B(q)=2t\int_{\underline y}^{\overline y}(1-2F_q(s))H_{h_q}(s)\,ds+o(t)\), uniformly on \(K\). The mean \(m(\nu_x)\) is affine in \((x,\pi,q)\) and is bounded away from zero. Applying the quotient rule to \(B(\nu_x)/(2m(\nu_x))\), and then to its support-coordinate gradient, gives \eqref{eq:objective_uniform_expansion} for Gini.

For Hoover, \(L(\nu_x)=(1-\pi)\sum_jp_j|z_j-m(x;\pi,q)|+\pi\ell_q(m(x;\pi,q))\), where \(\ell_q(m)=\int|y-m|\,dq(y)\) and \(m(x;\pi,q)=(1-\pi)\sum_jp_jz_j+\pi\int y\,dq(y)\). Condition~\textup{(b)} fixes the signs in the atom term, so that term is \(C^1\) in \(x\). The local continuity modulus at the mean gives \(\ell_q'(m)=2F_q(m)-1\) and \(|\ell_q(m+\Delta)-\ell_q(m)-\Delta(2F_q(m)-1)|\le2|\Delta|\omega_K^H(|\Delta|)\). The same integration-by-parts argument gives \(\ell_{q_t}(m)-\ell_q(m)=t\Lambda_m^{\mathrm{abs}}(h_q)+o(t)\), and the compact-set modulus gives the support-gradient perturbation \(2tH_{h_q}(m)+o(t)\), uniformly over the retained neighborhoods. Since the mean is affine and bounded away from zero, the quotient rule gives the value and gradient expansions for Hoover. In both cases, the displayed moduli and the fixed finite support bound give the common modulus for \(x\mapsto\dot\Phi_{k,x}(h)\) and \(x\mapsto\dot G_{k,x}(h)\), uniformly over \(h\in K\). This completes the verification of Assumption~\ref{ass:objective} for the two indices.
\end{proof}

\begin{corollary}\label{cor:gini_hoover_primitive_domains}
Suppose the feasible mean is uniformly bounded below by \(\underline m>0\). For Gini, condition~\textup{(a)} above is implied if the exact-observation CDF has a common local continuity modulus on neighborhoods of the possible optimizer support locations and the \(q\)-directions in every compact subset of \(\mathbb D_0\) have the same modulus there. For Hoover, condition~\textup{(b)} is implied if, in addition, the optimizer mean is uniformly separated from the candidate support set; in a design with a finite candidate support set it is enough that \(\inf_{z\in\mathcal B_n}|m(\nu_x)-z|\ge\delta\) uniformly over endpoint optimizers and large \(n\). Under these conditions, Proposition~\ref{prop:gini_hoover_examples} verifies the objective-specific clauses of Assumption~\ref{ass:objective} for the corresponding index.
\end{corollary}

\begin{proof}[Proof of Corollary~\ref{cor:gini_hoover_primitive_domains}]
For Gini, repeated optimizer atoms are merged before the endpoint neighborhoods are formed. The positive support-separation clause of Assumption~\ref{ass:endpoint_rank} then gives a neighborhood on which distinct optimizer support points remain separated. The assumed local continuity modulus for the exact-observation CDF, together with the same modulus for all direction functions in compact subsets of \(\mathbb D_0\), is exactly the modulus required in condition~\textup{(a)} of Proposition~\ref{prop:gini_hoover_examples}. The feasible mean lower bound supplies the denominator condition.

For Hoover, the additional separation condition keeps \(m(\nu_x)\) a positive distance from every candidate support point on the retained neighborhood. In a finite-interval design, the displayed lower bound over \(z\in\mathcal B_n\) gives this separation uniformly because every optimizer support point belongs to the candidate set of possible intervals after the finite-support reduction. The sign of \(z_j-m(\nu_x)\) is therefore locally constant for every optimizer support point. Applying the same CDF and direction-function continuity modulus at the local mean gives condition~\textup{(b)}. Proposition~\ref{prop:gini_hoover_examples} then verifies the objective-specific clauses of Assumption~\ref{ass:objective} for the corresponding index.
\end{proof}

\subsection{Input, grid, and approximation conditions}\label{subsec:appendix_approximation_primitives}

This subsection supplies sufficient conditions for the sampling, grid, finite-support, and omitted-row assumptions in the main text. The sampling conditions below are one convenient i.i.d.\ route; design-based releases can instead verify Assumption~\ref{ass:input_process} directly. Suppose the sampling units \(W_i\), \(i=1,\ldots,n\), are i.i.d.\ with law \(P_0\). Recompute \(\eta_m^*\) from an Efron resample of \(m\) sampling units drawn with replacement from \(W_1,\ldots,W_n\), and define \(\eta_n^*\) analogously when the ordinary bootstrap is used. Let the nonquantile coordinates of \((c,\gamma,\pi)\) be continuously differentiable functions of a finite vector \(P_0\psi\), with \(E_{P_0}\|\psi(W)\|^2<\infty\), and use the corresponding plug-in estimators. Let each cutoff in \(\tau\) be fixed or a population quantile of a scalar component of \(W\), with density continuous and bounded above and away from zero in a neighborhood of the target. If \(\pi_0>0\), estimate \(q_0(t)=P_0(D=1,Y\le t)/P_0(D=1)\) by its empirical analogue.

For a growing family of released coordinates, suppose the associated influence functions and indicator functions are contained in a measurable \(P_0\)-Donsker class with a square-integrable envelope, and that the coordinate maps have a uniform first-order expansion on that class. If the growing family includes quantile cutoffs, impose the density bounds and the Bahadur remainder uniformly over those cutoffs. In Scenario~2, containment lower-bound shares are averages of \(\mathbf 1\{D=0,\ L\ge a,\ U\le b\}\), while overlap upper-bound shares are averages of \(\mathbf 1\{D=0,\ L\le b,\ U\ge a\}\). The two indicator families are bounded VC classes on \(\mathbb R^2\), so their union is \(P_0\)-Donsker; any finite or increasing collection indexed by \(J_n\) is therefore covered by the growing-coordinate condition above.

\begin{lemma}\label{lem:input_process_primitives}
Under the i.i.d., finite-second-moment, quantile-density, and Donsker conditions in the preceding two paragraphs, Assumption~\ref{ass:input_process} holds.
\end{lemma}

\begin{proof}[Proof of Lemma~\ref{lem:input_process_primitives}]
Let \(\mathcal F\) be the union of the finite moment functions, the indicator functions used for released shares and cutoffs, the exact-observation half-line class, and any growing coordinate class in the statement. The maintained Donsker and envelope conditions imply \(\sqrt n(P_n-P_0)\Rightarrow\mathbb G_{P_0}\) in \(\ell^\infty(\mathcal F)\), and the Efron empirical process satisfies \(\sqrt m(P_m^*-P_n)\Rightarrow\mathbb G_{P_0}\) conditionally in probability for every \(m\to\infty\) with \(m/n\to0\), and also for the ordinary choice \(m=n\) when that bootstrap is used. The finite vector \(\psi\) is contained in \(L_2(P_0)\), so the multivariate central limit theorem for \(M_n=P_n\psi\) is the finite-dimensional marginal of the same empirical-process limit.

The nonquantile coordinates of \((c,\gamma,\pi)\) are continuously differentiable functions of the finite vector \(P_0\psi\) and of any indexed means covered by the uniform first-order expansion. The ordinary and conditional delta methods therefore give their sample, \(m\)-out-of-\(n\), and ordinary-bootstrap linear representations with the same derivative evaluated at \(P_0\). For Scenario~2, the containment lower-bound indicators \(\mathbf 1\{D=0,L\ge a,U\le b\}\) and overlap upper-bound indicators \(\mathbf 1\{D=0,L\le b,U\ge a\}\) form bounded VC-subgraph classes in \((D,L,U)\). The conditional overlap shares used when exact observations are separated are obtained by dividing the corresponding unconditional averages by \(P_n(D=0)=1-\hat\pi\), so they are covered by the same delta-method argument when \(\pi_0<1\). Their union with the other maintained classes is Donsker, so finite or increasing retained subcollections are covered by the same empirical-process limit.

Consider a cutoff \(Q_\alpha=F_R^{-1}(\alpha)\). The density condition, with the density bounded away from zero and continuous near \(Q_\alpha\), gives the Bahadur expansion \(\hat Q_\alpha-Q_\alpha=-[\hat F_R(Q_\alpha)-F_R(Q_\alpha)]/f_R(Q_\alpha)+o_p(n^{-1/2})\). The conditional version follows from the bootstrap empirical process and the same local density bound. If a growing family contains cutoffs, the assumed uniform Bahadur remainder gives this expansion uniformly over that family, so the cutoff coordinates can be stacked with the other coordinates without changing the limit space.

When \(\pi_0>0\), write \(p_0=P_0(D=1)\) and \(N_t(P)=P(D\mathbf 1(Y\le t))\). The map \(P\mapsto N_t(P)/P(D=1)\) is Hadamard differentiable at \(P_0\), uniformly in \(t\) over the half-line class, because \(p_0>0\). Its derivative sends an empirical-process direction \(h\) to \(p_0^{-1}\bigl(h(D\mathbf 1(Y\le t))-q_0(t)h(D)\bigr)\). Hence \(\hat q-q_0\), and its bootstrap analogues, converge in \(\ell^\infty(\mathcal F_0)\). If \(\pi_0=0\), the convention in Section~\ref{sec:asymptotics} fixes the \(q\)-coordinate, and its tangent component is zero.

All components are measurable functionals of the same empirical process, and the displays above are joint linearizations in the product norm defining \(\mathbb H\). The resulting limit is tight, mean zero, and Gaussian. By construction of the tangent set, its sample paths lie in \(\mathbb D_0\) almost surely. The conditional versions are obtained by replacing \(\sqrt n(P_n-P_0)\) with \(\sqrt m(P_m^*-P_n)\) or \(\sqrt n(P_n^*-P_n)\) in the same linear maps. These three joint weak convergence statements are exactly Assumption~\ref{ass:input_process}.
\end{proof}

Empirical-mass programs require compatibility of the masses pinned down by step equalities with the observation grid. The individual right-hand sides of general step restrictions need not themselves be integer multiples of \(1/N\).

\begin{lemma}\label{lem:grid_compatibility}
Under \eqref{eq:boundary_consistency}, if the included step equalities are evaluated at an empirical measure based on \(N\) observations in the integer normalization used in \eqref{eq:boundary_consistency}, then every cumulative mass at a relevant cutoff and every block mass determined by those equalities belongs to \(N^{-1}\mathbb Z\).
\end{lemma}

\begin{proof}[Proof of Lemma~\ref{lem:grid_compatibility}]
Let \(\hat m\) be the vector of empirical cell masses under any compatible assignment of boundary observations to adjacent cells. In the integer normalization, each retained step-equality value \(v_{u,n}^\top\hat m\) is an empirical average of an integer-valued cell indicator combination, and therefore belongs to \(N^{-1}\mathbb Z\). For a relevant boundary \(d\), \eqref{eq:boundary_consistency} gives \(e_{1:d}^\top\hat m=\sum_{u\in\mathcal L_{\mathrm{step,eq},n}} b_{d,u,n}\,v_{u,n}^\top\hat m\). The coefficients \(b_{d,u,n}\) are integers, so \(e_{1:d}^\top\hat m\in N^{-1}\mathbb Z\). This argument uses the equality-implied cumulative mass and is therefore independent of the side from which a boundary atom is represented. If a block is bounded by two relevant cutoffs \(d_1<d_2\), its mass is \((e_{1:d_2}-e_{1:d_1})^\top\hat m\), the difference of two grid-valued cumulative masses. Endpoint blocks use zero or one as the missing cumulative mass, both of which are in \(N^{-1}\mathbb Z\). Hence every block mass determined by the included step equalities is grid-valued.
\end{proof}

More generally, the discretization lemmas below require only that the cumulative and block masses implied by the step equalities be grid-valued. If released inputs do not have this property, Assumption~\ref{ass:implementation} allows an asymptotically negligible projection of those implied masses, translated back into the equality right-hand sides.

The bounded-support and omitted-inequality conditions have different roles. Assumption~\ref{ass:finite_complexity} controls the support dimension of the empirical problem; the solution-form results above verify it with zero approximation error in the principal designs. Assumption~\ref{ass:tail_existence} controls the rows not yet included in \(g(J_n)\).

The following proposition records when the bounded-support condition follows directly from the solution-form results in the main text.

\begin{prop}\label{prop:finite_complexity_primitives}
In Scenario~1A with \(D\) brackets and the objective classes covered by Proposition~\ref{prop:base_cornersol}, Assumption~\ref{ass:finite_complexity} holds with \(r_n=0\) before numerical error and with \(k_n\le2D\), up to the fixed zero-mass coordinates used for equality adjustment. In Scenario~1B with a fixed number \(q_1+q_2\) of auxiliary rows and a linear-fractional objective covered by Proposition~\ref{prop:Sc1BsolLF}, it holds with \(k_n\le2D+q_1+q_2\), again up to the fixed spare coordinates. In Scenario~2, when the set of interval endpoints \(\mathcal B\) has bounded cardinality and the hypotheses of Theorems~\ref{th:Sc3}--\ref{th:Sc3_LF} hold, it holds with \(k_n\le |\mathcal B|+1\), up to the fixed spare coordinates. If an inequality family grows outside these cases, Assumption~\ref{ass:finite_complexity} remains a separate bounded-effective-support condition.
\end{prop}

\begin{proof}[Proof of Proposition~\ref{prop:finite_complexity_primitives}]
Fix \(N\ge k_n\) and suppose \(\mathcal C_{J_n}^{(N,N)}(c,\theta)\) is nonempty. In Scenario~1A, Proposition~\ref{prop:base_cornersol} applies to the finite empirical endpoint problem after the observations are grouped by bracket cells. It gives an optimizer whose distinct positive support values lie in the finite set described there, so the number of distinct positive atoms is bounded by the displayed constant. In Scenario~1B, Proposition~\ref{prop:Sc1BsolLF} gives the bound \(2D+q_1+q_2\) on the number of distinct positive values after the fixed auxiliary rows are imposed. In Scenario~2, Theorems~\ref{th:Sc3}--\ref{th:Sc3_LF} give an optimizer supported on the set of interval endpoints \(\mathcal B\), with at most one additional exceptional value when the theorem allows it. The assumed uniform bound on \(|\mathcal B|\) therefore gives the stated support bound.

It remains to check that these solution-form reductions are reductions inside the empirical-mass program. In each cited result, the transformation preserves the released bracket masses, auxiliary row values, or interval-overlap restrictions that define feasibility. When several empirical observations or cells are collapsed to the same support value, the new atom mass is the sum of their original empirical masses. Sums of \(1/N\)-grid masses remain on the \(1/N\) grid. Hence an endpoint value attainable in \(\mathcal C_{J_n}^{(N,N)}(c,\theta)\) is also attainable in \(\mathcal C_{J_n}^{(k_n,N)}(c,\theta)\), after increasing \(k_n\) by the fixed number of spare zero-mass coordinates used for equality adjustment. Those zero-mass coordinates do not alter the represented measure, feasibility, or objective. The reverse inclusion is automatic because \(\mathcal C_{J_n}^{(k_n,N)}\subseteq\mathcal C_{J_n}^{(N,N)}\). Thus the endpoint vectors agree, so \(r_n=0\) before numerical and input-projection errors. If a growing inequality family is outside the cited finite-support characterizations, the proposition does not supply a bounded effective support; that remaining requirement is precisely Assumption~\ref{ass:finite_complexity}.
\end{proof}

The next lemma gives sufficient conditions for Assumption~\ref{ass:tail_existence}, first for uniformly Lipschitz indexed rows and then for the moving interval rows that arise in Scenario~2. For moving interval rows, the Lipschitz condition in part~\textup{(i)} is stronger than needed because the optimization measure can place mass near a moving cutoff. The monotonicity of interval indicators gives an alternative sufficient condition for the overlap inequalities used in Scenario~2; it requires approximation of the released lower and upper interval probabilities, not Lipschitz continuity of \(\mu([a,b])\) uniformly over all feasible \(\mu\).

\begin{lemma}\label{lem:tail_sufficient_primitives}
Each of the following conditions is sufficient for Assumption~\ref{ass:tail_existence}.
\begin{enumerate}
\item[(i)] Let \((\mathcal T,d_{\mathcal T})\) be compact, let \(\mathcal T_0\subset\mathcal T\) be countable and dense, and index \(g(\infty)\) by \(\mathcal T_0\). Suppose that the Lipschitz bound \(|G_t(\mu;\theta)-G_s(\mu;\theta)|+|c_t^2-c_s^2|\le L_{\mathcal T}d_{\mathcal T}(t,s)\) holds uniformly over \(\eta\in\mathcal N_\eta\), \(\mu\in\mathcal M\), and \(s,t\in\mathcal T_0\). If every \(t\in\mathcal T_0\) is within \(\Delta_{J_n}\) of an index retained in \(g(J_n)\), then Assumption~\ref{ass:tail_existence} holds with \(\kappa_{J_n}=L_{\mathcal T}\Delta_{J_n}\). In particular, a mesh \(\Delta_{J_n}=o(n^{-1/2})\) is sufficient. If \(g(\infty)\) is finite and eventually included, the same conclusion holds with \(\kappa_{J_n}=0\).
\item[(ii)] Let upper rows have the form \(\mu([a,b])-U(a,b)\le0\) and lower rows have the form \(-\mu([a,b])+L(a,b)\le0\), for \((a,b)\) in a compact interval-index set. Suppose the retained grid has mesh \(\Delta_{J_n}\). For every omitted \([a,b]\), let \([a^-,b^+]\) be a retained outer interval with \(a^-\le a\le b\le b^+\) and endpoint distance at most \(\Delta_{J_n}\), and, when nonempty, let \([a^+,b^-]\) be a retained inner interval with \(a\le a^+\le b^-\le b\) and the same endpoint distance. If, uniformly on the local nuisance neighborhood, \(0\le U(a^-,b^+)-U(a,b)\le C\Delta_{J_n}\) and \(0\le L(a,b)-L(a^+,b^-)\le C\Delta_{J_n}\), with the second lower bound interpreted as zero when the inner interval is empty, then Assumption~\ref{ass:tail_existence} holds for these interval rows with \(\kappa_{J_n}=C\Delta_{J_n}\). A finite set of possible intervals gives \(\kappa_{J_n}=0\) once every interval is retained.
\end{enumerate}
\end{lemma}

\begin{proof}
Part~\textup{(i)}. Fix a local input \(\eta=(c,\theta,\pi,q)\), a retained-feasible measure \(\mu\in\mathcal C_{J_n}(c,\theta)\), and an omitted index \(t\in\mathcal T_0\). By the mesh condition, there is a retained index \(s\in g(J_n)\) with \(d_{\mathcal T}(t,s)\le\Delta_{J_n}\). Feasibility of the retained row gives \(G_s(\mu;\theta)\le c_s^2\). Therefore
\[
\begin{aligned}
G_t(\mu;\theta)-c_t^2
&=G_s(\mu;\theta)-c_s^2+\bigl[G_t(\mu;\theta)-G_s(\mu;\theta)\bigr]+\bigl[c_s^2-c_t^2\bigr]\\
&\le |G_t(\mu;\theta)-G_s(\mu;\theta)|+|c_s^2-c_t^2|\\
&\le L_{\mathcal T}\Delta_{J_n}.
\end{aligned}
\]
The Lipschitz hypothesis already bounds the sum of the two absolute values, so no extra factor is introduced. Taking the positive part and then the suprema over omitted rows, retained-feasible measures, and local inputs gives Assumption~\ref{ass:tail_existence} with \(\kappa_{J_n}=L_{\mathcal T}\Delta_{J_n}\). If the full family is finite and eventually included, the innermost supremum in Assumption~\ref{ass:tail_existence} is over the empty set and is defined as zero.

Part~\textup{(ii)}. Fix a local input and a measure \(\mu\) feasible for the retained interval rows. For an omitted upper row \([a,b]\), choose the retained outer interval \([a^-,b^+]\). Monotonicity of closed-interval indicators gives \(\mu([a,b])\le\mu([a^-,b^+])\), and retained feasibility gives \(\mu([a^-,b^+])\le U(a^-,b^+)\). Hence \(\mu([a,b])-U(a,b)\le U(a^-,b^+)-U(a,b)\le C\Delta_{J_n}\). This proves the required bound for every omitted overlap upper-bound row.

For an omitted containment lower-bound row with a nonempty retained inner interval, \([a^+,b^-]\subseteq[a,b]\). Thus \(\mu([a,b])\ge\mu([a^+,b^-])\), and retained feasibility of the inner lower row gives \(\mu([a^+,b^-])\ge L(a^+,b^-)\). Therefore \(L(a,b)-\mu([a,b])\le L(a,b)-L(a^+,b^-)\le C\Delta_{J_n}\). If there is no nonempty retained inner interval, the convention in the statement is that the inner lower probability is zero. Since \(\mu([a,b])\ge0\), the displayed hypothesis gives \(L(a,b)-\mu([a,b])\le L(a,b)\le C\Delta_{J_n}\). The bounds are uniform over the local nuisance neighborhood by assumption. Taking positive parts and then the suprema over omitted intervals, retained-feasible measures, and local inputs verifies Assumption~\ref{ass:tail_existence}. If the set of possible intervals is finite, every interval row is retained for all sufficiently large \(n\), and the omitted-row supremum is zero.
\end{proof}

The countable-constraint endpoint is defined by an infimum or supremum and need not have a global optimizer. For a fixed support size and cell assignment, the local analysis instead uses the closed coordinate program obtained by evaluating each row with its retained one-sided affine formula. An adjacent boundary branch may describe a one-sided feasible limit rather than a second probability distribution.

\begin{lemma}\label{lem:finite_program_attainment}
Under Assumptions~\ref{ass:support},~\ref{ass:restriction_geometry}, and~\ref{ass:objective}, every nonempty closed cell-assigned finite-\(J\), finite-support coordinate program attains its lower and upper values. The same conclusion holds after imposing an empirical mass grid.
\end{lemma}

\begin{proof}[Proof of Lemma~\ref{lem:finite_program_attainment}]
Fix \(J\), a support size \(a\), and one admitted assignment of support coordinate \(j\) to partition cell \(d_j\). Let \(I_j=[t_{d_j-1,n}(\theta),t_{d_j,n}(\theta)]\) denote the closed coordinate copy of that half-open cell. The coordinate space for this branch is a closed subset of the compact product \(\Delta^{a-1}\times I_1\times\cdots\times I_a\). Step restrictions are evaluated using the one-sided cell value fixed by the branch, affine restrictions are continuous in masses and locations on the closed cell copy, and support-position bounds are closed inequalities. Hence the feasible set of the branch is closed and compact. If the branch also imposes the empirical mass grid, the mass coordinates are restricted to the closed finite set \(\Delta^{a-1}\cap b^{-1}\mathbb Z_+^a\).

Every admitted branch represents either genuine feasible measures or one-sided \(\Wone\)-limits of such measures under a compatible boundary convention. The local objective domain \(\mathcal O\) was defined to contain the closure of all represented measures generated by the finite and full feasible sets on the local input neighborhood. Assumption~\ref{ass:objective} therefore makes the objective continuous on the compact branch. Weierstrass' theorem gives the lower and upper extrema on that branch. A finite union over admitted support-cell assignments is again compact, so the same conclusion holds for the finite union of branches used by a closed cell-assigned finite program. If all retained branches correspond to actual measures, this proves attainment for the measure program itself. If a branch is retained only as a one-sided closure, the lemma asserts attainment for the closed coordinate program; the separate attainment of the reduced endpoint values used in the local sensitivity proof is imposed in Assumption~\ref{ass:endpoint_rank}. The full countable-constraint problem is not covered by this compactness argument and need not attain its value.
\end{proof}

\subsection{Finite rank and slack conditions}\label{subsec:appendix_regularity}
This subsection makes the endpoint diagnostics in Assumptions~\ref{ass:feasibility_rank} and~\ref{ass:endpoint_rank} finite-dimensional. The matrices below are the objects checked after row reduction and support cleaning: equality-repair blocks for Assumption~\ref{ass:feasibility_rank} and active-gradient matrices for Assumption~\ref{ass:endpoint_rank}. For endpoint sign \(\circ\in\{\inf,\sup\}\), let \(\mathcal S_n^\circ(\eta)\) be the set of optimizer measures for \(V_{J_n}^{k_n,\circ}(\eta)\), and write \(\phi_n^\circ(\eta):=V_{J_n}^{k_n,\circ}(\eta)\).

After row reduction, write the step equalities as \(\mathcal L_{\mathrm{step,eq},n}=\{r_1,\ldots,r_s\}\), and select mass indices \(j_0,\ldots,j_s\). Define the square matrix \(B_{\mathrm{step}}(z,\theta):=\bigl(1,f_{r_1,n}(z_{j_t};\theta),\ldots,f_{r_s,n}(z_{j_t};\theta)\bigr)_{t=0}^s\). Write the remaining affine equalities as \(\mathcal L_{\mathrm{aff,eq},n}=\{\ell_1,\ldots,\ell_m\}\). For selected interior-location indices \(i_1,\ldots,i_m\), define
\[
A_{\mathrm{aff}}(z,\theta)
:=\bigl(\partial_y f_{\ell_r,n}(z_{i_t};\theta)\bigr)_{t,r},
\qquad
S_{\mathrm{aff}}(x,\theta)
:=\operatorname{diag}(p_{i_1},\ldots,p_{i_m})A_{\mathrm{aff}}(z,\theta),
\]
and let \(C_{\mathrm{aff}}(z,\theta):=(f_{\ell_r,n}(z_{j_t};\theta))_{r,t}\). The step block differentiates the simplex row and the row-reduced step equalities with respect to the selected masses. The affine-location block differentiates the affine equalities with respect to selected within-cell support positions, so selected locations must be interior points of their retained cells. With selected mass coordinates ordered before selected locations, the combined equality Jacobian is
\begin{equation}\label{eq:combined_equality_jacobian}
J_{\mathrm{eq}}(x,\theta)
:=
\begin{pmatrix}
B_{\mathrm{step}}(z,\theta)^\top&0\\
C_{\mathrm{aff}}(z,\theta)&S_{\mathrm{aff}}(x,\theta)^\top
\end{pmatrix}.
\end{equation}
Empty affine or step blocks are omitted. The lower-left block records that affine equalities also change when selected masses change; the block-triangular form is why the step and within-cell slope calculations can be checked separately.

The class of augmented representations is defined without imposing rank or slack. Fix a uniform integer \(\bar a\), included in the choice of \(k_n\). For \(N\ge k_n\), let \(\mathfrak X_n^{\mathrm{aug}}(N)\) contain every compatible, cell-assigned \(N\)-coordinate representation \(x=(p,z)\in\Delta^{N-1}\times\YY^N\) of a measure feasible for the \(J_n\)-problem. Zero masses, repeated locations, and compatible adjacent-cell assignments are allowed. At most \(\bar a\) zero-mass coordinates may be placed at cell-interior locations for equality adjustment; any remaining padding coordinates repeat existing locations. Assumption~\ref{ass:feasibility_rank} requires each feasible measure with at most \(N\) support points to have at least one representation in this class, selected step-mass and affine-location coordinates with uniformly invertible blocks, and a same-support strictly feasible reweighting that places uniformly positive mass on every selected affine-location coordinate.

\paragraph{Local parameterizations.}
Fix \(n\), an endpoint \(\circ\), and an optimizer \(\mu^*=\sum_{j=1}^k p_j^*\delta_{z_j^*}\) at \(\eta_0\), after repeated support points and zero masses have been removed. A compatible local parameterization assigns each \(z_j^*\) to a partition cell whose closure contains it, evaluates step restrictions by the corresponding one-sided cell value, and retains the assignments that satisfy the row-reduced restrictions at \(x^*:=(p^*,z^*)\). The local variables are \(x=(p_1,\ldots,p_k,z_1,\ldots,z_k)\), with the simplex restriction, the cell-specific support bounds, the equalities, and the included inequalities evaluated using the affine formula on the retained cells. There are finitely many compatible assignments for each optimizer.

The familiar designs reduce the combined equality-rank calculation to elementary finite matrices.

The following proposition collects two finite-dimensional sufficient checks for Assumption~\ref{ass:feasibility_rank}. The first verifies the same-support repair clause from a finite repair support. The second records the elementary equality-Jacobian calculations used in the familiar designs.

\begin{prop}\label{prop:feasibility_rank_diagnostics}
The following two statements are sufficient for the indicated clauses of Assumption~\ref{ass:feasibility_rank}.
\begin{enumerate}
\item[(i)] Suppose that the effective support bound has been enlarged by a fixed number of spare coordinates, and that, uniformly over large \(n\) and local inputs, there is a finite repair support \(R_n\), with bounded cardinality and all points in cell interiors. Suppose a probability vector on \(R_n\) satisfies the equality rows and every inequality in the full family with slack at least \(\underline s>0\). Suppose also that, whenever \(R_n\) is appended as the spare zero-mass part of an augmented representation, selected equality coordinates can be chosen so that the Jacobian in \eqref{eq:combined_equality_jacobian} has a uniformly bounded inverse, the selected affine-adjustment coordinates receive mass bounded below under the repair probability vector, and the selected repair locations are a uniformly positive distance from the boundaries of their assigned cells. Then, after this fixed enlargement of \(k_n\), the augmented-representation and same-support repair clauses of Assumption~\ref{ass:feasibility_rank} hold.
\item[(ii)] Under Assumption~\ref{ass:restriction_geometry}, the following calculations are sufficient for the equality-Jacobian clause of Assumption~\ref{ass:feasibility_rank}. In Scenario~1A, the nonoverlapping bracket-share rows give a uniformly invertible \(B_{\mathrm{step}}\) after the simplex redundancy and equality-implied zero cells are removed. With \(m\) additional affine equalities, it is enough to select \(m\) interior support points whose masses under the strictly feasible reweighting are bounded away from zero and for which \(A_{\mathrm{aff}}\) has a uniformly bounded inverse. In the baseline Scenario~2 formulation, the combined equality Jacobian contains only the simplex block.
\end{enumerate}
\end{prop}

\begin{proof}
Part~\textup{(i)}. Fix a large \(n\), a local input, and a feasible measure represented by \(x^0=(p^0,z^0)\). Append the points in \(R_n\) as the spare coordinates of the augmented representation, assigning them zero mass. Let \(p^R\) be the probability vector on this augmented support that places the repair weights on \(R_n\) and zero mass on the original support coordinates. Both \(p^0\) and \(p^R\) satisfy the equality rows with the same right-hand side. For every included inequality row \(u\), feasibility gives \(G_u(p^0,z^0;\theta)-c_u^2\le0\), while the repair vector gives \(G_u(p^R,z^R;\theta)-c_u^2\le-\underline s\). Hence, for any fixed \(\alpha\in(0,1)\), \(p^{\mathrm{sl}}:=(1-\alpha)p^0+\alpha p^R\) satisfies all equalities and every included inequality with slack at least \(\alpha\underline s\). The selected affine-adjustment masses are at least \(\alpha\underline p_R\), where \(\underline p_R>0\) is the uniform lower bound supplied by the hypothesis for the repair vector. The support locations selected for affine adjustment are repair points with a uniform boundary margin in their assigned cells.

It remains to verify the local equality-repair part. Write the simplex and row-reduced equality residuals in the selected coordinates, with the current right-hand side subtracted. The derivative of this residual with respect to the selected masses and selected interior locations is exactly \(J_{\mathrm{eq}}\) in \eqref{eq:combined_equality_jacobian}. Given a small equality-right-hand-side perturbation \(\Delta c\), solve the linear system \(J_{\mathrm{eq}}d=\Delta c\). The inverse bound gives \(\|d\|\le C\|\Delta c\|\). Step restrictions are constant within retained cells, and affine restrictions are affine within retained cells; therefore the correction changes the equality residuals by exactly \(\Delta c\), not just to first order. For \(\|\Delta c\|\) small enough, the selected masses remain nonnegative because they start with the lower bound \(\alpha\underline p_R\), and the selected locations remain in their cells because they start in cell interiors with a uniform boundary margin. The row sup-norm, slope, and total-variation bounds in Assumption~\ref{ass:restriction_geometry} imply that every included inequality residual changes by at most \(C'\|\Delta c\|\). Reducing the perturbation radius so that \(C'\|\Delta c\|\le\alpha\underline s/2\) preserves strict feasibility. These constants are uniform because \(|R_n|\), the equality dimension, the inverse bound, and the row smoothness constants are uniform. This proves the required augmented representation and same-support repair clauses.

Part~\textup{(ii)}. Consider first the nonoverlapping bracket-share equalities. Remove equality-implied zero cells, and delete one share row because the shares sum to one together with the simplex row. Select one mass coordinate in each remaining bracket cell. With these coordinates ordered by cell, the step map sends the selected masses \(m=(m_1,\ldots,m_D)\) to their total mass and the first \(D-1\) bracket masses, up to a permutation of rows and columns. Its inverse maps \((t,s_1,\ldots,s_{D-1})\) to \((s_1,\ldots,s_{D-1},t-\sum_{d<D}s_d)\). All inverse entries are in \(\{-1,0,1\}\), and the number of retained equality rows is uniformly bounded by Assumption~\ref{ass:restriction_geometry}; hence \(B_{\mathrm{step}}^{-1}\) is uniformly bounded in any fixed finite-dimensional norm.

Now add \(m\) affine equality rows and select \(m\) interior support locations. The combined Jacobian has the block-triangular form in \eqref{eq:combined_equality_jacobian}. The lower-right block is \(S_{\mathrm{aff}}^\top=A_{\mathrm{aff}}^\top\operatorname{diag}(p_{i_1},\ldots,p_{i_m})\). If the selected masses are at least \(\underline p>0\) and \(A_{\mathrm{aff}}^{-1}\) is uniformly bounded, then \(S_{\mathrm{aff}}^{-1}=A_{\mathrm{aff}}^{-1}\operatorname{diag}(p_{i_1}^{-1},\ldots,p_{i_m}^{-1})\) is uniformly bounded. Since \(C_{\mathrm{aff}}\) is uniformly bounded by the row sup-norm clause of Assumption~\ref{ass:restriction_geometry}, the inverse
\[
J_{\mathrm{eq}}^{-1}
=
\begin{pmatrix}
B_{\mathrm{step}}^{-\top}&0\\
-S_{\mathrm{aff}}^{-\top}C_{\mathrm{aff}}B_{\mathrm{step}}^{-\top}&S_{\mathrm{aff}}^{-\top}
\end{pmatrix}
\]
is uniformly bounded. Empty blocks are omitted from the display, so the argument also covers the case with no affine equality rows. In the baseline Scenario~2 formulation, the only equality is total mass. The selected step matrix is then the \(1\times1\) matrix \((1)\), and the equality-rank condition is automatic. Auxiliary equalities in Scenario~2, if imposed, are not covered by this vacuous calculation and must satisfy the same finite block check.
\end{proof}

The next lemma records the uniform constants supplied by the feasibility condition.

\begin{lemma}\label{lem:equality_adjustment}
Under Assumptions~\ref{ass:support},~\ref{ass:restriction_geometry},~\ref{ass:finite_complexity}, and~\ref{ass:feasibility_rank}, there are \(r_{\mathrm{reg}}>0\), positive constants \(\underline s,\underline p,\underline\delta\), and finite constants \(\bar C_{\mathrm{eq}},\bar C_{\mathrm{slope}}\) such that, uniformly over all sufficiently large \(n\) and \(\|\eta-\eta_0\|_\infty\le r_{\mathrm{reg}}\):
\begin{enumerate}
\item[(i)] \(\mathcal C_\infty(c,\theta)\) contains a measure whose slack in every row-reduced inequality is at least \(\underline s\).
\item[(ii)] Every measure in \(\mathcal C_{J_n}^{(N)}(c,\theta)\), \(N\ge k_n\), has a representation \(x=(p,z)\in\mathfrak X_n^{\mathrm{aug}}(N)\) and a same-support vector \(p^{\mathrm{sl}}\) satisfying the equalities and every included inequality with slack at least \(\underline s\). At \(x^{\mathrm{sl}}=(p^{\mathrm{sl}},z)\), the selected equality Jacobian satisfies \(\|J_{\mathrm{eq}}^{-1}\|\le\bar C_{\mathrm{eq}}\), every selected affine-adjustment mass is at least \(\underline p\), and every selected location is at least \(\underline\delta\) from its cell boundary.
\item[(iii)] For the selected affine-location block, \(\|A_{\mathrm{aff}}^{-1}\|_\infty+\|A_{\mathrm{aff}}^{-\top}\|_\infty\le\bar C_{\mathrm{slope}}\), whenever that block is nonempty.
\end{enumerate}
\end{lemma}

\begin{proof}[Proof of Lemma~\ref{lem:equality_adjustment}]
Part~\textup{(i)} is the full-system Slater clause of Assumption~\ref{ass:feasibility_rank}. The measure \(\mu_n^{\mathrm{sl},\infty}(\eta)\) has a cell-moment vector in \(\operatorname{relint}\mathcal P_n^H(c,\theta)\), satisfies the row-reduced equalities, and has at least \(\underline s\) slack in every normalized inequality row. Reducing the neighborhood of \(\eta_0\), if necessary, gives the same constant for every local input in the stated ball.

Part~\textup{(ii)} is the augmented-representation and same-support strict-feasibility clause of Assumption~\ref{ass:feasibility_rank}, again with the neighborhood reduced once so that a single collection of constants applies. The original representation may contain zero-mass padding coordinates. The lower mass bound is required only for selected affine-adjustment coordinates after the same-support slack vector has been chosen; nonselected zero-mass padding coordinates are not used in the equality correction.

For part~\textup{(iii)}, evaluate \eqref{eq:combined_equality_jacobian} at \(x^{\mathrm{sl}}\). Its lower-right block is \(S_{\mathrm{aff}}^\top\). Write \(D_p^{\mathrm{sl}}:=\operatorname{diag}(p_{i_1}^{\mathrm{sl}},\ldots,p_{i_m}^{\mathrm{sl}})\), so \(S_{\mathrm{aff}}=D_p^{\mathrm{sl}}A_{\mathrm{aff}}\). The bound on \(J_{\mathrm{eq}}^{-1}\) implies a bound on \(S_{\mathrm{aff}}^{-1}\) and \(S_{\mathrm{aff}}^{-\top}\). Since
\[
A_{\mathrm{aff}}^{-1}=S_{\mathrm{aff}}^{-1}\operatorname{diag}(p_{i_1}^{\mathrm{sl}},\ldots,p_{i_m}^{\mathrm{sl}}),
\qquad
A_{\mathrm{aff}}^{-\top}=S_{\mathrm{aff}}^{-\top}\operatorname{diag}(p_{i_1}^{\mathrm{sl}},\ldots,p_{i_m}^{\mathrm{sl}}),
\]
and every probability mass is at most one, the Euclidean operator norms of \(A_{\mathrm{aff}}^{-1}\) and \(A_{\mathrm{aff}}^{-\top}\) are uniformly bounded. The affine equality dimension is uniformly bounded, so equivalence of norms in these finite dimensions gives the stated max-norm bound.
\end{proof}

The second lemma records the rank and slack properties at endpoint optimizers. For a restriction row \(u\) and local input \(\eta=(c,\theta,\pi,q)\), write
\[
\nabla_x R_{u,n}(x^*;\eta)
=
\left(
 f_{u,n}(z_1^*;\theta),\ldots,f_{u,n}(z_k^*;\theta),
 p_1^*\partial_y f_{u,n}(z_1^*;\theta),\ldots,
 p_k^*\partial_y f_{u,n}(z_k^*;\theta)
\right),
\]
where \(\partial_y f_{u,n}\) is taken within the retained cell and the right-hand side is moved to the left in the residual \(R_{u,n}\). Let \(A_G(x^*;\eta)\) be the included inequalities binding at \(x^*\). The matrix \(\mathcal A_n(x^*;\eta)\) stacks the simplex row, a basis for the equality gradients, the gradients in \(A_G(x^*;\eta)\), and the active support-position bounds, after duplicate and uniformly locally implied rows have been removed. In baseline calculations, write \(\mathcal A_n(x^*):=\mathcal A_n(x^*;\eta_0)\).

\begin{lemma}\label{lem:endpoint_active_rank}
Under Assumptions~\ref{ass:restriction_geometry},~\ref{ass:objective},~\ref{ass:finite_complexity}, and~\ref{ass:endpoint_rank}, there are positive constants \(\underline p_\star,\underline\delta_{\mathrm{sep}},\underline\sigma_A,\underline s_{\mathrm{in}},\underline\delta_{\mathrm{bd}}\), a finite constant \(\bar C_\Lambda\), and a neighborhood of \(\eta_0\) such that the following statements hold uniformly over all sufficiently large \(n\), both endpoints, local inputs in that neighborhood, endpoint optimizers, and compatible reduced representations. Every optimizer mass is at least \(\underline p_\star\), distinct support points are separated by at least \(\underline\delta_{\mathrm{sep}}\), every inactive nonimplied inequality has slack at least \(\underline s_{\mathrm{in}}\), every nonboundary support point is at least \(\underline\delta_{\mathrm{bd}}\) from the adjacent cell boundaries, and \(\sigma_{\min}(\mathcal A_n(x^*;\eta)\mathcal A_n(x^*;\eta)^\top)\ge\underline\sigma_A^2\). At \(\eta_0\), the Lagrange multiplier associated with the retained active rows is unique and has norm at most \(\bar C_\Lambda\).
\end{lemma}

\begin{proof}[Proof of Lemma~\ref{lem:endpoint_active_rank}]
The positive-mass, support-separation, inactive-slack, boundary-distance, and singular-value bounds are the corresponding finite-coordinate clauses of Assumption~\ref{ass:endpoint_rank}. Because the optimizer representations are compact and the bounds in that assumption are uniform over all large \(n\), both endpoint signs, and all compatible local cell assignments, one may reduce the neighborhood of \(\eta_0\) and choose the constants in the statement common to all these objects.

It remains only to record the multiplier bound used later. Fix a baseline optimizer representation \(x^*\) for endpoint \(\circ\) at \(\eta_0\). On its retained cell-assigned branch, the objective and all retained active constraint functions are continuously differentiable in \(x\). The rows of \(\mathcal A_n(x^*)\) are linearly independent by the displayed singular-value bound, so the linear independence constraint qualification holds for the finite-dimensional program that minimizes \(\sigma_\circ\Phi_k(x;\pi_0,q_0)\) subject to the local equality rows, active inequality rows, and active support-position bounds. The Karush--Kuhn--Tucker conditions therefore give a multiplier \(\Lambda\) satisfying \(\nabla_x(\sigma_\circ\Phi_k(x^*;\pi_0,q_0))+\mathcal A_n(x^*)^\top\Lambda=0\), with the nonnegative sign on active inequality components under the residual convention used to form \(\mathcal A_n\). Since \(\mathcal A_n(x^*)\) has full row rank, the multiplier is unique and equals
\[
\Lambda
=-[\mathcal A_n(x^*)\mathcal A_n(x^*)^\top]^{-1}
  \mathcal A_n(x^*)\nabla_x(\sigma_\circ\Phi_k(x^*;\pi_0,q_0)).
\]
Assumption~\ref{ass:objective} bounds the support-coordinate gradient of \(\Phi_k\) uniformly on the retained optimizer neighborhoods, while the singular-value bound gives \(\|[\mathcal A_n(x^*)\mathcal A_n(x^*)^\top]^{-1}\mathcal A_n(x^*)\|\le\underline\sigma_A^{-1}\). These two bounds imply \(\|\Lambda\|\le\bar C_\Lambda\) after enlarging \(\bar C_\Lambda\) to cover the finitely many endpoint signs and compatible row conventions.
\end{proof}

\begin{lemma}\label{lem:endpoint_isolation_sufficient}
Fix an endpoint sign. Suppose that, for all sufficiently large \(n\), the reduced finite endpoint program has finitely many feasible active-set systems after row reduction, every optimal active set satisfies the active-rank, positive-mass, inactive-slack, and boundary-separation clauses of Assumption~\ref{ass:endpoint_rank}, and every feasible nonoptimal active set has objective value at least \(\gamma>0\) worse than the endpoint value after applying the sign convention \(\sigma_\circ\). Then the objective-isolation clause of Assumption~\ref{ass:endpoint_rank} holds for that endpoint. A sufficient finite diagnostic is that the endpoint value remains at least \(\gamma\) worse when the program is re-solved subject to distance at least \(\varepsilon\) from the optimizer set, for each fixed \(\varepsilon>0\).
\end{lemma}

\begin{proof}[Proof of Lemma~\ref{lem:endpoint_isolation_sufficient}]
Fix \(\varepsilon>0\). Work in the finite union of closed cell-assigned coordinate branches used for the reduced endpoint program. Lemma~\ref{lem:finite_program_attainment} gives compactness of each branch, and the finite active-set hypothesis gives a finite union. The subset of feasible reduced coordinate points whose represented measures are at \(\Wone\)-distance at least \(\varepsilon\) from the optimizer set is closed in this compact union, because \(x\mapsto\mu_x\) is continuous in \(\Wone\) on bounded-support coordinate spaces and the optimizer set is compact. Hence the signed objective gap attains its minimum on this exclusion set whenever the set is nonempty.

If that minimum were zero along a subsequence, compactness would give feasible points \(x_{n_r}\) at distance at least \(\varepsilon\) from the optimizer set and with signed objective values converging to the endpoint value. Passing to a convergent subsequence within one active-set system gives a feasible limit that attains the endpoint value. Its active-set system is therefore optimal, so the represented limiting measure belongs to the optimizer set, contradicting the maintained distance at least \(\varepsilon\). Equivalently, under the finite active-set hypothesis, a zero limiting gap outside the optimizer neighborhood would identify a feasible nonoptimal active system with endpoint value, which is ruled out by the assumed \(\gamma\)-separation. Thus the exclusion set has a strictly positive signed objective gap. The diagnostic stated in the lemma computes this compact minimum directly for each fixed \(\varepsilon\), so a positive diagnostic lower bound is sufficient for the objective-isolation clause.
\end{proof}

The next elementary fact converts Wasserstein localization of regular finite-support measures into localization of their support coordinates. For reduced atomic measures \(\mu=\sum_{j=1}^k p_j\delta_{z_j}\) and \(\nu=\sum_{j=1}^\ell q_j\delta_{w_j}\), define
\[
d_{\mathrm{at}}(\mu,\nu)
:=
\begin{cases}
\displaystyle
\min_{\rho\in\mathfrak S_k}\max_{1\le j\le k}
\{|p_j-q_{\rho(j)}|+|z_j-w_{\rho(j)}|\},& k=\ell,\\[2mm]
+\infty,& k\ne\ell,
\end{cases}
\]
where \(\mathfrak S_k\) is the set of permutations of \(\{1,\ldots,k\}\).

\begin{lemma}\label{lem:atomic_matching}
Fix \(\bar k<\infty\), \(\underline p>0\), and \(\underline\delta>0\). Let \(\mu_r\) and \(\nu_r\) be reduced probability measures with at most \(\bar k\) atoms, every atom mass at least \(\underline p\), and every pair of distinct support points within either measure separated by at least \(\underline\delta\). If \(\Wone(\mu_r,\nu_r)\to0\), then \(d_{\mathrm{at}}(\mu_r,\nu_r)\to0\).
\end{lemma}

\begin{proof}[Proof of Lemma~\ref{lem:atomic_matching}]
Write \(\mu_r=\sum_{i=1}^{k_r}p_{ri}\delta_{z_{ri}}\) and \(\nu_r=\sum_{j=1}^{\ell_r}q_{rj}\delta_{w_{rj}}\), and let \(\Gamma_r\) be an optimal coupling, which exists because \(\YY\) is compact. Set \(a:=\underline\delta/3\). For all sufficiently large \(r\), \(\Wone(\mu_r,\nu_r)<\underline p a\). If a support point \(z_{ri}\) had no \(w_{rj}\) within distance \(a\), then the mass \(p_{ri}\ge\underline p\) starting at \(z_{ri}\) would have to move at least distance \(a\), giving transportation cost at least \(\underline p a\), a contradiction. The same argument with the roles of the two measures reversed shows that every \(w_{rj}\) is within distance \(a\) of some \(z_{ri}\).

The separation condition makes this nearby point unique. Indeed, two distinct support points of \(\mu_r\) cannot both lie within distance \(a\) of the same \(w_{rj}\), and one \(z_{ri}\) cannot lie within distance \(a\) of two distinct support points of \(\nu_r\), because \(2a<\underline\delta\). Thus \(k_r=\ell_r\) for all large \(r\), and, after relabeling, \(|z_{rj}-w_{rj}|<a\) for every \(j\). For this matching, any mass transported from \(z_{rj}\) to \(w_{ri}\), \(i\ne j\), travels at least \(2\underline\delta/3\); the same lower bound applies to mass entering \(w_{rj}\) from \(z_{ri}\), \(i\ne j\). Let \(o_{rj}:=\sum_{i\ne j}\Gamma_r\{(z_{rj},w_{ri})\}\) and \(i_{rj}:=\sum_{i\ne j}\Gamma_r\{(z_{ri},w_{rj})\}\). Since these are off-diagonal flows, \(o_{rj}+i_{rj}\le3\Wone(\mu_r,\nu_r)/(2\underline\delta)\). The mass-balance identity gives \(|p_{rj}-q_{rj}|=|o_{rj}-i_{rj}|\le3\Wone(\mu_r,\nu_r)/(2\underline\delta)\). Moreover, the matched flow satisfies \(\Gamma_r\{(z_{rj},w_{rj})\}=p_{rj}-o_{rj}\ge\underline p-3\Wone(\mu_r,\nu_r)/(2\underline\delta)\), which is at least \(\underline p/2\) for all large \(r\). The cost on this matched flow is at least \((\underline p/2)|z_{rj}-w_{rj}|\), so \(|z_{rj}-w_{rj}|\le2\Wone(\mu_r,\nu_r)/\underline p\). The bounds are uniform over \(j\), and therefore \(d_{\mathrm{at}}(\mu_r,\nu_r)\to0\).
\end{proof}

Shrink \(\mathcal N_\eta\), if necessary, so that the attainment and compactness clause of Assumption~\ref{ass:endpoint_rank}, Lemmas~\ref{lem:equality_adjustment} and~\ref{lem:endpoint_active_rank}, and the common objective moduli in Assumption~\ref{ass:objective} hold on one neighborhood. Fix \(0<r_0<\tilde r_0\) so that \(\mathcal N_0:=\{\eta:\|\eta-\eta_0\|_\infty\le r_0\}\) lies in the interior of \(\widetilde{\mathcal N}_0:=\{\eta:\|\eta-\eta_0\|_\infty\le\tilde r_0\}\), both contained in that neighborhood. Recall that \(\bar k:=\sup_n k_n<\infty\). A row is included in the \(n\)-th problem when its index lies in \(h(J_n)\) or \(g(J_n)\).

The next subsection uses these constants to compare the unrestricted finite-\(J_n\) problem, its bounded-support version, its empirical-mass version, and the full population value. The local-sensitivity subsection then applies the envelope argument to the bounded-support program, and the final proofs transfer the first-stage limit to the endpoint estimator.

\subsection{Approximation of the finite and full programs}\label{subsec:appendix_finite_full_approximation}
The equality-repair argument invoked in this subsection was established in the preceding finite-rank and slack subsection.

\paragraph{Finite-support and mass-grid approximation.}
The next results implement the deterministic approximation chain from the unrestricted finite-\(J_n\) problem to its bounded-support and empirical-mass versions.

\begin{lemma}\label{lem:empirical}
Fix \(c,\theta,J\). An empirical measure based on \(N_2\) observations and at most \(N_1\) distinct values belongs to \(\mathcal C_J^{(N_1,N_2)}(c,\theta)\) if and only if it is the empirical measure of an \(N_2\)-vector satisfying the restrictions in \(\mathcal C_J(c,\theta)\).
\end{lemma}

\begin{proof}[Proof of Lemma~\ref{lem:empirical}]
Write \(\mathcal C^{N_1,N_2}:=\mathcal C_J^{(N_1,N_2)}(c,\theta)\). If \(y=(y_1,\ldots,y_{N_2})\in\YY^{N_2}\) takes \(m\le N_1\) distinct values \(z_1,\ldots,z_m\) with multiplicities \(r_1,\ldots,r_m\), then its empirical measure has support \(z_j\) and weights \(r_j/N_2\). Padding with repeated support points and zero masses to length \(N_1\) gives a weight vector in \(\Delta^{N_1-1}\cap N_2^{-1}\mathbb Z_+^{N_1}\). Because every restriction in \(\mathcal C_J(c,\theta)\) is a restriction on integrals with respect to the empirical measure, the vector \(y\) satisfies those restrictions if and only if \(\mu_y\in\mathcal C^{N_1,N_2}\).

Conversely, take \(\mu\in\mathcal C^{N_1,N_2}\). After deleting zero-mass entries, write \(\mu=\sum_{j=1}^m(r_j/N_2)\delta_{z_j}\), where \(m\le N_1\), \(r_j\in\mathbb Z_+\), and \(\sum_{j=1}^m r_j=N_2\). Listing each \(z_j\) exactly \(r_j\) times produces an \(N_2\)-vector whose empirical measure is \(\mu\). Since the restrictions depend only on that empirical measure, the vector satisfies the restrictions in \(\mathcal C_J(c,\theta)\). This proves both directions.
\end{proof}

\begin{lemma}\label{lem:largeN}
Under Assumptions~\ref{ass:support},~\ref{ass:restriction_geometry}, and~\ref{ass:objective}, for every integer \(N\ge s_n+1\),
\[
\sup_{\substack{\eta=(c,\theta,\pi,q)\in\mathcal N_0:\ \mathcal C_{J_n}(c,\theta)\neq\emptyset}}
\Big|
\sup_{\mu\in\mathcal C_{J_n}(c,\theta)}F(\mu;\pi,q)
-
\sup_{\mu\in\mathcal C_{J_n}^{(N)}(c,\theta)}F(\mu;\pi,q)
\Big|
\ \le\ \bar C\frac{s_n}{N},
\]
where \(\bar C:=4L\diam(\YY)<\infty\). An analogous bound holds with the outermost \(\sup\) replaced by \(\inf\).
\end{lemma}

\begin{proof}[Proof of Lemma~\ref{lem:largeN}]
Fix \(\eta=(c,\theta,\pi,q)\in\mathcal N_0\) with \(\mathcal C_{J_n}(c,\theta)\neq\varnothing\), and let \(N\ge s_n+1\). Write the partition as \(I_{1,n}(\theta)=[t_{0,n}(\theta),t_{1,n}(\theta)]\) and \(I_{d,n}(\theta)=(t_{d-1,n}(\theta),t_{d,n}(\theta)]\) for \(d=2,\ldots,s_n\). Set \(\alpha:=\lfloor (N-1)/s_n\rfloor\ge1\). For each cell, divide \((t_{d-1,n}(\theta),t_{d,n}(\theta)]\) into \(\alpha\) right-closed subcells \(J_{i,d}:=(\xi_{i-1,d},\xi_{i,d}]\), \(i=1,\ldots,\alpha\), with equal lengths, where \(\xi_{0,d}:=t_{d-1,n}(\theta)\) and \(\xi_{\alpha,d}:=t_{d,n}(\theta)\). The possible atom at the global lower endpoint \(t_{0,n}(\theta)\) is kept separate.

Take any \(\mu\in\mathcal C_{J_n}(c,\theta)\). For each \((i,d)\), let \(w_{i,d}:=\mu(J_{i,d})\). If \(w_{i,d}>0\), set \(\mu_{i,d}:=\mu(\cdot\cap J_{i,d})/w_{i,d}\) and \(\bar x_{i,d}:=\int x\,d\mu_{i,d}(x)\); because the left endpoint of \(J_{i,d}\) is not included, except for the separately handled global lower endpoint, \(\bar x_{i,d}\) belongs to the closed coordinate copy of the same subcell. If \(w_{i,d}=0\), choose any \(\bar x_{i,d}\in J_{i,d}\). Define
\[
\mu_N
:=\mu(\{t_{0,n}(\theta)\})\delta_{t_{0,n}(\theta)}
+\sum_{d=1}^{s_n}\sum_{i=1}^{\alpha}w_{i,d}\delta_{\bar x_{i,d}}.
\]
Then \(\mu_N\) is a probability measure with at most \(1+\alpha s_n\le N\) support points.

We next check feasibility. Fix an equality row or included inequality row \(\ell\in h(J_n)\cup g(J_n)\). If \(\ell\in\mathcal L_{\mathrm{step},n}\), then \(f_{\ell,n}(\cdot;\theta)\) is constant on every partition cell and hence on every subcell, so replacing \(\mu\) by the subcell point masses leaves the row integral unchanged. If \(\ell\in\mathcal L_{\mathrm{aff},n}\), then on each partition cell \(f_{\ell,n}(y;\theta)=a_{\ell,d,n}(\theta)y+b_{\ell,d,n}(\theta)\). The construction preserves the mass and first moment inside each subcell, and the atom at \(t_{0,n}(\theta)\) is unchanged. Therefore \(\int f_{\ell,n}\,d\mu_N=\int f_{\ell,n}\,d\mu\) for every retained row. Hence \(\mu_N\in\mathcal C_{J_n}(c,\theta)\), and the support bound gives \(\mu_N\in\mathcal C_{J_n}^{(N)}(c,\theta)\).

Let \(\Delta_N\) be the maximum subcell length. The measure
\[
\gamma:=\mu(\{t_{0,n}(\theta)\})\delta_{(t_{0,n}(\theta),t_{0,n}(\theta))}
+\sum_{d=1}^{s_n}\sum_{i=1}^{\alpha}w_{i,d}\,\mu_{i,d}\otimes\delta_{\bar x_{i,d}}
\]
is a coupling of \(\mu\) and \(\mu_N\). Since both coordinates of each nontrivial component lie in the same subcell, \(\Wone(\mu,\mu_N)\le\Delta_N\). Assumption~\ref{ass:objective} gives \(|F(\mu;\pi,q)-F(\mu_N;\pi,q)|\le L\Delta_N\).

Let \(V(\pi,q)\) be the supremum of \(F(\cdot;\pi,q)\) over \(\mathcal C_{J_n}(c,\theta)\), and let \(V_N(\pi,q)\) be the same supremum over \(\mathcal C_{J_n}^{(N)}(c,\theta)\). Since the restricted feasible set is contained in the unrestricted one, \(V_N(\pi,q)\le V(\pi,q)\). Conversely, for any \(\varepsilon>0\), choose \(\mu^\varepsilon\in\mathcal C_{J_n}(c,\theta)\) with \(F(\mu^\varepsilon;\pi,q)\ge V(\pi,q)-\varepsilon\). The preceding construction gives \(\mu_N^\varepsilon\in\mathcal C_{J_n}^{(N)}(c,\theta)\) and therefore \(V(\pi,q)-V_N(\pi,q)\le\varepsilon+L\Delta_N\). Letting \(\varepsilon\downarrow0\) yields the value bound. Finally, \(\Delta_N\le\diam(\YY)/\alpha\), and \(\alpha\ge (N-1)/(2s_n)\) for \(N\ge s_n+1\). Thus \(\Delta_N\le4s_n\diam(\YY)/N\), after enlarging the constant to cover the harmless case \(N=s_n+1\). Taking the supremum over admissible inputs proves the displayed inequality. The infimum case follows by applying the same argument to \(-F\), which satisfies the same \(\Wone\)-Lipschitz bound.
\end{proof}

\begin{lemma}\label{lem:quant}
Under Assumptions~\ref{ass:support},~\ref{ass:restriction_geometry},~\ref{ass:objective},~\ref{ass:finite_complexity}, and~\ref{ass:feasibility_rank}, let \(N_1\ge k_n\), let \(N_2\) be sufficiently large, and let \(\eta=(c,\theta,\pi,q)\in\widetilde{\mathcal N}_0\) satisfy \(\mathcal C_{J_n}^{(N_1)}(c,\theta)\neq\varnothing\). Suppose every block mass implied by the included step equalities belongs to \(N_2^{-1}\mathbb Z\). Then \(\mathcal C_{J_n}^{(N_1,N_2)}(c,\theta)\neq\varnothing\) and \(\|V_{J_n}^{N_1}(\eta)-V_{J_n}^{N_1,N_2}(\eta)\|_2\le \sqrt{2}\,LK/N_2\), where \(K\) is the finite constant constructed in the repair claim in the proof.
\end{lemma}

\begin{proof}[Proof of Lemma~\ref{lem:quant}]
We first record the repair step used to place the masses on the empirical grid.

\smallskip\noindent\emph{Claim.}
Under Assumptions~\ref{ass:support},~\ref{ass:restriction_geometry},~\ref{ass:finite_complexity}, and~\ref{ass:feasibility_rank}, for all large \(n\), there exist \(N_{2,0},K<\infty\) such that the following holds uniformly over \(N_1\ge k_n\) and \(\eta=(c,\theta,\pi,q)\in\widetilde{\mathcal N}_0\): if \(N_2\ge N_{2,0}\), every block mass implied by the included step equalities belongs to \(N_2^{-1}\mathbb Z\), and \(\mu\in\mathcal C_{J_n}^{(N_1)}(c,\theta)\), then there exists \(\tilde\mu\in\mathcal C_{J_n}^{(N_1,N_2)}(c,\theta)\) with \(\Wone(\mu,\tilde\mu)\le K/N_2\).

\smallskip\noindent\emph{Proof of the claim.}
Fix \(n\), \(N_1\), \(N_2\), \(\eta\), and \(\mu\) as in the statement. Choose an adjustment-ready representation \(\mu=\sum_{j=1}^{N_1}p_j\delta_{z_j}\) from Lemma~\ref{lem:equality_adjustment}. Let \(m_0:=m_{\mathrm{aff},n}\), and let \(j_1,\ldots,j_{m_0}\) denote the selected affine-adjustment coordinates, with the selected set interpreted as empty when \(m_0=0\). Let \(p^{\mathrm{sl}}\) be the same-support slack vector and \(\mu^{\mathrm{sl}}:=\sum_jp_j^{\mathrm{sl}}\delta_{z_j}\). If \(m_0>0\), set \(C_A:=m_0\bar C_{\mathrm{slope}}\), \(C_r:=\bar V_f(m_0+1)\), \(C_w:=C_AC_r\), and \(C_I:=\bar V_f(m_0+1)+\bar f_\partial m_0C_w\). If \(m_0=0\), set \(C_A=C_w=0\) and \(C_I:=\bar V_f\). Choose \(a>0\) so large that \(a\underline p>2(m_0+1)\), \(2C_w/(a\underline p)<\underline\delta\), and \(a\underline s\ge C_I\). Then choose \(N_{2,0}\) so that \(a/N_2<1\) for all \(N_2\ge N_{2,0}\). Set \(\alpha:=a/N_2\) and \(\mu_1:=(1-\alpha)\mu+\alpha\mu^{\mathrm{sl}}=\sum_{j=1}^{N_1}p_j^{(1)}\delta_{z_j}\). The measure \(\mu_1\) satisfies all equalities, every included inequality has slack at least \(\alpha\underline s\), \(\Wone(\mu,\mu_1)\le\alpha\diam(\YY)\), and every selected affine-adjustment coordinate has mass at least \(\alpha\underline p\).

Let \(\tau_0<\cdots<\tau_B\) be the ordered distinct collection of \(\underline y\), \(\overline y\), and all jump points of included step-function equality rows. Write \(K_1:=[\tau_0,\tau_1]\) and \(K_b:=(\tau_{b-1},\tau_b]\) for \(b=2,\ldots,B\). If a cumulative boundary row is not itself retained after row reduction, use the bounded integer relation in \eqref{eq:boundary_consistency} to express its right-hand side from the retained step equalities. Since \(\mu_1\) satisfies the step equalities, \(\mu_1(K_b)\) is the block mass implied by those equalities, and the hypothesis gives \(\mu_1(K_b)\in N_2^{-1}\mathbb Z\) for every \(b\).

Round masses within each block while preserving the block total. If block \(K_b\) contains no atom of \(\mu_1\), leave it unchanged. Otherwise, list its atoms as \((z_{b,r},p_{b,r}^{(1)},i_{b,r})_{r=1}^{q_b}\) in increasing order of location, where \(i_{b,r}\) is the original coordinate label, and set \(P_{b,r}^{(1)}:=\sum_{u=1}^r p_{b,u}^{(1)}\). Define \(S_{b,0}:=0\). For \(r<q_b\), set
\[
S_{b,r}:=
\begin{cases}
S_{b,r-1}+\lfloor N_2p_{b,r}^{(1)}\rfloor/N_2,
& i_{b,r}\in\{j_1,\ldots,j_{m_0}\},\\
\lfloor N_2P_{b,r}^{(1)}\rfloor/N_2,
& i_{b,r}\notin\{j_1,\ldots,j_{m_0}\}.
\end{cases}
\]
Then put \(p_{b,r}^{(2)}:=S_{b,r}-S_{b,r-1}\) for \(r<q_b\) and \(p_{b,q_b}^{(2)}:=\mu_1(K_b)-S_{b,q_b-1}\). The sequence \((S_{b,r})\) is nondecreasing because each \(S_{b,r}\) is a grid point not exceeding \(P_{b,r}^{(1)}\) and not below the preceding grid point. Since the block total is on the \(1/N_2\) grid, all new masses are nonnegative multiples of \(1/N_2\) and sum to \(\mu_1(K_b)\). If \(e_{b,r}:=P_{b,r}^{(1)}-S_{b,r}\), then \(0\le e_{b,r}\le(m_0+1)/N_2\) for \(r<q_b\) and \(e_{b,q_b}=0\). Therefore the rounded measure \(\mu_2:=\sum_jp_j^{(2)}\delta_{z_j}\) preserves every step-equality block mass exactly and satisfies \(\sup_y|\mu_2((-\infty,y])-\mu_1((-\infty,y])|\le(m_0+1)/N_2\) and \(\Wone(\mu_2,\mu_1)\le(m_0+1)\diam(\YY)/N_2\). The bounded-variation part of Assumption~\ref{ass:restriction_geometry} gives, for every affine equality row, a residual \(r_\ell:=\int f_{\ell,n}(y;\theta)\,d\mu_2(y)-c_\ell\) satisfying \(\|r\|_\infty\le C_r/N_2\). For each selected affine-adjustment coordinate, the rounding rule gives \(p_{j_t}^{(2)}\ge p_{j_t}^{(1)}-(m_0+1)/N_2\ge a\underline p/(2N_2)\).

If \(m_0=0\), set \(\tilde\mu:=\mu_2\). If \(m_0>0\), write \(A:=A_{\mathrm{aff}}(z,\theta)\) and solve \(A^\top w=-r\). Lemma~\ref{lem:equality_adjustment} gives \(\|w\|_\infty\le C_w/N_2\). Move selected atom \(j_t\) by \(\Delta_t:=w_t/p_{j_t}^{(2)}\). The selected-mass lower bound gives \(\max_t|\Delta_t|\le2C_w/(a\underline p)<\underline\delta\), so all moved atoms remain in their retained cells. Because the relevant rows are affine within cells, \(\sum_t p_{j_t}^{(2)}\partial_y f_{\ell,n}(z_{j_t};\theta)\Delta_t=(A^\top w)_\ell=-r_\ell\), and the affine equalities are restored exactly. The movements do not change step restrictions.

The only remaining restrictions to check are inequalities. The rounding step changes any retained row by at most \(\bar V_f(m_0+1)/N_2\), and the affine-location movement changes nonstep rows by at most \(\bar f_\partial\|w\|_1\le\bar f_\partial m_0C_w/N_2\). By the choice of \(a\), this total change is at most \(\alpha\underline s\). Since \(\mu_1\) had slack \(\alpha\underline s\), all included inequalities remain feasible. The final measure has at most \(N_1\) support coordinates and grid masses in \(N_2^{-1}\mathbb Z_+\), so \(\tilde\mu\in\mathcal C_{J_n}^{(N_1,N_2)}(c,\theta)\). Finally, \(\Wone(\mu,\tilde\mu)\le[a\diam(\YY)+(m_0+1)\diam(\YY)+m_0C_w]/N_2\le K/N_2\), for a finite constant \(K\) that is uniform because \(m_0\) is uniformly bounded.

\smallskip
Choose any \(\mu\in\mathcal C_{J_n}^{(N_1)}(c,\theta)\). The repair claim gives \(\tilde\mu\in\mathcal C_{J_n}^{(N_1,N_2)}(c,\theta)\), so the grid-mass feasible set is nonempty. This set is contained in \(\mathcal C_{J_n}^{(N_1)}(c,\theta)\), so the unrestricted-mass supremum is at least the grid-mass supremum. Let \(v\) and \(v_g\) denote these two suprema. For \(\varepsilon>0\), choose \(\mu^\varepsilon\in\mathcal C_{J_n}^{(N_1)}(c,\theta)\) with \(F(\mu^\varepsilon;\pi,q)\ge v-\varepsilon\). Applying the repair claim to \(\mu^\varepsilon\) gives \(\tilde\mu^\varepsilon\in\mathcal C_{J_n}^{(N_1,N_2)}(c,\theta)\) and \(\Wone(\mu^\varepsilon,\tilde\mu^\varepsilon)\le K/N_2\). Therefore \(v-\varepsilon\le F(\mu^\varepsilon;\pi,q)\le F(\tilde\mu^\varepsilon;\pi,q)+LK/N_2\le v_g+LK/N_2\). Letting \(\varepsilon\downarrow0\) proves \(0\le v-v_g\le LK/N_2\). The same argument applied to \(-F\) gives the corresponding bound for the infimum endpoint. Combining the two scalar bounds gives the displayed Euclidean bound.
\end{proof}

For the next result, partition \(\YY\) at the jump points of the included step-equality rows, and write \(m=(m_1,\ldots,m_B)\) for the resulting block masses. Cumulative rows not retained after row reduction are interpreted through the bounded integer combinations in \eqref{eq:boundary_consistency}. For two right-hand-side vectors \(c\) and \(\bar c\) that agree on affine equalities and inequalities, let \(m\) and \(\bar m\) be the compatible block-mass vectors, require the same zero pattern, and set \(\Delta:=\sum_b|\bar m_b-m_b|\).

\begin{lemma}\label{lem:step_rhs_value_perturb}
Under Assumptions~\ref{ass:support},~\ref{ass:restriction_geometry},~\ref{ass:objective},~\ref{ass:finite_complexity}, and~\ref{ass:feasibility_rank}, there are \(\Delta_0>0\) and \(C<\infty\) such that, uniformly over large \(n\), \(N\ge k_n\), and \(\eta=(c,\theta,\pi,q),\bar\eta=(\bar c,\theta,\pi,q)\in\widetilde{\mathcal N}_0\), if \(\Delta\le\Delta_0\) and either finite-support feasible set is nonempty, then both are nonempty and \(\|V_{J_n}^{N}(\eta)-V_{J_n}^{N}(\bar\eta)\|_2\le \sqrt{2}LC\Delta\).
\end{lemma}

\begin{proof}[Proof of Lemma~\ref{lem:step_rhs_value_perturb}]
Let \(\tau_0<\cdots<\tau_B\) be the ordered endpoints and jump points of the included step equalities, with \(K_1=[\tau_0,\tau_1]\) and \(K_b=(\tau_{b-1},\tau_b]\) for \(b\ge2\). If \(B=1\), set \(m_1=\bar m_1=1\). Otherwise, let \(u_b(y)=\mathbf 1[y\le\tau_b]\), interpret its right-hand side through \eqref{eq:boundary_consistency} when the cumulative row is not retained, and set \(m_1=c_{u_1}\), \(m_b=c_{u_b}-c_{u_{b-1}}\) for \(2\le b<B\), and \(m_B=1-c_{u_{B-1}}\), with \(\bar m\) defined analogously from \(\bar c\). For every retained step equality \(u\), its right-hand side is \(c_u=\sum_bv_{u,b}m_b\), and similarly \(\bar c_u=\sum_bv_{u,b}\bar m_b\). The hypothesis imposes the same zero pattern for \(m\) and \(\bar m\), and \(\Delta=\sum_b|\bar m_b-m_b|\).

By symmetry, suppose \(\mathcal C_{J_n}^{(N)}(c,\theta)\neq\varnothing\). If \(\Delta=0\), then the step right-hand sides agree and hence \(\eta=\bar\eta\) in the coordinates that differ, so there is nothing to prove. We prove the one-sided bound for the supremum endpoint; the infimum endpoint follows by replacing \(F\) with \(-F\). Fix \(\varepsilon>0\), and choose an adjustment-ready representation \(\mu=\sum_{j=1}^Np_j\delta_{z_j}\in\mathcal C_{J_n}^{(N)}(c,\theta)\), selected affine-adjustment coordinates, and a same-support slack vector \(p^{\mathrm{sl}}\) from Lemma~\ref{lem:equality_adjustment}, with \(F(\mu;\pi,q)\ge V_{J_n}^{N,\sup}(\eta)-\varepsilon\). Let \(\mu^{\mathrm{sl}}:=\sum_jp_j^{\mathrm{sl}}\delta_{z_j}\). Set \(C_r:=\bar V_f\). If \(m_{\mathrm{aff},n}>0\), set \(C_A:=m_{\mathrm{aff},n}\bar C_{\mathrm{slope}}\), \(C_w:=C_AC_r\), and \(C_I:=\bar V_f+\bar f_\partial m_{\mathrm{aff},n}C_w\); otherwise set \(C_A=C_w=0\) and \(C_I:=\bar V_f\). Choose \(a>0\) so large that \(a\underline p>2\), \(2C_w/(a\underline p)<\underline\delta\), and \(a\underline s\ge C_I\). Let \(\alpha:=a\Delta\), and choose \(\Delta_0\) so that \(\alpha<1\) whenever \(\Delta\le\Delta_0\). Define \(\mu_1:=(1-\alpha)\mu+\alpha\mu^{\mathrm{sl}}\). Then \(\mu_1\in\mathcal C_{J_n}^{(N)}(c,\theta)\), every included inequality has slack at least \(\alpha\underline s\), and each selected affine-adjustment coordinate has mass at least \(\alpha\underline p\).

Since \(\mu_1\) satisfies the step equalities under \(c\), \(\mu_1(K_b)=m_b\) for every \(b\). For each block with \(m_b>0\), let \(\nu_b:=\mu_1(\cdot\cap K_b)/m_b\). If \(m_b=0\), choose any probability measure supported on \(K_b\); it is multiplied by \(\bar m_b=0\) and is immaterial. Define \(\mu_2:=\sum_{b=1}^B\bar m_b\nu_b\). Then \(\mu_2\) uses the same positive support atoms as \(\mu_1\) in every block with positive mass, has block masses \(\bar m_b\), and therefore satisfies all included step equalities under \(\bar c\). Coupling the common mass within each block at the same locations and moving only the excess block mass gives \(\Wone(\mu_1,\mu_2)\le\diam(\YY)\Delta\). The same block calculation gives \(\sup_y|\mu_2((-\infty,y])-\mu_1((-\infty,y])|\le\Delta\), so bounded variation implies \(|\int f_{\ell,n}\,d\mu_2-\int f_{\ell,n}\,d\mu_1|\le C_r\Delta\) for every affine row.

The affine equality right-hand sides are the same under \(c\) and \(\bar c\). Thus the residual \(r_\ell:=\int f_{\ell,n}(y;\theta)\,d\mu_2(y)-c_\ell\) satisfies \(\|r\|_\infty\le C_r\Delta\). If selected coordinate \(j_t\) lies in block \(K_{b(t)}\), then \(m_{b(t)}>0\) and the rescaling within that block gives \(|p_{j_t}^{(2)}-p_{j_t}^{(1)}|\le |\bar m_{b(t)}-m_{b(t)}|\le\Delta\). Therefore \(p_{j_t}^{(2)}\ge a\underline p\Delta-\Delta\ge a\underline p\Delta/2\).

If \(m_{\mathrm{aff},n}=0\), set \(\tilde\mu:=\mu_2\). If \(m_{\mathrm{aff},n}>0\), write \(A:=A_{\mathrm{aff}}(z,\theta)\) and solve \(A^\top w=-r\). Lemma~\ref{lem:equality_adjustment} gives \(\|w\|_\infty\le C_w\Delta\). Move selected atom \(j_t\) by \(\Delta_t^{\mathrm{aff}}:=w_t/p_{j_t}^{(2)}\). The preceding lower bound on selected masses gives \(\max_t|\Delta_t^{\mathrm{aff}}|\le2C_w/(a\underline p)<\underline\delta\), so all selected atoms remain in their cells. The affine equalities are restored exactly, and step restrictions are unchanged because the selected movements stay within cells.

The block-mass perturbation changes any included row by at most \(\bar V_f\Delta\), and the affine-location movement changes nonstep rows by at most \(\bar f_\partial m_{\mathrm{aff},n}C_w\Delta\). This total is at most \(C_I\Delta\le\alpha\underline s\), so the included inequalities remain feasible for \(\bar c\). Hence \(\tilde\mu\in\mathcal C_{J_n}^{(N)}(\bar c,\theta)\). Moreover, \(\Wone(\mu,\tilde\mu)\le\alpha\diam(\YY)+\diam(\YY)\Delta+m_{\mathrm{aff},n}C_w\Delta\le C\Delta\), with \(C\) uniform. Assumption~\ref{ass:objective} gives \(F(\mu;\pi,q)\le F(\tilde\mu;\pi,q)+LC\Delta\). Letting \(\varepsilon\downarrow0\) yields \(V_{J_n}^{N,\sup}(\eta)\le V_{J_n}^{N,\sup}(\bar\eta)+LC\Delta\). Reversing the roles of \(c\) and \(\bar c\) gives the reverse inequality and also shows nonemptiness in the opposite direction. Applying the same argument to \(-F\) gives the infimum bound, and the Euclidean norm contributes the factor \(\sqrt{2}\).
\end{proof}

\begin{lemma}\label{lem:unif_kn_vs_inf}
Under Assumptions~\ref{ass:support},~\ref{ass:restriction_geometry},~\ref{ass:objective},~\ref{ass:finite_complexity}, and~\ref{ass:feasibility_rank}, \(\sup_{\eta\in\mathcal N_0}\|V_{J_n}(\eta)-V_{J_n}^{k_n}(\eta)\|_2=o(n^{-1/2})\). Moreover, \(\mathcal C_{J_n}^{(k_n)}(c,\theta)\ne\varnothing\) for all \(\eta=(c,\theta,\pi,q)\in\mathcal N_0\) and all sufficiently large \(n\).
\end{lemma}

\begin{proof}[Proof of Lemma~\ref{lem:unif_kn_vs_inf}]
We prove the claim for the supremum coordinate; the infimum coordinate follows by replacing \(F\) with \(-F\). Set \(N_1:=\lceil2n(s_n+1)^2\rceil\) and \(N_2:=nN_1\). For all sufficiently large \(n\), \(N_1\ge n\ge k_n\), \(N_1\ge s_n+1\), \(s_n/N_1\le[2n(s_n+1)]^{-1}\), and \(N_2^{-1}=O(n^{-2})\).

Fix \(\eta=(c,\theta,\pi,q)\in\mathcal N_0\). Let \(\tau_0<\cdots<\tau_B\) be the ordered distinct collection of \(\underline y\), \(\overline y\), and the jump points of the included step-equality rows, and set \(K_1=[\tau_0,\tau_1]\) and \(K_b=(\tau_{b-1},\tau_b]\) for \(b\ge2\). If \(B=1\), set \(m_1=1\). Otherwise, for \(b<B\), let \(u_b(y)=\mathbf 1[y\le\tau_b]\), interpreting its right-hand side through \eqref{eq:boundary_consistency} when the cumulative row is not retained, and define \(m_1=c_{u_1}\), \(m_b=c_{u_b}-c_{u_{b-1}}\) for \(2\le b<B\), and \(m_B=1-c_{u_{B-1}}\). Lemma~\ref{lem:equality_adjustment}\textup{(i)} gives a measure in \(\mathcal C_\infty(c,\theta)\), so these block masses are nonnegative and sum to one.

Choose \(b_*\) among the positive blocks so that \(m_{b_*}\) is largest. For \(b\ne b_*\), set \(\bar m_b=0\) when \(m_b=0\) and \(\bar m_b=N_2^{-1}\max(1,\lfloor N_2m_b\rfloor)\) otherwise. Set \(\bar m_{b_*}=1-\sum_{b\ne b_*}\bar m_b\). Since \(m_{b_*}\ge B^{-1}\) and the upward rounding error outside \(b_*\) is at most \((B-1)/N_2\), \(\bar m_{b_*}>0\) for all sufficiently large \(n\). The vector \(\bar m\) has the same zero pattern as \(m\), lies on the \(1/N_2\) grid, and satisfies \(\Delta_n(\eta):=\sum_{b=1}^B|\bar m_b-m_b|\le2(B-1)/N_2\le2s_n/N_2=o(n^{-1/2})\) uniformly over \(\eta\in\mathcal N_0\), because \(B\le s_n+1\).

For each retained step equality \(u\), let \(v_{u,b}\) be its value on \(K_b\) and set \(\bar c_u:=\sum_bv_{u,b}\bar m_b\). Keep the affine-equality and inequality right-hand sides unchanged, and write \(\bar\eta=(\bar c,\theta,\pi,q)\). The step block masses implied by \(\bar c\) are \(\bar m\), hence lie on the \(1/N_2\) grid. Also \(\|\bar c-c\|_\infty\le\bar f\Delta_n(\eta)\), so \(\bar\eta\in\widetilde{\mathcal N}_0\) uniformly for all sufficiently large \(n\).

Let
\begin{align*}
A_{1n}&:=V_{J_n}^{\sup}(\eta)-V_{J_n}^{N_1,\sup}(\eta),&
A_{2n}&:=|V_{J_n}^{N_1,\sup}(\eta)-V_{J_n}^{N_1,\sup}(\bar\eta)|,\\
A_{3n}&:=V_{J_n}^{N_1,\sup}(\bar\eta)-V_{J_n}^{N_1,N_2,\sup}(\bar\eta),&
A_{4n}&:=V_{J_n}^{N_2,N_2,\sup}(\bar\eta)-V_{J_n}^{k_n,N_2,\sup}(\bar\eta),\\
A_{5n}&:=|V_{J_n}^{k_n,\sup}(\bar\eta)-V_{J_n}^{k_n,\sup}(\eta)|.
\end{align*}
The inclusions implied by \(N_1\le N_2\) give \(V_{J_n}^{N_1,N_2,\sup}(\bar\eta)\le V_{J_n}^{N_2,N_2,\sup}(\bar\eta)\) and \(V_{J_n}^{k_n,N_2,\sup}(\bar\eta)\le V_{J_n}^{k_n,\sup}(\bar\eta)\). The two omitted telescoping terms are therefore nonpositive, and \(0\le V_{J_n}^{\sup}(\eta)-V_{J_n}^{k_n,\sup}(\eta)\le A_{1n}+A_{2n}+A_{3n}+A_{4n}+A_{5n}\).

Lemma~\ref{lem:largeN} gives \(A_{1n}\le\bar C s_n/N_1=o(n^{-1/2})\). It also constructs an \(N_1\)-point feasible measure because \(\mathcal C_\infty(c,\theta)\subseteq\mathcal C_{J_n}(c,\theta)\). Lemma~\ref{lem:step_rhs_value_perturb}, applied with \(N=N_1\), gives \(A_{2n}\le\sqrt{2}LC\Delta_n(\eta)=o(n^{-1/2})\) and makes \(\mathcal C_{J_n}^{(N_1)}(\bar c,\theta)\) nonempty. Lemma~\ref{lem:quant} then gives \(A_{3n}\le\sqrt{2}LK/N_2=o(n^{-1/2})\) and produces a feasible measure in \(\mathcal C_{J_n}^{(N_1,N_2)}(\bar c,\theta)\). Since \(N_1\le N_2\), this measure also belongs to \(\mathcal C_{J_n}^{(N_2,N_2)}(\bar c,\theta)\). Assumption~\ref{ass:finite_complexity}, with \(N=N_2\), gives \(A_{4n}\le r_n=o(n^{-1/2})\) and \(\mathcal C_{J_n}^{(k_n,N_2)}(\bar c,\theta)\ne\varnothing\). Thus \(\mathcal C_{J_n}^{(k_n)}(\bar c,\theta)\ne\varnothing\), and a second application of Lemma~\ref{lem:step_rhs_value_perturb}, now with \(N=k_n\) and the roles of \(c\) and \(\bar c\) reversed, gives \(A_{5n}\le\sqrt{2}LC\Delta_n(\eta)=o(n^{-1/2})\). This last application also gives \(\mathcal C_{J_n}^{(k_n)}(c,\theta)\ne\varnothing\).

All bounds are uniform over \(\eta\in\mathcal N_0\). Their sum proves the supremum-coordinate result. Applying the same argument to \(-F\) proves the infimum-coordinate result and therefore the vector bound.
\end{proof}

\paragraph{Approximation of the full inequality system.}
Assumption~\ref{ass:tail_existence} and the full-system Slater point convert the maximum violation of an omitted row into a value error.

\begin{lemma}\label{lem:tail-measure}
Under Assumptions~\ref{ass:support},~\ref{ass:restriction_geometry},~\ref{ass:objective},~\ref{ass:tail_existence}, and~\ref{ass:feasibility_rank}, for all sufficiently large \(n\), \(\sup_{\eta\in\mathcal N_0}\|V_{J_n}(\eta)-V_\infty(\eta)\|_2\le\sqrt{2}L\diam(\YY)\kappa_{J_n}/\underline s\).
\end{lemma}

\begin{proof}[Proof of Lemma~\ref{lem:tail-measure}]
Fix \(n\) large enough that all equality rows are included, and fix \(\eta=(c,\theta,\pi,q)\in\mathcal N_0\). Let \(\mu^{\mathrm{sl},\infty}:=\mu_n^{\mathrm{sl},\infty}(\eta)\) be the full-system Slater measure from Assumption~\ref{ass:feasibility_rank}. It belongs to \(\mathcal C_\infty(c,\theta)\), satisfies all equalities, and has slack at least \(\underline s\) in every normalized inequality row. Since \(\mathcal C_\infty(c,\theta)\subseteq\mathcal C_{J_n}(c,\theta)\), the finite-\(J_n\) feasible set is nonempty.

Consider the supremum endpoint and fix \(\varepsilon>0\). Choose \(\mu^n\in\mathcal C_{J_n}(c,\theta)\) with \(F(\mu^n;\pi,q)\ge V_{J_n}^{\sup}(\eta)-\varepsilon\). Assumption~\ref{ass:tail_existence} gives, for every omitted inequality row \(u\), \(G_u(\mu^n;\theta)-c_u^2\le\kappa_{J_n}\); retained inequalities are already satisfied. Set \(\kappa:=\kappa_{J_n}\) and \(\alpha:=\kappa/(\underline s+\kappa)\), with \(\alpha=0\) if \(\kappa=0\), and define \(\tilde\mu:=(1-\alpha)\mu^n+\alpha\mu^{\mathrm{sl},\infty}\). Equalities are preserved by linearity. For a retained inequality, both measures are feasible and the Slater measure has slack, so the mixture is feasible. For an omitted row, \(G_u(\tilde\mu;\theta)-c_u^2\le(1-\alpha)\kappa-\alpha\underline s=0\). Thus \(\tilde\mu\in\mathcal C_\infty(c,\theta)\).

The mixture moves the measure by no more than \(\alpha\Wone(\mu^n,\mu^{\mathrm{sl},\infty})\). This displacement is bounded by \(\alpha\diam(\YY)\), and therefore by \(\diam(\YY)\kappa_{J_n}/\underline s\). The \(\Wone\)-Lipschitz property of \(F\) implies that the objective loss from replacing \(\mu^n\) by \(\tilde\mu\) is at most \(L\diam(\YY)\kappa_{J_n}/\underline s\). The measure \(\tilde\mu\) is full-feasible, so letting \(\varepsilon\downarrow0\) gives the stated supremum endpoint bound. Its lower bound follows from \(\mathcal C_\infty(c,\theta)\subseteq\mathcal C_{J_n}(c,\theta)\).

For the lower endpoint, apply the same argument to \(-F\). Equivalently, start from an \(\varepsilon\)-minimizer in \(\mathcal C_{J_n}(c,\theta)\), mix it with the Slater measure using the same \(\alpha\), and use the \(\Wone\)-Lipschitz bound for \(F\). This gives \(0\le V_\infty^{\inf}(\eta)-V_{J_n}^{\inf}(\eta)\le L\diam(\YY)\kappa_{J_n}/\underline s\). Combining the two scalar bounds gives the lemma's Euclidean bound, uniformly over \(\eta\in\mathcal N_0\).
\end{proof}

\subsection{Local sensitivity and differentiability}\label{subsec:appendix_local_sensitivity}
This subsection proves the local expansion used in Proposition~\ref{prop:hdd_limit_phi_n} once the finite diagnostics hold. The baseline objective gap in Assumption~\ref{ass:endpoint_rank} localizes optimizers under small input perturbations. The argument first proves a uniform active-set repair: an endpoint optimizer at one nearby input can be moved to satisfy the active restrictions at another nearby input with coordinate displacement proportional to the input displacement. The repair is then combined with the objective-gap condition and the atomic matching lemma. If a support point lies on a partition boundary, all adjacent-cell representations that are compatible with the retained step equalities are kept, so a nearby optimizer is not lost by the local parameterization.

\begin{lemma}\label{lem:optimizer_localization}
Under Assumptions~\ref{ass:support},~\ref{ass:restriction_geometry},~\ref{ass:objective},~\ref{ass:finite_complexity},~\ref{ass:feasibility_rank}, and~\ref{ass:endpoint_rank}, for every compact \(K\subset\mathbb D_0\), there is \(N_K\) such that
\[
\sup_{\substack{n\ge N_K,\ \circ\in\{\inf,\sup\},\ h\in K:\\
                  \eta_0+t h\in\mathcal N_0}}
\ \sup_{\mu\in\mathcal S_n^\circ(\eta_0+t h)}
\ \inf_{\nu\in\mathcal S_n^\circ(\eta_0)}
 d_{\mathrm{at}}(\mu,\nu)
\longrightarrow0
\]
as \(t\downarrow0\). For all sufficiently small \(t\), every optimizer in the display therefore has a compatible reduced representation in one of the retained neighborhoods \(\mathfrak X_n\).
\end{lemma}

\begin{proof}[Proof of Lemma~\ref{lem:optimizer_localization}]
Lemma~\ref{lem:unif_kn_vs_inf} gives nonempty reduced feasible sets on \(\mathcal N_0\) for all sufficiently large \(n\), and Assumption~\ref{ass:endpoint_rank} gives attainment of both reduced endpoints. We first record the repair used twice below. Fix such an \(n\), an endpoint \(\circ\), a source input \(\eta^s\in\mathcal N_0\), a target input \(\eta^t\in\mathcal N_0\), and a compatible reduced representation \(x^s\) of an endpoint optimizer at \(\eta^s\). Repeated atoms are merged, zero masses are removed, and redundant or uniformly locally implied active rows are deleted as in Assumption~\ref{ass:endpoint_rank}. Let \(C_s(x;\eta)\) stack, with the signs used in the local coordinate problem, the simplex residual, the row-reduced equality residuals, the active nonimplied inequality residuals, and the active support-position residuals at \(x^s\). Thus \(C_s(x^s;\eta^s)=0\). Write \(Q_s:=D_xC_s(x^s;\eta^s)\) and \(R_s:=Q_s^\top(Q_sQ_s^\top)^{-1}\). The active-rank clause gives \(\|R_s\|\le \underline\sigma_A^{-1}\), uniformly over the source optimizer, \(n\), and \(\circ\).

Set \(d:=\|\eta^t-\eta^s\|_\infty\). The common input derivatives of the restriction rows and support bounds imply \(\|C_s(x^s;\eta^t)\|\le C d\). Let \(\mathcal U_s:=\operatorname{range}(Q_s^\top)\), and define \(T_s(u):=u-R_sC_s(x^s+u;\eta^t)\) on \(\mathcal U_s\). Since \(Q_sR_s=I\) on the residual space, \(R_sQ_s\) is the identity on \(\mathcal U_s\). The common modulus of \(D_xC_s\) therefore gives, after reducing the optimizer neighborhoods once and for all, \(\|T_s(u)-T_s(v)\|\le\|u-v\|/2\) whenever \(u,v\in\mathcal U_s\) and \(\|u\|\) and \(\|v\|\) are bounded by a fixed small radius. Also \(\|T_s(0)\|\le C_1d\). For \(d\) small, \(T_s\) maps the ball \(\{u\in\mathcal U_s:\|u\|\le2C_1d\}\) into itself, so Banach's fixed-point theorem gives \(u_s\) in that ball with \(T_s(u_s)=u_s\). The identity \(T_s(u_s)=u_s\) gives \(R_sC_s(x^s+u_s;\eta^t)=0\), and applying \(Q_s\) gives \(C_s(x^s+u_s;\eta^t)=0\). Thus \(x^t:=x^s+u_s\) satisfies every row in the active system at the target input and \(\|x^t-x^s\|\le C_2d\).

The repair remains feasible for the full retained local problem. Positive masses and inactive support bounds remain feasible because the source optimizer has common mass and boundary margins. Active inequalities are imposed as equalities in \(C_s\), while inactive nonimplied inequalities retain at least half of their source slack for small \(d\). A deleted uniformly locally implied row is a bounded nonnegative linear combination of retained inequality residuals plus a bounded linear combination of equality residuals on the same neighborhood and for the perturbed right-hand sides, so it is feasible whenever the retained system is feasible. The repaired point is therefore feasible at \(\eta^t\), belongs to the same compatible cell assignment, and is within \(C_2d\) of the source optimizer. The same argument applies with \(\eta^s\) and \(\eta^t\) interchanged whenever the new source point is an endpoint optimizer, because Assumption~\ref{ass:endpoint_rank} is imposed on the whole neighborhood \(\mathcal N_0\).

Suppose the localization conclusion fails for a compact \(K\). Then, after passing to a subsequence, there are indices \(n_r\to\infty\), one endpoint \(\circ\), numbers \(t_r\downarrow0\), directions \(h_r\in K\) with \(h_r\to h\in K\), and optimizer measures \(\mu_r\in\mathcal S_{n_r}^\circ(\eta_r)\), where \(\eta_r:=\eta_0+t_rh_r\), such that \(\inf_{\nu\in\mathcal S_{n_r}^\circ(\eta_0)}d_{\mathrm{at}}(\mu_r,\nu)\ge\varepsilon\) for some \(\varepsilon>0\). Let \(x_r\) be a compatible reduced representation of \(\mu_r\). Choose any compatible reduced baseline optimizer representation \(x_r^0\in\mathcal S_{n_r}^\circ(\eta_0)\). Repairing \(x_r^0\) from \(\eta_0\) to \(\eta_r\) gives a feasible point \(\tilde x_r^0\) with \(\|\tilde x_r^0-x_r^0\|=O(t_r)\). Since \(x_r\) is optimal at \(\eta_r\) for the signed minimization problem \(\sigma_\circ\Phi\), the objective expansion and input-continuity clause of Assumption~\ref{ass:objective} give \(\sigma_\circ\Phi(x_r;\eta_r)\le\sigma_\circ\Phi(\tilde x_r^0;\eta_r)=\sigma_\circ V_{J_{n_r}}^{k_{n_r},\circ}(\eta_0)+O(t_r)\), where \(\Phi(x;\eta)\) abbreviates \(\Phi_k(x;\pi,q)\) on the relevant support dimension.

Repair \(x_r\) in the reverse direction, from \(\eta_r\) to \(\eta_0\), and call the resulting baseline-feasible representation \(y_r\). The repair bound gives \(\|y_r-x_r\|=O(t_r)\), hence \(\Wone(\mu_{y_r},\mu_r)=O(t_r)\) because the support has at most \(\bar k\) atoms and lies in \(\YY\). Baseline feasibility gives \(\sigma_\circ V_{J_{n_r}}^{k_{n_r},\circ}(\eta_0)\le\sigma_\circ\Phi(y_r;\eta_0)\). The objective Lipschitz and input-continuity bounds give the reverse inequality up to \(o(1)\):
\[
\begin{aligned}
0
&\le \sigma_\circ\Phi(y_r;\eta_0)
      -\sigma_\circ V_{J_{n_r}}^{k_{n_r},\circ}(\eta_0) \\
&\le \sigma_\circ\Phi(x_r;\eta_r)
      -\sigma_\circ V_{J_{n_r}}^{k_{n_r},\circ}(\eta_0)
      +L\Wone(\mu_{y_r},\mu_r)
      +\omega_F(d_{\pi q}((\pi_r,q_r),(\pi_0,q_0))) \\
&=o(1).
\end{aligned}
\]
Thus the signed baseline objective gap of \(\mu_{y_r}\) converges to zero. By the isolation clause of Assumption~\ref{ass:endpoint_rank}, \(\inf_{\nu\in\mathcal S_{n_r}^\circ(\eta_0)}\Wone(\mu_{y_r},\nu)\to0\). Since \(\Wone(\mu_{y_r},\mu_r)=O(t_r)\), the perturbed optimizer \(\mu_r\) is also \(\Wone\)-close to the baseline optimizer set. Choose \(\nu_r\in\mathcal S_{n_r}^\circ(\eta_0)\) within \(r^{-1}\) of this infimum. The common mass and support-separation bounds in Assumption~\ref{ass:endpoint_rank} apply to \(\mu_r\) and \(\nu_r\), so Lemma~\ref{lem:atomic_matching} yields \(d_{\mathrm{at}}(\mu_r,\nu_r)\to0\), contradicting the displayed separation.

It remains only to connect the matched optimizer to the retained coordinate neighborhoods. A matched atom that is a positive distance from every cutoff remains in the same cell for small \(t_r\). If a baseline atom is on a cutoff, the matched perturbed atom lies on one adjacent side, or remains on the cutoff. Feasibility of the perturbed representation, the boundary consistency relation for step equalities, and continuity of the one-sided affine residuals imply that this adjacent assignment satisfies the retained equalities and all locally active or implied inequalities in the limit. Such adjacent assignments are exactly the compatible boundary representations included in \(\mathfrak X_n\). Hence every perturbed optimizer belongs to a retained neighborhood for all sufficiently small \(t\), completing the proof.
\end{proof}

The next lemma is the finite-dimensional envelope calculation used for the reduced value functions. For endpoint sign \(\circ\), write a retained local signed problem as
\[
v_{n,r}^\circ(\eta)
:=
\inf\{\sigma_\circ\Phi_k(x;\pi,q):A_{n,r}(x;\eta)=0,\ B_{n,r}(x;\eta)\le0,\ x\in X_{n,r}\},
\]
where \(\sigma_{\inf}=1\), \(\sigma_{\sup}=-1\), \(X_{n,r}\) is a compact coordinate neighborhood whose artificial boundary is separated from the relevant optimizer set, redundant rows have been removed, and support-position bounds are included among the inequalities. Let \(S_{n,r}^\circ\) be its optimizer set at \(\eta_0\). For \(x\in S_{n,r}^\circ\), let \(C_{n,r,x}\) stack the equalities and the inequalities and support bounds active at \(x\), and let \(\Lambda_{n,r}^\circ(x)\) be the unique multiplier for this active system. With the convention \(L=f+\lambda^\top C\), where \(f=\sigma_\circ\Phi_k\), define \(D_\eta L_{n,r}^\circ(x,\lambda;\eta_0)[h]:=\sigma_\circ\dot\Phi_{k,x}(h)+\lambda^\top\dot C_{n,r,x}(h)\), where \(\dot C_{n,r,x}(h)\) stacks the input derivatives of the active rows.

\begin{lemma}\label{lem:uniform_local_sensitivity}
Under Assumptions~\ref{ass:support},~\ref{ass:restriction_geometry},~\ref{ass:objective},~\ref{ass:finite_complexity},~\ref{ass:feasibility_rank}, and~\ref{ass:endpoint_rank}, for every compact \(K\subset\mathbb D_0\), there are \(N_K\) and \(\rho_K(t)\downarrow0\) such that, uniformly over \(n\ge N_K\), endpoints, retained local parameterizations, and paths \(\eta_t=\eta_0+t h_t\) with \(h_t\in K\) and \(h_t\to h\),
\[
\left|
\frac{v_{n,r}^\circ(\eta_t)-v_{n,r}^\circ(\eta_0)}{t}
-
\min_{x\in S_{n,r}^\circ}
D_\eta L_{n,r}^\circ(x,\Lambda_{n,r}^\circ(x);\eta_0)[h_t]
\right|
\le \rho_K(t).
\]
Replacing \(h_t\) by \(h\) in the derivative term changes the right side by at most \(C_K\|h_t-h\|\). Multiplication by \(\sigma_\circ\) gives the corresponding expansion for the original lower or upper endpoint coordinate.
\end{lemma}

\begin{proof}[Proof of Lemma~\ref{lem:uniform_local_sensitivity}]
All coordinate dimensions are bounded by \(2\bar k\). After redundant active rows are removed, the number of rows in \(C_{n,r,x}\) is no larger than this dimension. Lemma~\ref{lem:endpoint_active_rank} gives common lower bounds on positive masses, inactive slack, nonboundary-cell distance, and the smallest active-Jacobian singular value at every baseline optimizer. Shrinking the retained neighborhoods and using the common gradient modulus in \eqref{eq:restriction_uniform_expansion}, the same singular-value bound holds with one half of the baseline constant throughout the neighborhoods. Hence each active derivative matrix has a right inverse with norm bounded uniformly in \(n\), \(r\), \(\circ\), and \(x\in S_{n,r}^\circ\). Lemma~\ref{lem:endpoint_active_rank} also gives a unique multiplier \(\Lambda_{n,r}^\circ(x)\) with uniformly bounded norm.

Fix \(x\in S_{n,r}^\circ\), write \(Q_x:=D_xC_{n,r,x}(x;\eta_0)\), and define the minimum-norm solution of the linearized active restrictions by \(d_x(h_t):=-Q_x^\top(Q_xQ_x^\top)^{-1}\dot C_{n,r,x}(h_t)\). The derivative maps in \eqref{eq:restriction_uniform_expansion} are uniformly bounded on \(K\), so \(\|d_x(h_t)\|\le C_0\). The definition gives \(Q_xd_x(h_t)+\dot C_{n,r,x}(h_t)=0\). Applying \eqref{eq:restriction_uniform_expansion} to \(x+t d_x(h_t)\) and \(\eta_t\) therefore gives \(\|C_{n,r,x}(x+t d_x(h_t);\eta_t)\|\le t\rho_{K,C_0}^R(t)\). The contraction repair from the proof of Lemma~\ref{lem:optimizer_localization}, now started from \(x+t d_x(h_t)\), gives an exact point \(x_t=x+t d_x(h_t)+t e_t\) satisfying \(C_{n,r,x}(x_t;\eta_t)=0\) and \(\|e_t\|\le C\rho_{K,C_0}^R(t)\). All inactive restrictions remain slack, so \(x_t\) is feasible for the retained local problem at \(\eta_t\).

Let \(f_{n,r}^\circ\) denote the signed local objective, and set \(\Delta f_t:=f_{n,r}^\circ(x_t,\eta_t)-f_{n,r}^\circ(x,\eta_0)\). By \eqref{eq:objective_uniform_expansion}, \(\Delta f_t=tD_xf_{n,r}^\circ(x,\eta_0)[d_x(h_t)]+t\sigma_\circ\dot\Phi_{k,x}(h_t)+t o_K(1)\), uniformly over the retained neighborhoods. Stationarity for the active Lagrangian gives \(D_xf_{n,r}^\circ(x,\eta_0)+\Lambda_{n,r}^\circ(x)^\top Q_x=0\). Since \(Q_xd_x(h_t)=-\dot C_{n,r,x}(h_t)\), the leading term equals \(tD_\eta L_{n,r}^\circ(x,\Lambda_{n,r}^\circ(x);\eta_0)[h_t]\), up to the common \(t o_K(1)\) remainder. Choosing \(x\in S_{n,r}^\circ\) to minimize the Lagrangian input derivative and using feasibility of \(x_t\) proves the upper directional bound
\[
\limsup_{t\downarrow0}
\sup
\left[
\frac{v_{n,r}^\circ(\eta_t)-v_{n,r}^\circ(\eta_0)}{t}
-
\min_{x\in S_{n,r}^\circ}D_\eta L_{n,r}^\circ(x,\Lambda_{n,r}^\circ(x);\eta_0)[h_t]
\right]
\le0,
\]
where the supremum is over the uniformity indices in the lemma.

For the lower bound, view the equality and inequality restrictions as the closed convex set \(\{0\}^{q_E}\times\mathbb R_-^{q_I}\). Because \(Q_x\) has full row rank, the map \(d\mapsto Q_xd\) is onto the active residual space; hence there is a direction solving \(Q_xd=(0,-\mathbf 1)\), where the zero coordinates correspond to equalities and the negative coordinates to active inequalities. This gives Robinson's constraint qualification, equivalently the Mangasarian--Fromovitz condition, with constants independent of the retained problem. The active-rank lemma gives singleton multiplier sets. Lemma~\ref{lem:optimizer_localization} gives the required restricted inf-compactness: for perturbations \(\eta_t\) with \(h_t\in K\), every optimizer of a retained local problem remains in the retained compact optimizer neighborhoods. The preceding exact repair verifies inner semicontinuity of the feasible set in each tangent direction. The objective and restrictions are \(C^1\) in the coordinate vector on the retained neighborhoods, and \eqref{eq:objective_uniform_expansion} and \eqref{eq:restriction_uniform_expansion} identify their input derivatives along \(h_t\). Thus the hypotheses of the directional value theorem for finite-dimensional constrained programs, \citet[Theorem~4.26]{BonnansShapiro2000}, hold for each retained problem, and the scalar directional derivative of the signed local value is the displayed minimum of the Lagrangian input derivatives over \(S_{n,r}^\circ\).

It remains to justify that the remainder in the lower estimate is common across the sequence of retained problems. If no common modulus existed, then for some \(\varepsilon>0\) there would be retained problems, directions \(h_j\in K\), and \(t_j\downarrow0\) for which the lower error exceeded \(\varepsilon t_j\). Because \(\bar k<\infty\) and the active-row count is bounded by the coordinate dimension, pass to a subsequence on which the coordinate dimension, equality-row count, active-inequality count, support-bound pattern, and one-sided boundary conventions are fixed. Translate one baseline optimizer in each selected problem to the origin. The common margins allow all inactive rows to be discarded on a common closed coordinate ball, while the active objective and active residual maps extend to that ball with uniformly bounded values, uniformly bounded derivatives, and a common derivative modulus. Arzel\`a--Ascoli applied to the maps and to their derivatives gives \(C^1\) limits on the ball; the limiting derivative is the derivative of the limiting map by the fundamental theorem of calculus. Compactness of \(K\) gives \(h_j\to h\) along a further subsequence, and the uniformly bounded multipliers have a convergent subsequence. The common moduli for \(x\mapsto\dot\Phi_{k,x}(h)\) and \(x\mapsto\dot C_{n,r,x}(h)\) give uniform limits of the input-derivative maps as well.

The limiting active matrix still has full row rank, so the limiting problem satisfies Robinson's constraint qualification and has the limiting unique multiplier. Feasible sets for the selected finite problems converge to the limiting feasible set in Hausdorff distance on the coordinate ball: limits of feasible sequences are feasible by uniform convergence, and any feasible point of the limit can be repaired to nearby feasible points of the selected problems using the same right-inverse contraction applied to the rows active at that point. Uniform convergence of the objectives gives convergence of local values and of optimizer sets. Applying \citet[Theorem~4.26]{BonnansShapiro2000} to the limiting problem yields the same lower directional estimate as the limit of the selected Lagrangian derivatives. Transferring this estimate back through the uniform \(C^1\) convergence contradicts the assumed \(\varepsilon t_j\) lower-error violation. Hence the lower and upper estimates share a common modulus \(\rho_K(t)\downarrow0\). Finally, the multiplier and input-derivative maps are uniformly bounded and linear in \(h\), so replacing \(h_t\) by \(h\) changes the derivative term by at most \(C_K\|h_t-h\|\).
\end{proof}

\begin{lemma}\label{lem:phi_n-ii-iv}
Under Assumptions~\ref{ass:support},~\ref{ass:restriction_geometry},~\ref{ass:objective},~\ref{ass:finite_complexity},~\ref{ass:feasibility_rank}, and~\ref{ass:endpoint_rank}, fix \(\circ\in\{\inf,\sup\}\) and let \(\phi_n^\circ(\eta):=V_{J_n}^{k_n,\circ}(\eta)\).
\begin{enumerate}
\item[(i)] For all sufficiently large \(n\), \(\phi_n^\circ\) is Hadamard directionally differentiable at \(\eta_0\) tangentially to \(\mathbb D_0\); write its derivative as \((\phi_n^\circ)'_{\eta_0}\).
\item[(ii)] For every compact \(K\subset\mathbb D_0\), there exists \(N_K\) such that
\[
\sup_{n\ge N_K}\sup_{h\in K}
\left|
\frac{\phi_n^\circ(\eta_0+t h)-\phi_n^\circ(\eta_0)}{t}
-(\phi_n^\circ)'_{\eta_0}(h)
\right|\to0
\]
as \(t\downarrow0\), for \(t\) small enough that every displayed input lies in the local parameter domain.
\item[(iii)] There are \(N_L\) and \(L<\infty\), independent of \(n\) and \(\circ\), such that \(|(\phi_n^\circ)'_{\eta_0}(h)-(\phi_n^\circ)'_{\eta_0}(h')|\le L\|h-h'\|\) for all \(n\ge N_L\) and \(h,h'\in\mathbb D_0\).
\item[(iv)] If the reduced endpoint problem has a unique optimizer measure and all retained local parameterizations of that measure give the same linear envelope derivative, then \(\phi_n^\circ\) is Hadamard differentiable. The agreement condition is automatic when the optimizer has no atom on a partition boundary at which a retained row changes form.
\end{enumerate}
\end{lemma}

\begin{proof}[Proof of Lemma~\ref{lem:phi_n-ii-iv}]
Fix \(n\) large enough that the conclusions of Lemmas~\ref{lem:optimizer_localization} and~\ref{lem:uniform_local_sensitivity} hold. For endpoint \(\circ\), let \(w_n^\circ:=\sigma_\circ\phi_n^\circ\) be the signed endpoint value, so both lower and upper endpoints are written as minimization problems. The compactness clause of Assumption~\ref{ass:endpoint_rank} gives a compact set of compatible baseline optimizer representations. Choose a finite family \(\mathfrak C_n^\circ\) of retained local parameterizations whose neighborhoods cover this set. The number of such neighborhoods may depend on \(n\), but each member satisfies the same mass, slack, rank, and expansion bounds.

Lemma~\ref{lem:optimizer_localization} implies that, for \(h\) in a fixed compact subset of \(\mathbb D_0\) and \(t\) small, every optimizer of the reduced signed endpoint at \(\eta_0+th\) lies in one of these neighborhoods. Hence
\[
w_n^\circ(\eta_0+th)=\min_{r\in\mathfrak C_n^\circ}v_{n,r}^\circ(\eta_0+th),
\qquad
w_n^\circ(\eta_0)=v_{n,r}^\circ(\eta_0)\quad\text{for every }r\in\mathfrak C_n^\circ.
\]
Lemma~\ref{lem:uniform_local_sensitivity} gives, uniformly over \(r\in\mathfrak C_n^\circ\),
\[
v_{n,r}^\circ(\eta_0+th)
=v_{n,r}^\circ(\eta_0)
+t\ell_{n,r}^\circ(h)+t e_{n,r,t}(h),
\qquad
\sup_{r,h}|e_{n,r,t}(h)|\to0,
\]
where \(\ell_{n,r}^\circ(h):=\min_{x\in S_{n,r}^\circ}D_\eta L_{n,r}^\circ(x,\Lambda_{n,r}^\circ(x);\eta_0)[h]\). Since \(|\min_r(a_r+e_r)-\min_ra_r|\le\max_r|e_r|\) for every finite family, the signed derivative exists and equals \(\min_{r\in\mathfrak C_n^\circ}\ell_{n,r}^\circ(h)\). Therefore
\[
(\phi_n^\circ)'_{\eta_0}(h)
=
\sigma_\circ\min_{r\in\mathfrak C_n^\circ}\min_{x\in S_{n,r}^\circ}
D_\eta L_{n,r}^\circ(x,\Lambda_{n,r}^\circ(x);\eta_0)[h].
\]
The same display, with the common remainder from Lemma~\ref{lem:uniform_local_sensitivity}, gives the compact-uniform expansion in part~\textup{(ii)} and the Hadamard directional differentiability in part~\textup{(i)}, including paths \(h_t\to h\).

The input-derivative maps have uniformly bounded operator norms, and the multipliers are uniformly bounded. Hence each \(\ell_{n,r}^\circ\) is Lipschitz on \(\mathbb D_0\) with a constant independent of \(n\), \(r\), and \(\circ\). A finite minimum of functions with a common Lipschitz constant has the same Lipschitz constant, and multiplication by \(\sigma_\circ\) does not change it. This proves part~\textup{(iii)}.

Under the condition in part~\textup{(iv)}, all retained local parameterizations that can be active at the baseline represent the same optimizer measure and have the same linear envelope derivative. The finite minimum in the displayed formula is then that single linear map. Part~\textup{(ii)} therefore gives a Hadamard expansion with a continuous linear derivative, which is Hadamard differentiability of \(\phi_n^\circ\).
\end{proof}

\subsection{Proofs of the main asymptotic propositions}\label{subsec:appendix_main_asymptotic_proofs}

\begin{proof}[Proof of Proposition~\ref{prop:hdd_limit_phi_n}]
\smallskip\noindent\emph{Step 1. Construct the coordinatewise derivative.}
For \(\circ\in\{\inf,\sup\}\), write \(\phi_n^\circ(\eta):=V_{J_n}^{k_n,\circ}(\eta)\) and \(\delta_n^\circ:=\sup_{\eta\in\mathcal N_0}|\phi_n^\circ(\eta)-V_\infty^\circ(\eta)|\). Lemma~\ref{lem:unif_kn_vs_inf} and Lemma~\ref{lem:tail-measure} imply \(\delta_n^\circ\to0\); in fact the convergence is \(o(n^{-1/2})\), but only uniform convergence is needed for differentiability. For a compact \(K\subset\mathbb D_0\), let
\[
\omega_{K,N}^\circ(t)
:=
\sup_{n\ge N}\sup_{h\in K}
\left|
\frac{\phi_n^\circ(\eta_0+t h)-\phi_n^\circ(\eta_0)}{t}
-(\phi_n^\circ)'_{\eta_0}(h)
\right|,
\]
where \(N\) is large enough for Lemma~\ref{lem:phi_n-ii-iv}\textup{(ii)}. Then \(\omega_{K,N}^\circ(t)\to0\) as \(t\downarrow0\).

Fix \(h\in\mathbb D_0\), and take \(N_h\) corresponding to \(K=\{h\}\). For \(n,m\ge N_h\), set \(\Delta_{n,m}^\circ:=\delta_n^\circ+\delta_m^\circ\). If \(\Delta_{n,m}^\circ=0\), then \(\phi_n^\circ\) and \(\phi_m^\circ\) agree with \(V_\infty^\circ\) on \(\mathcal N_0\), so their directional derivatives at \(\eta_0\) in direction \(h\) are equal. Otherwise set \(t_{n,m}:=\min(\sqrt{\Delta_{n,m}^\circ},r_0/(1+\|h\|))\). For all large \(n,m\), this number is positive, tends to zero, and \(\eta_0+t_{n,m}h\in\mathcal N_0\). Since \(\sup_{\eta\in\mathcal N_0}|\phi_n^\circ(\eta)-\phi_m^\circ(\eta)|\le\Delta_{n,m}^\circ\), Lemma~\ref{lem:phi_n-ii-iv}\textup{(ii)} gives \(|(\phi_n^\circ)'_{\eta_0}(h)-(\phi_m^\circ)'_{\eta_0}(h)|\le2\omega_{\{h\},N_h}^\circ(t_{n,m})+2\Delta_{n,m}^\circ/t_{n,m}\) whenever \(\Delta_{n,m}^\circ>0\). The right side converges to zero. Thus \((\phi_n^\circ)'_{\eta_0}(h)\) is Cauchy in \(\mathbb R\), and we define \((V_\infty^\circ)'_{\eta_0}(h):=\lim_{n\to\infty}(\phi_n^\circ)'_{\eta_0}(h)\). The Lipschitz bound in Lemma~\ref{lem:phi_n-ii-iv}\textup{(iii)} passes to the limit, so \((V_\infty^\circ)'_{\eta_0}\) is Lipschitz on \(\mathbb D_0\), with a constant that is common to both endpoint coordinates.

\smallskip\noindent\emph{Step 2. Verify the Hadamard expansion of the full value function.}
Fix \(h\in\mathbb D_0\), let \(t_k\downarrow0\), and let \(h_k\to h\) in \(\mathbb D_0\), with \(\eta_0+t_kh_k\) in the local parameter domain. The set \(K:=\{h\}\cup\{h_k:k\ge1\}\) is compact. Choose \(N_K\) as in Lemma~\ref{lem:phi_n-ii-iv}\textup{(ii)} and then integers \(n_k\to\infty\) satisfying \(n_k\ge N_K\) and \(\delta_{n_k}^\circ\le t_k^2\). For large \(k\), \(\eta_0+t_kg\in\mathcal N_0\) for every \(g\in K\). Therefore
\[
\begin{aligned}
&\left|
\frac{V_\infty^\circ(\eta_0+t_kh_k)-V_\infty^\circ(\eta_0)}{t_k}
-(V_\infty^\circ)'_{\eta_0}(h)
\right| \\
&\quad\le
\frac{2\delta_{n_k}^\circ}{t_k}
+
\omega_{K,N_K}^\circ(t_k)
+
\left|(\phi_{n_k}^\circ)'_{\eta_0}(h_k)-(V_\infty^\circ)'_{\eta_0}(h)\right|.
\end{aligned}
\]
The first term is at most \(2t_k\), the second tends to zero by Lemma~\ref{lem:phi_n-ii-iv}\textup{(ii)}, and the last is bounded by \(L\|h_k-h\|+|(\phi_{n_k}^\circ)'_{\eta_0}(h)-(V_\infty^\circ)'_{\eta_0}(h)|\), which tends to zero by Step~1 and the common Lipschitz bound. Hence \(V_\infty^\circ\) is Hadamard directionally differentiable at \(\eta_0\) tangentially to \(\mathbb D_0\). Stacking the lower and upper coordinates gives the derivative \(V'_{\eta_0}:\mathbb D_0\to\mathbb R^2\), and the coordinatewise Lipschitz bounds imply Lipschitz continuity of the stacked derivative.

\smallskip\noindent\emph{Step 3. Treat the uniqueness case.}
If the uniqueness and local-representation agreement condition in the proposition statement holds for an endpoint coordinate, then Lemma~\ref{lem:phi_n-ii-iv}\textup{(iv)} gives a linear derivative \((\phi_n^\circ)'_{\eta_0}\) for all sufficiently large \(n\). The pointwise limit \((V_\infty^\circ)'_{\eta_0}\) is therefore linear. If the condition holds for both endpoint coordinates, the stacked derivative \(V'_{\eta_0}\) is a continuous linear map on \(\mathbb D_0\). The Hadamard directional expansion in Step~2 is then a Hadamard expansion with a continuous linear derivative, so \(V_\infty\) is Hadamard differentiable at \(\eta_0\) tangentially to \(\mathbb D_0\).
\end{proof}

\begin{proof}[Proof of Proposition~\ref{prop:asymptotics1}]
Let \(\mathcal E_n:=\{\hat\eta\in\mathcal N_0\}\). Since \(\sqrt n(\hat\eta-\eta_0)=O_p(1)\) and \(\mathcal N_0\) is a fixed neighborhood of \(\eta_0\), \(P(\mathcal E_n)\to1\). On this event, write the implemented endpoint as \(\hat V^{(n)}(\hat\eta)=V_{J_n}^{n,n}(\hat\eta)+\varepsilon_n\). The exact decomposition is
\begin{align*}
\hat V^{(n)}(\hat\eta)-V_\infty(\eta_0)
&=\varepsilon_n
 +\bigl[V_{J_n}^{n,n}(\hat\eta)-V_{J_n}^{k_n,n}(\hat\eta)\bigr]
 +\bigl[V_{J_n}^{k_n,n}(\hat\eta)-V_{J_n}^{k_n}(\hat\eta)\bigr]\\
&\quad
 +\bigl[V_{J_n}^{k_n}(\hat\eta)-V_{J_n}(\hat\eta)\bigr]
 +\bigl[V_{J_n}(\hat\eta)-V_\infty(\hat\eta)\bigr]
 +\bigl[V_\infty(\hat\eta)-V_\infty(\eta_0)\bigr].
\end{align*}
The implementation error is \(o_p(n^{-1/2})\) by Assumption~\ref{ass:implementation}. Lemma~\ref{lem:unif_kn_vs_inf} gives nonemptiness of the \(k_n\)-support feasible set and the bound \(\|V_{J_n}^{k_n}(\hat\eta)-V_{J_n}(\hat\eta)\|_2=o_p(n^{-1/2})\), uniformly on \(\mathcal N_0\).

The block masses implied by the included step equalities are on the \(1/n\) grid by Lemma~\ref{lem:grid_compatibility}, or by the projection allowed in Assumption~\ref{ass:implementation}. Since \(n\ge k_n\) eventually, Lemma~\ref{lem:quant} with \((N_1,N_2)=(k_n,n)\) gives \(\|V_{J_n}^{k_n,n}(\hat\eta)-V_{J_n}^{k_n}(\hat\eta)\|_2=O_p(n^{-1})\) and nonemptiness of the corresponding grid-feasible set. This feasible set is contained in the \((n,n)\) grid-feasible set. Assumption~\ref{ass:finite_complexity}, applied with \(N=n\), gives \(\|V_{J_n}^{n,n}(\hat\eta)-V_{J_n}^{k_n,n}(\hat\eta)\|_2\le r_n=o(n^{-1/2})\).

Finally, Lemma~\ref{lem:tail-measure} gives \(\|V_{J_n}(\hat\eta)-V_\infty(\hat\eta)\|_2=o_p(n^{-1/2})\), uniformly on \(\mathcal N_0\). Thus \(\hat V^{(n)}(\hat\eta)-V_\infty(\eta_0)=V_\infty(\hat\eta)-V_\infty(\eta_0)+o_p(n^{-1/2})\). Assumption~\ref{ass:input_process} gives \(\sqrt n(\hat\eta-\eta_0)\Rightarrow Z\) in \(\mathbb H\), with \(Z\in\mathbb D_0\) almost surely. Proposition~\ref{prop:hdd_limit_phi_n} gives Hadamard directional differentiability of \(V_\infty\) at \(\eta_0\) tangentially to \(\mathbb D_0\). The directional delta method, \citet[Theorem~2.1]{FangSantos2019}, yields \(\sqrt n\,[V_\infty(\hat\eta)-V_\infty(\eta_0)]\Rightarrow V'_{\eta_0}(Z)\). Combining this limit with the preceding \(o_p(n^{-1/2})\) reduction gives the stated limit for \(\hat V^{(n)}(\hat\eta)\).
\end{proof}

\begin{proof}[Proof of Proposition~\ref{prop:bootstrap}]
\smallskip\noindent\emph{Step 1. Reduce the bootstrap root to the full value function.}
Write \(\hat\eta_n:=\hat\eta\), and let \(\mathcal E_{m,n}:=\{\hat\eta_n\in\mathcal N_0,\ \eta_m^*\in\mathcal N_0\}\). Assumption~\ref{ass:input_process} implies \(P(\hat\eta_n\in\mathcal N_0)\to1\) and \(P^*(\eta_m^*\in\mathcal N_0)\to1\) in probability. On \(\mathcal E_{m,n}\), Lemma~\ref{lem:unif_kn_vs_inf} gives \(\|V_{J_n}^{k_n}(\eta_m^*)-V_{J_n}(\eta_m^*)\|_2=o(n^{-1/2})\) uniformly, and therefore this term is \(o_{P^*}(m^{-1/2})\) after multiplication by \(\sqrt m\), because \(m/n\to0\). The bootstrap step-equality block masses are on the \(1/m\) grid, either exactly or after the negligible projection in Assumption~\ref{ass:implementation}. Since \(m\to\infty\) and \(\sup_nk_n<\infty\), Lemma~\ref{lem:quant} gives \(\|V_{J_n}^{k_n,m}(\eta_m^*)-V_{J_n}^{k_n}(\eta_m^*)\|_2=O(m^{-1})=o(m^{-1/2})\). Assumption~\ref{ass:finite_complexity} gives \(\|V_{J_n}^{m,m}(\eta_m^*)-V_{J_n}^{k_n,m}(\eta_m^*)\|_2\le r_n=o(n^{-1/2})=o(m^{-1/2})\), again using \(m/n\to0\). Assumption~\ref{ass:implementation} gives \(\|\varepsilon_m^*\|_2=o_{P^*}(m^{-1/2})\). Combining the four bounds gives \(\sqrt m\|\hat V_m^*-V_{J_n}(\eta_m^*)\|_2=o_{P^*}(1)\). The same deterministic bounds applied to the centering statistic, with sample grid size \(n\) and implementation error \(\varepsilon_n\), give \(\sqrt m\|\hat V^{(n)}(\hat\eta_n)-V_{J_n}(\hat\eta_n)\|_2=o_p(1)\), because every sample approximation error is \(o_p(n^{-1/2})\) or \(O_p(n^{-1})\) and \(m/n\to0\). Lemma~\ref{lem:tail-measure} gives \(\sup_{\eta\in\mathcal N_0}\|V_{J_n}(\eta)-V_\infty(\eta)\|_2=o(n^{-1/2})\), so \(\sqrt m\) times this tail error is \(o(1)\). Therefore \(R_m^*=\sqrt m\,[V_\infty(\eta_m^*)-V_\infty(\hat\eta_n)]+o_{P^*}(1)\) conditionally in probability.

\smallskip\noindent\emph{Step 2. Apply the numerical delta method for the directionally differentiable map.}
Set \(a_m:=m^{-1/2}\), \(H_m^*:=\sqrt m(\eta_m^*-\hat\eta_n)\), and \(\widehat D_m(h):=[V_\infty(\hat\eta_n+a_mh)-V_\infty(\hat\eta_n)]/a_m\) when the displayed inputs lie in the local parameter domain, extending the map outside that domain arbitrarily. The extension is immaterial because \(\eta_m^*=\hat\eta_n+a_mH_m^*\) is admissible on \(\mathcal E_{m,n}\). Proposition~\ref{prop:hdd_limit_phi_n} gives Hadamard directional differentiability of \(V_\infty\) at \(\eta_0\), and \(\|\hat\eta_n-\eta_0\|/a_m=O_p(\sqrt{m/n})=o_p(1)\). Thus the centering point is closer to \(\eta_0\) than the numerical-difference scale. The compact-uniform numerical derivative result for Hadamard directionally differentiable maps, \citet[Supplementary Appendix, Lemma~S.3.8]{FangSantos2019}, implies that \(\widehat D_m\) converges to \(V'_{\eta_0}\) along admissible tangent sequences, uniformly on compact subsets of \(\mathbb D_0\) in probability. Assumption~\ref{ass:input_process} gives \(H_m^*\Rightarrow Z\) conditionally in probability in \(\mathbb H\), with \(Z\in\mathbb D_0\) almost surely. The conditional numerical delta method, \citet[Theorem~3.2]{FangSantos2019}, yields \(\widehat D_m(H_m^*)\Rightarrow V'_{\eta_0}(Z)\) conditionally in probability. Since \(\widehat D_m(H_m^*)=\sqrt{m}\,[V_\infty(\eta_m^*)-V_\infty(\hat\eta_n)]\) on the admissible event, Step~1 and conditional Slutsky's theorem prove the \(m\)-out-of-\(n\) claim.

\smallskip\noindent\emph{Step 3. The ordinary bootstrap under full differentiability.}
Under the additional assumptions in the second sentence of Proposition~\ref{prop:bootstrap}, repeat Step~1 with \(m=n\) and \(\eta_n^*\) in place of \(\eta_m^*\). The finite-complexity error is \(r_n=o(n^{-1/2})\), the mass-grid error is \(O(n^{-1})\), Lemmas~\ref{lem:unif_kn_vs_inf} and~\ref{lem:tail-measure} give the remaining \(o(n^{-1/2})\) deterministic errors, and Assumption~\ref{ass:implementation} makes the ordinary-bootstrap numerical and projection errors \(o_{P^*}(n^{-1/2})\). Hence \(\sqrt n\,[\hat V_n^*-\hat V^{(n)}(\hat\eta_n)]=\sqrt n\,[V_\infty(\eta_n^*)-V_\infty(\hat\eta_n)]+o_{P^*}(1)\). The ordinary-bootstrap clause of Assumption~\ref{ass:input_process} gives \(\sqrt n(\eta_n^*-\hat\eta_n)\Rightarrow Z\) conditionally in probability. Because \(V_\infty\) is Hadamard differentiable at \(\eta_0\) tangentially to \(\mathbb D_0\), the ordinary bootstrap delta method, \citet[Theorem~3.1]{FangSantos2019}, gives the stated conditional weak limit. Full differentiability is used only in this ordinary-bootstrap step; the first part uses the \(m\)-out-of-\(n\) numerical derivative because the derivative may be nonlinear.
\end{proof}

\section{Additional imputation exercises for Application 1 (wealth data) }\label{app:additional_wealth_imputation}

We conduct an additional parametric imputation exercise for Application 1. The purpose is to compare the sharp bounds with inequality estimates obtained when the interval observations are completed using a smooth parametric model rather than midpoint or hot-deck imputation. For simplicity, we focus on the narrow savings definition. Let $y_i$ denote the (potentially latent) savings value and let $[\underline a_i,\overline a_i]$ denote the observed interval for observation \(i\). Exact observations are treated as degenerate intervals. We fit a parametric distribution \(F_\theta\) to the mixed exact and interval-censored data by maximum likelihood. Exact observations contribute density terms \(f_\theta(y_i)\), while interval observations contribute probability masses
\(
F_\theta(\overline a_i)-F_\theta(\underline a_i).
\)
When an upper endpoint equal to \(300{,}000\) is used only as an artificial closure of an open top interval, we also consider a version in which the corresponding observation is treated as right-censored, so that \(F_\theta(\overline a_i)\) is replaced by one. Given the fitted distribution \(F_{\widehat\theta}\), we generate \(R=500\) completed datasets. Exact observations are kept fixed. For each  replication \(r=1,\ldots,R\) we get a completed dataset and  compute the usual sample Gini coefficient. Unlike the bootstrap analysis, this exercise does not resample observations. It isolates only the variation induced by drawing unobserved within-interval values from the fitted parametric distribution.

We implement two versions of the exercise. The first is a like-for-like imputation exercise that retains the artificial upper endpoint \(300{,}000\) as a finite upper bound. Thus all imputed observations remain inside the same finite intervals used in the main analysis. The second treats the top-coded observations as genuinely open-ended and therefore allows draws potentially above \(300{,}000\), with the magnitude of those draws determined entirely by the fitted parametric tail.

\begin{figure}[tbp]
\centering
\begin{minipage}{0.48\textwidth}
\centering
\includegraphics[width=\textwidth]{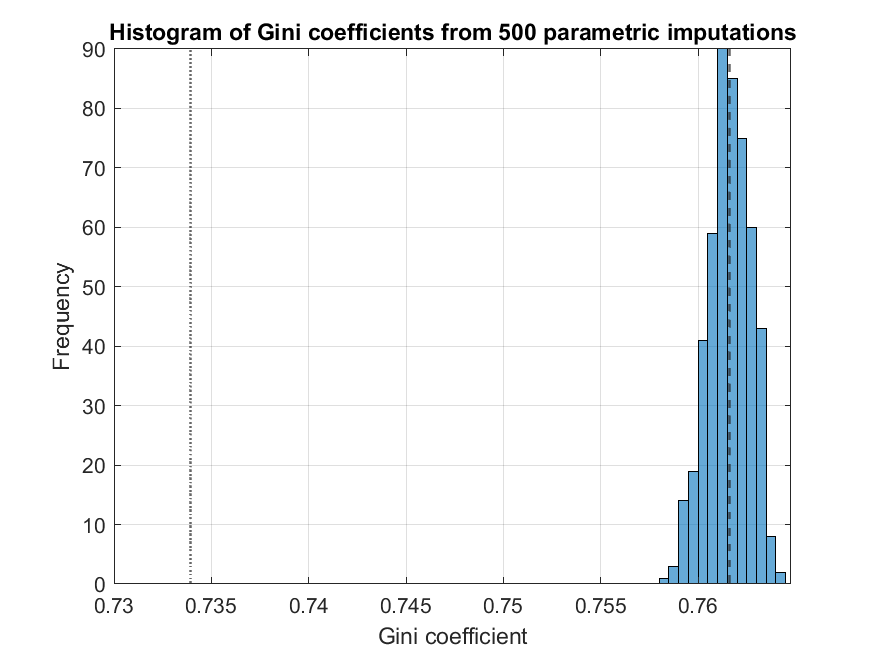}
\caption*{Shifted-lognormal imputation}
\end{minipage}
\hfill
\begin{minipage}{0.48\textwidth}
\centering
\includegraphics[width=\textwidth]{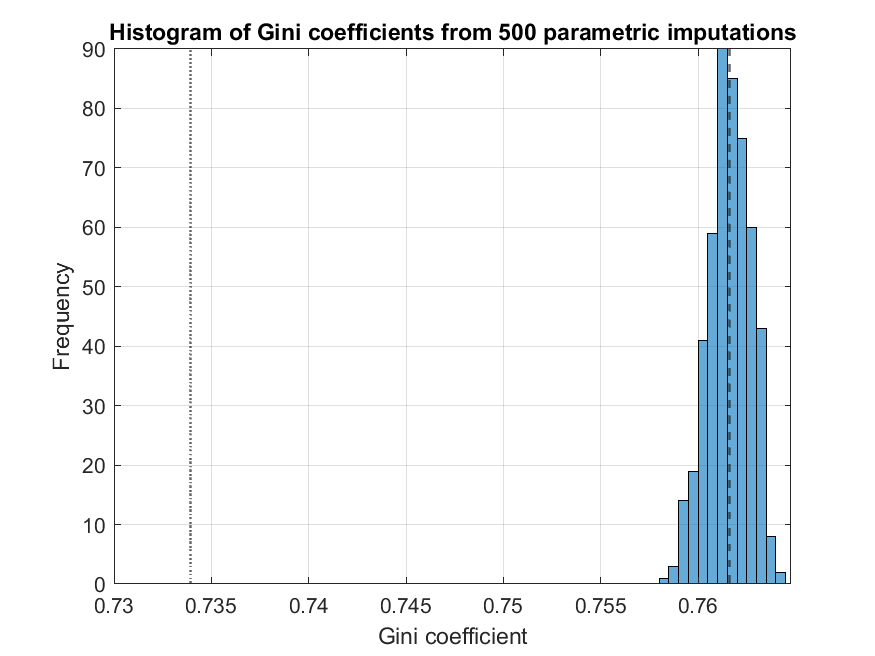}
\caption*{Generalized Pareto imputation}
\end{minipage}
\caption{Wealth application: Distribution of the Gini coefficient for the narrow savings definition across \(R=500\) parametric imputations, retaining the artificial upper endpoint \(300{,}000\) for open-ended intervals.}
 \label{fig:sav_imputationTC} 
\end{figure}

The results for the finite-top version are shown in Figure~\ref{fig:sav_imputationTC}. The resulting Gini coefficients vary over a narrow range. Under the shifted-lognormal specification, the 500 imputed Gini values lie in $[0.7581,0.7644]$, with standard deviation $0.0011$. Under the generalized Pareto specification, they lie in $[0.7558,0.7616]$, with standard deviation $0.0010$. These findings are similar to the hot-deck imputation results: once the top endpoint is kept finite, parametric imputation produces a tight distribution of point estimates relative to the sharp identified range.

\begin{figure}[tbp]
\centering
\begin{minipage}{0.48\textwidth}
\centering
\includegraphics[width=\textwidth]{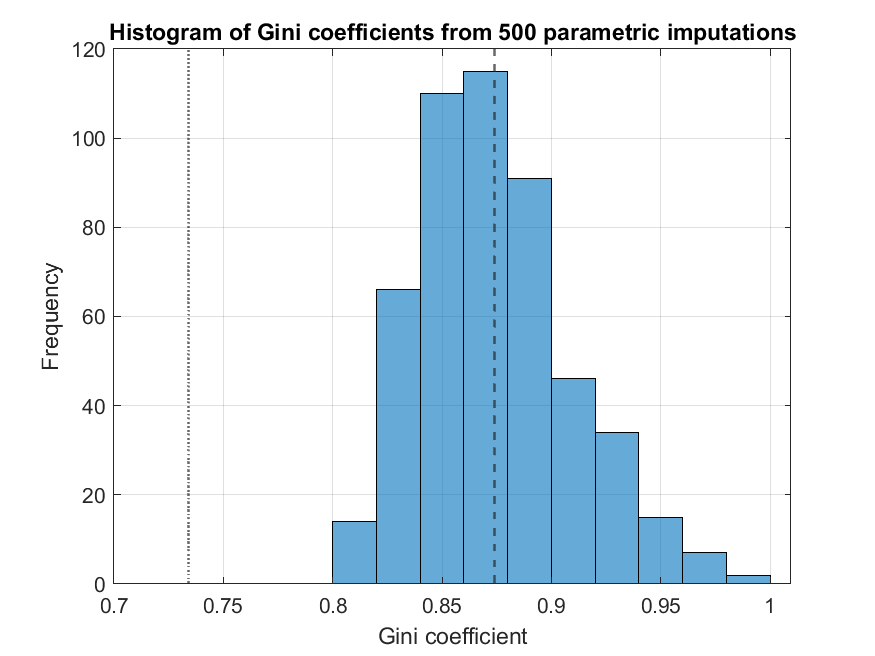}
\caption*{Shifted-lognormal imputation}
\end{minipage}
\hfill
\begin{minipage}{0.48\textwidth}
\centering
\includegraphics[width=\textwidth]{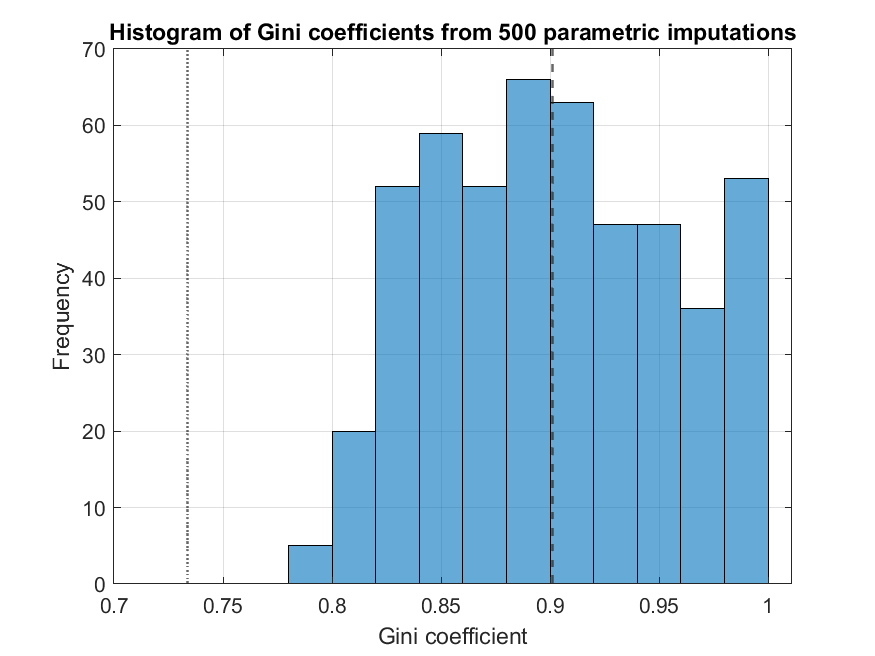}
\caption*{Generalized Pareto imputation}
\end{minipage}
\caption{Wealth application: Distribution of the Gini coefficient for the narrow savings definition across \(R=500\) parametric imputations, treating open-ended intervals as right-censored.}
 \label{fig:sav_imputation} 
\end{figure}

The open-top version gives much more extreme results, as shown in Figure \ref{fig:sav_imputation}. This is not surprising, since treating the top interval as right-censored allows arbitrarily large draws. The fitted distributions imply very dispersed upper tails. In the generalized Pareto specification, the fitted shape parameter implies an infinite mean. In the shifted-lognormal specification, all polynomial moments are finite, but the distribution is heavy-tailed in the sense that its moment generating function is infinite for every positive argument. In our data, the fitted scale parameter is large, ($\widehat\sigma=2.9123$), implying that the 99.9th percentile of \(1+y\) is approximately \(8{,}000\) times its median. As a result, the open-top specification can generate extremely large draws for observations in the top interval and consequently very high imputed Gini coefficients.

We therefore interpret the open-top parametric exercise as a tail-extrapolation diagnostic rather than as our preferred imputation rule. Without auxiliary information on means, higher moments, top shares,  the data do not discipline the conditional distribution within the open upper tail. The resulting estimates are therefore driven by functional-form extrapolation rather than by information contained in the interval data alone. Our preferred parametric comparison is the finite-top version, which completes the data within the same bounded intervals used by the midpoint and hot-deck procedures.

\end{document}